\documentclass{elsarticle}

\usepackage{amsmath,amssymb}

\usepackage[hang,small,bf]{caption}
\usepackage[subrefformat=parens]{subcaption}
\usepackage{hyperref}
\usepackage{ulem}
\usepackage{xcolor}

\journal{Computer Methods in Applied Mechanics and Engineering}

\begin{document}


\begin{abstract}
Reduced-order models (ROMs) that capture changes in fluid systems due to variations in parameters, such as the Reynolds number or the shape of a stationary body placed in the flow, are attracting increasing attention in engineering applications. In this study, we identify linear operators that characterize the behavior of fluid systems across a wide parameter range by using flow field datasets at several representative parameter values. We then comprehensively assess the applicability of ROMs constructed through the interpolation of these operators. Specifically, we consider two intrusive operator-based ROMs: one derived from Galerkin projection and the other based on operator inference using dynamic mode decomposition (DMD).
The performance of these ROMs is evaluated for flows around circular and elliptical cylinders over a range of Reynolds numbers and aspect ratios. The Galerkin-based ROM successfully predicts only the eigenvalue and eigenmode corresponding to the fundamental frequency of the Kármán vortex shedding, while other frequencies are not captured and show a decaying behavior. In contrast, the DMD-based ROM accurately predicts both the fundamental frequency and its higher harmonics.
Furthermore, visualization of the linear operator matrix elements reveals that interpolation fails when the subspace includes bases contaminated by numerical errors. However, by carefully selecting the subspace dimension and reference conditions, it is possible to accurately predict eigenvalues and corresponding modes even under conditions where multiple low-frequency modes exist outside the harmonic structure of the fundamental frequency. These findings underscore the robustness of the DMD-based parametric operator ROM approach.

\end{abstract}

\title{On the conservation of physical properties in operator interpolation of parameterized hydrodynamic systems}

\author[1]{Yuto Nakamura\corref{cor1}
}
\ead{yuto.nakamura.t4@dc.tohoku.ac.jp}
\author[2]{Shintaro Sato
}
\author[3]{Naofumi Ohnishi
}
\affiliation[1]{organization={Department of Aerospace Engineering, Tohoku University},
            city={Sendai},
            postcode={980-8579},
            state={Miyagi},
            country={Japan}}
\cortext[cor1]{Corresponding author}

\date{\today}
\maketitle



\section{\label{sec:level1}Introduction}
A central objective in engineering, physics, and the mathematical sciences is to identify and estimate the dynamics underlying physical phenomena from observational data. In fluid systems, these dynamics are often governed by spatially and temporally dependent linear or nonlinear operators. Analyzing the spectral properties of such operators offers key insights into flow behavior and stability\cite{mezic2005spectral, mypaper_6, freeman_2024, Luchini_2014, Ohmichi_D,Ranjan_2020,mauroy2016global, mypaper_7}. In practice, observations from experiments or simulations are typically obtained at discrete points in space and time under specific physical conditions, such as Reynolds number, Mach number, or boundary conditions\cite{Satomanifold_2024, du5181891interpolation, wang2025kernelridgeregressioncombining, mypaper4, hess2023data, andreuzzi2023dynamic, mfPOD, cylinder4}. This requires a discrete framework to estimate the governing dynamics. While recent advances in data-driven modeling\cite{Satomanifold_2024, mypaper4, POD1, POD15, DMD, fukagata2025compressingfluidflowsnonlinear, staggl2025reducedbasismodelcompressible, shift, freeman_2024} have facilitated the analysis of spatiotemporally discrete data, a general approach for predicting how operators or their spectral characteristics vary with physical parameters (e.g., Reynolds number or geometry) is still lacking. This study contributes to both the theoretical foundation and engineering applications of such predictions by introducing methods for estimating governing operators under varying flow conditions, using operator information obtained from a limited set of reference datasets.

The operator perspective is an important framework in fluid flow, rooted in operator theory first introduced by Koopman in 1931 for the analysis of Hamiltonian systems\cite{koopman1931hamiltonian}. The resulting operator, now known as the Koopman operator, describes the evolution of observables along trajectories of a nonlinear dynamical system, mapping each observable to its future state or its temporal derivative.
 Subsequent studies\cite{koopman1932dynamical, mezic2005spectral, nadkarni1998spectral} have demonstrated that the spectral properties of the Koopman operator encode essential information about the temporal evolution of observables and can uncover coherent structures in complex systems. Since its original formulation, the Koopman framework has been extended in both theoretical and applied contexts, leading to developments such as point-spectrum-based stability analysis\cite{Stuart_1958}, resolvent operator formulations\cite{trefethen1993hydrodynamic, susuki2021koopman}, and modal decomposition techniques\cite{arnoldi1951principle, DMD2}. 

Among various physical systems, fluid systems present a particular challenge due to their high dimensionality and inherent nonlinearity. Consequently, operator-theoretic approaches in fluid mechanics were historically limited to relatively simple configurations, such as flow past a cylinder\cite{GSA1, Mittal_2007, variousshapes1, LSA1, GIANNETTI_2007, barkley2006linear}, boundary layers\cite{malik1990numerical, zhou2021rayleigh}, and channel flows\cite{izumi2000linear}. Over the past two decades, however, these limitations have been largely overcome through advances in computational power and data-driven methods for analyzing complex, high-dimensional systems.
A key development in this context is dynamic mode decomposition (DMD)\cite{DMD, Tu_2013}, a technique that extracts coherent structure from a flow dataset. DMD constructs a best-fit linear operator within an optimally chosen $r$-dimensional subspace of the full dataset, thereby providing an approximation of the Koopman operator acting directly on the observable, typically the state vector itself\cite{brunton2021modern}. Building on this foundation, numerous DMD variants\cite{DMD3} have been introduced to improve robustness, accuracy, and physical interpretability.

DMD also has emerged as a powerful tool for fast prediction of fluid flows\cite{mypaper_6, du5181891interpolation, hess2023data, zhang2017model, lin2023dynamic, TERRAGNI2025113997}. In this framework, the eigenvalues and eigenmodes obtained from DMD are used to characterize and predict the evolution of coherent structures within the flow. These quantities capture the underlying spatio-temporal patterns in the dataset and can be used to reconstruct or forecast the flow field over time. Because the dominant flow dynamics are often governed by a small number of coherent modes, significantly fewer than the total degrees of freedom in full-order numerical simulations, this approach enables efficient modeling and rapid prediction. As a result, reduced-order models (ROMs) based on DMD have found widespread application in fluid dynamics, supporting both flow control strategies and the physical interpretation of unsteady flow phenomena.

The most fundamental type of ROM is the projection-based ROM constructed using proper orthogonal decomposition (POD)\cite{POD0, POD2, POD_snap, Holmes}. POD is a technique that identifies an optimal orthonormal basis, commonly referred to as POD modes or POD basis vectors, that best represent a given dataset of flow fields. These basis vectors span the most efficient low-dimensional subspace for approximating the data. By projecting the governing equations onto this subspace, one obtains a reduced set of equations, a procedure known as Galerkin projection\cite{Satomanifold_2024, shift, POD_Galerkin, Noack_1994, cylinder7, pressure2, 12ICCFD, GP_deane, mypaper3}. When POD basis vectors are used for this projection, the resulting model is commonly referred to as a POD–Galerkin model. Numerical integration of these reduced-order equations allows for efficient prediction of the temporal evolution of the flow field.
However, it is well known that even when the POD basis vectors accurately capture the original dataset, the resulting ROM may fail to reproduce key features of the full-order system, such as oscillation frequencies or amplitudes\cite{Satomanifold_2024, shift, GP_deane, mypaper3, schlegel2015long}. 
 This limitation arises because the Galerkin projection does not necessarily preserve the essential dynamical properties of the original high-dimensional system.

Approaches similar to the POD–Galerkin model have been adopted in other modal-analysis-based and projection-based ROMs. For instance, some projection-based ROMs\cite{mypaper_6, zhang2017model, frame2024spacetimemodelreductionfrequency, cong2019model, Aitzhan_2025} employ non-orthogonal bases instead of the orthonormal basis. In parallel, non-intrusive models that do not directly solve the governing equations, but instead learn system behavior from data, have also been developed\cite{POD_ANN, kani2017dr}.
More recently, non-intrusive ROMs based on the projection framework have gained significant attention. A common approach in these models is to use autoencoders\cite{hinton2006reducing}, including convolutional neural networks (CNNs)\cite{CNN}, to compress high-dimensional flow field data into a low-dimensional latent space and predict its evolution within that space\cite{fukagata2025compressingfluidflowsnonlinear, Murata_2019, hasegawa, Hasegawa_2020, solera2024beta, fukami2024data, Fukami:2023}.

Although non-intrusive models generally achieve higher predictive accuracy than methods based on numerical integration of the projected governing equations, they often lack interpretability, as the underlying physical mechanisms are not explicitly represented. Recent studies have sought to address this limitation by incorporating physically meaningful outputs into non-intrusive frameworks. For example, Fukami and Taira\cite{Fukami:2023} introduced an architecture that predicts lift coefficients directly from the compressed latent space in their analysis of gust-vortex interactions around an airfoil. This addition aids in interpreting the flow dynamics captured within the reduced representation.
The dimensionality reduction employed in such non-intrusive models can be interpreted as a nonlinear form of projection. In this sense, projection-based concepts continue to provide a theoretical foundation, even in state-of-the-art non-intrusive modeling approaches.

An important performance metric for ROMs is their ability to accurately predict flow behavior under conditions different from those used in the training data, such as variations in Reynolds number\cite{Satomanifold_2024, hasegawa} or object geometry\cite{Hasegawa_2020, mypaper2}. However, in the POD–Galerkin model, accurate prediction under such untrained conditions is generally difficult\cite{shift, GP_deane}. This limitation stems from the fact that POD basis vectors are constructed to optimally represent flow fields under specific training conditions, but are not necessarily suitable for other parameter values. 
To address this challenge, several studies have proposed interpolating subspaces, typically represented by POD basis vectors, across different flow conditions\cite{Satomanifold_2024, ELOMARI2025113564, lieu2007aerodynamic, amsallem2008interpolation}, using techniques such as Grassmann or Stiefel manifold interpolation\cite{zimmermann2017matrix, zimmermann2019manifold}. The goal is to construct new basis vectors or a subspace that captures the essential flow features at intermediate or untrained parameter values. By projecting the governing equations onto the interpolated subspace via Galerkin projection and integrating the resulting reduced-order equations in time, the flow field under new conditions can be predicted.
While this method retains physical interpretability by maintaining a direct link to the governing equations, it remains unclear how subspace interpolation quantitatively affects the structure of the reduced operators or the resulting dynamics. Most existing studies assess predictive performance based on the outcome of time integration, without examining whether the reduced operators themselves are consistent with those of the original full-order system.

In this paper, the focus is not on prediction through time integration, but rather on how the properties of operators in subspaces derived from the governing equations change under manifold interpolation.
Focusing on operator-based methods and motivated by the success of subspace interpolation in the POD–Galerkin model, several studies have extended this concept to DMD-based approaches\cite{du5181891interpolation, hess2023data, duan2024non, andreuzzi2023dynamic}. In this context, interpolating DMD modes and their associated eigenvalues provides direct insight into both the spatial structure and temporal evolution of the flow\cite{du5181891interpolation}. However, because DMD modes are inherently determined by the linear operator that approximates the system’s dynamics, such interpolation should be interpreted as a form of operator interpolation.
For instance, Hess et al.\cite{hess2023data} proposed a method for interpolating operators obtained via DMD under different training conditions. Their approach involves interpolating both the subspace that optimally represents the data in the DMD sense and the reduced operator projected onto this subspace. Related strategies have also been proposed in the context of non-intrusive modeling\cite{andreuzzi2023dynamic, duan2024non}.

Since the eigenvalues and eigenvectors of system operators encode fundamental physical properties of the flow within the Koopman operator framework, interpreting the physical meaning of these spectral components can yield valuable insights into the underlying system dynamics in operator-based ROMs. A major strength of ROMs such as POD and DMD lies in this interpretability. Indeed, recent developments in highly accurate non-intrusive models have often been informed by insights derived from these modal frameworks. Consequently, in operator-based ROMs, the ability to retain and interpret physical properties from interpolated operators is expected to play a critical role in building more robust and reliable models.
Despite this potential, prior studies on operator-based ROMs\cite{du5181891interpolation, hess2023data, duan2024non, andreuzzi2023dynamic} have focused primarily on the reconstruction of flow fields using modal information, with limited attention to whether the physical properties encoded in the operators, such as stability characteristics or energy transfer mechanisms, are preserved under interpolation.

In this study, we investigate whether the physical properties originally encoded in the system operator can be preserved through interpolation. In the context of fluid dynamics, such properties include the effects of viscosity, nonlinear energy transfer, and energy cascades from fundamental frequencies to their harmonics\cite{freeman_2024, mypaper_6}. We consider two types of operator-based ROMs. The first is based on interpolation of low-dimensional operators obtained by projecting the governing equations using the POD–Galerkin framework. The second approach involves interpolation of low-dimensional linear operators derived from DMD, formulated as an extension of the methodology proposed by Hess et al.\cite{hess2023data}. 
From an operator-theoretic standpoint, the POD–Galerkin-based model emphasizes consistency with the governing equations, whereas the DMD-based model prioritizes optimal approximation of the dataset.
By comparing these two types of ROMs, we examine how the method of constructing reduced operators influences their physical properties. This comparison aims to address the fundamental question of which operator construction approach is more suitable for developing physically meaningful and reliable operator-based ROMs.

The comparison of operator-based ROMs is conducted for three flow scenarios. The first scenario involves the growth or decay of flow oscillations about a steady base flow around a two-dimensional circular cylinder. This case serves as a classical testbed for stability analysis\cite{GSA1, shift, mypaper_6, Mittal_2007, Mittal_2005, GIANNETTI_2007, barkley2006linear, Ohmichi_D, mypaper_7}, where operator-theoretic methods have been extensively applied in fluid dynamics.
The second scenario considers periodic flow around a two-dimensional circular cylinder. This configuration is a widely used benchmark in ROM studies and differs from the first scenario in that nonlinear effects give rise to harmonics of the fundamental frequency\cite{shift, mypaper_6, Satomanifold_2024, mypaper3, GP_deane, Hasegawa_2020, zhang2017model, du5181891interpolation, 12ICCFD}. In both the first and second scenarios, interpolation is performed along the Reynolds number.
The third scenario focuses on flows around elliptical cylinders with varying aspect ratios. In these cases, wake vortex streets may emerge that are not harmonically related to the fundamental frequency, resulting in qualitative changes in the operator eigenvalues and eigenvectors. These structures, identified in previous studies as secondary vortex streets\cite{taneda1959downstream, Jiang_2019, Johnson_2004, shi2020wakes, variousshapes2}, are characterized by low-frequency oscillations in the wake, often occurring at frequencies below half of the fundamental frequency.

The remainder of this paper is organized as follows. 
Section~\ref{fundamental} outlines the fundamental methodologies employed in this study, including POD, DMD, Galerkin projection, and subspace interpolation techniques (Grassmann manifold interpolation).
Section~\ref{ROM} introduces two operator interpolation approaches constructed based on the methods presented in Section~\ref{fundamental}.
Section~\ref{result} presents a comparative analysis of the ROMs across the three flow scenarios described earlier.

\section{Review of fundamental method and its theory}\label{fundamental}
This section provides a brief overview of a widely used method from previous studies to aid the reader in understanding the methods introduced in this paper.
\subsection{POD}
The goal of POD is to extract $r$ orthonormal basis vectors $\boldsymbol{\phi}_i(\boldsymbol{x})\ (i = 1, 2, \cdots, r) \in \mathbb{R}^{N}$ that optimally represent the given flow dataset $\boldsymbol{u}(\boldsymbol{x}, t_j)\ (j = 1, 2, \cdots, M) \in \mathbb{R}^{N}$, where $M$ is the number of snapshots in the dataset and $r < M$.
Here, $N$ denotes the dimension of the vector space, whose inner product is defined as
\begin{equation}
\begin{split}
 \langle \,\boldsymbol{a}(\boldsymbol{x}), \boldsymbol{b}(\boldsymbol{x}) \rangle = \{\boldsymbol{a}(\boldsymbol{x})\}^{\top}J \boldsymbol{b}(\boldsymbol{x}) = \sum_{j=1}^n \sum_{i=1}^{n_d} {a}_i (\boldsymbol{x}_j){b}_i(\boldsymbol{x}_j) \Delta x_j,
 \end{split}
\end{equation}
where $\langle \cdot, \cdot \rangle$ denotes the inner product, $n$ is the number of grid points, $n_d$ is the number of state variables, $\boldsymbol{x}_j$ is the coordinate of the $j$th grid point, $\Delta x_j$ is the Jacobian at that point,  $J \in \mathbb{R}^{N \times N}$ is the weight matrix derived from inner product, and superscript $\top$ denotes transpose. Here, $a_i(\boldsymbol{x}_j)$ and $b_i(\boldsymbol{x}_j)$ are elements of the $N$-dimensional vectors $\boldsymbol{a}(\boldsymbol{x})$ and $\boldsymbol{b}(\boldsymbol{x})$, with the subscripts $i$ and $j$ indicating the $i$th component of the state variable at the $j$th grid point. The vector space dimension is given by $N = n \times n_d$. For example, if the velocity vector from a two-dimensional numerical simulation is used as the dataset (state variable), then $n_d = 2$. Mathematically, the set of orthonormal bases obtained from POD is given by
\begin{equation}
 U = [\boldsymbol{\phi}_1, \, \boldsymbol{\phi}_2, \, \cdots \,, \boldsymbol{\phi}_r]  \in \mathbb{R}^{N \times r},  
\end{equation}
which solves the following minimization problem:
\begin{equation}
 U = \underset{U} {\operatorname{argmin}}  \sum^M_{j=1} \left\| \boldsymbol{u}(\boldsymbol{x}, t_j) -  \sum^r_{k=1} \langle \boldsymbol{u}(\boldsymbol{x}, t_j) , \boldsymbol{\phi}_k \rangle\boldsymbol{\phi}_k(\boldsymbol{x})\right\|^2,
\end{equation}
where $\| \cdot \|$ denotes the vector norm defined as
\begin{equation}
\begin{split}
 &\| \,\boldsymbol{a}(\boldsymbol{x}) \, \|^2 = \langle \,\boldsymbol{a}(\boldsymbol{x}), \boldsymbol{a}(\boldsymbol{x}) \, \rangle.
 \end{split}
\end{equation}

The solution of the minimization problem is $r$ eigenvectors with the largest eigenvalues in the following eigenvalue problem
\begin{equation}
  XX^\top JU = U\Lambda,
\end{equation}
where
\begin{equation}
\begin{split}
 X &= [\boldsymbol{u}(\boldsymbol{x}, t_1), \, \boldsymbol{u}(\boldsymbol{x}, t_2), \, \cdots \,, \boldsymbol{u}(\boldsymbol{x}, t_M)]  \in \mathbb{R}^{N \times M}, \\
\Lambda &= \text{diag}[\lambda_1, \, \lambda_2, \, \cdots \,, \lambda_M], \,\,\, (\lambda_1 \geq \lambda_2 \geq \cdots \geq \lambda_M).
 \end{split}
\end{equation}
We can also compute $U$ from the left eigenvector of SVD with respect to $X$. In this study, the algorithm based on the snapshot POD\cite{POD_snap} is used to compute eigenvectors and eigenvalues for POD with low computational memory.

The minimization problem of POD also means that the subspace $\mathcal{S} \subset \mathbb{R}^{N}$ spanned by $U$ is the optimal subspace to represent the dataset. The optimal representation of a dataset in its subspace is as follows
\begin{equation}
  X \rightarrow U^\top JX  \in \mathbb{R}^{r \times M}.
\end{equation}
For simplicity, we denote $J^{1/2}X$ as $X$ in the following, implying that the spatial weight derived from the inner product is implicitly considered in the product between $N$-dimensional vectors or matrices.

\subsection{POD-Galerkin projection approach}
Galerkin projection\cite{POD_Galerkin, GP_deane} is a method used to derive governing equations for variables projected onto a low-dimensional subspace, based on the full-order governing equations defined in a finite- or infinite-dimensional space. We begin by considering a general governing equation for $N$-dimensional vector valiable $\boldsymbol{u}(\boldsymbol{x}, t) \in \mathbb{R}^{N}$ given by
\begin{equation}
\frac{\partial \boldsymbol{u}}{\partial t}= \boldsymbol{\mathcal{F}}(\boldsymbol{u}),
   \label{general_eq}
\end{equation}
where $\boldsymbol{\mathcal{F}}(\boldsymbol{u}): \mathbb{R}^{N} \rightarrow \mathbb{R}^{N}$ is the generalized operator.
Assume that the projected variable is $\boldsymbol{u}_r \in \mathbb{R}^{r}$, which satisfies the relation
\begin{equation}
\boldsymbol{u}_r= \boldsymbol{\mathcal{P}}_r(\boldsymbol{u}),
   \label{projection_GP_0}
\end{equation}
where $\boldsymbol{\mathcal{P}}_r(\boldsymbol{u}): \mathbb{R}^{N} \rightarrow \mathbb{R}^{r} (\boldsymbol{u} \rightarrow \boldsymbol{u}_r)$ denotes the projection operator.
We consider the $N$-dimensional representation of $\boldsymbol{u}_r$ as
\begin{equation}
\boldsymbol{v}= \boldsymbol{\mathcal{P}}^r(\boldsymbol{u}_r),
   \label{projection_GP}
\end{equation}
and 
\begin{equation}
 \boldsymbol{\mathcal{P}}(\boldsymbol{u}) =\boldsymbol{\mathcal{P}}^r\left\{\boldsymbol{\mathcal{P}}_r(\boldsymbol{u})\right\},
   \label{projection_GP}
\end{equation}
where $\boldsymbol{v} \in \mathbb{R}^{N}$ and $\boldsymbol{\mathcal{P}}^r(\boldsymbol{u}): \mathbb{R}^{r} \rightarrow \mathbb{R}^{N} (\boldsymbol{u}_r \rightarrow \boldsymbol{v})$. Note that $\boldsymbol{v}$ and $\boldsymbol{u}_r$ represent the same state variable expressed in different basis vectors.
Here, the governing equation is transformed as
 \begin{equation}
\frac{\partial \boldsymbol{\mathcal{P}}(\boldsymbol{u})}{\partial t}= \boldsymbol{\mathcal{F}}\left\{\boldsymbol{\mathcal{P}}(\boldsymbol{u})\right\}.
   \label{general_eq_projection}
\end{equation}
Strictly speaking, the projected variable $\boldsymbol{\mathcal{P}}(\boldsymbol{u})$ does not exactly satisfy the original governing equations. However, the governing dynamics can still be well approximated through projection, particularly in terms of the system operators.

Here, in the context of POD, the projection operator is represented by the projection matrix $UU^\top $, such that
 \begin{equation}
\boldsymbol{\mathcal{P}}(\boldsymbol{u})=UU^\top \boldsymbol{u},
   \label{general_eq_projection_operator}
\end{equation}
where $U$ is the matrix whose columns are the POD basis vectors.
Since the POD basis $U$ is time-independent, the time derivative of the left-hand side of Eq. (\ref{general_eq_projection}) can be written as
 \begin{equation}
\frac{\partial UU^\top \boldsymbol{u}}{\partial t}= U\frac{\partial U^\top \boldsymbol{u}}{\partial t},
   \label{general_eq_timeder}
\end{equation}
Multiplying both sides of Eq. (\ref{general_eq_projection}) from the left by $U^\top $, and applying Eq. (\ref{general_eq_timeder}), yields
\begin{equation}
\frac{\partial U^\top \boldsymbol{u}}{\partial t}= U^\top  \boldsymbol{\mathcal{F}}(UU^\top \boldsymbol{u}),
   \label{general_eq_projection_last}
\end{equation}
which governs the time evolution of the low-dimensional variables, since the POD basis is orthonormal.
Let a new variable $\boldsymbol{\tilde{u}} \in \mathbb{R}^{r}$ be the low-dimensional representation of $\boldsymbol{u}$ as follows
\begin{equation}
\boldsymbol{\tilde{u}} = U^\top \boldsymbol{u}.
   \label{low-rank-u_GP}
\end{equation}
The governing equations for $\boldsymbol{\tilde{u}}$ are given by 
\begin{equation}
\frac{\partial \boldsymbol{\tilde{u}}}{\partial t}= \boldsymbol{\tilde{\mathcal{F}}}(\boldsymbol{\tilde{u}}),
   \label{general_eq_projection_last}
\end{equation}
where the low-dimensional operator $\boldsymbol{\tilde{\mathcal{F}}}: \mathbb{R}^{r} \rightarrow \mathbb{R}^{r}$ is defined as
\begin{equation}
\boldsymbol{\tilde{\mathcal{F}}}(\boldsymbol{\tilde{u}})= U^\top  \boldsymbol{\mathcal{F}}(U\boldsymbol{\tilde{u}}).
   \label{general_eq_projection_last}
\end{equation}
These are the governing equations for the low-dimensional representation of variables.

For the incompressible Navier-Stokes equations, $\boldsymbol{\tilde{\mathcal{F}}}$ is written by the nonlinear term $\mathcal{N}$ and linear term $\mathcal{L}$ as follows
\begin{equation}
\boldsymbol{\tilde{\mathcal{F}}}(\boldsymbol{\tilde{u}})=\boldsymbol{\mathcal{N}}(\boldsymbol{\tilde{u}}) + \boldsymbol{\mathcal{L}}(\boldsymbol{\tilde{u}}).
\end{equation}
Note that for incompressible fluid systems, introducing a base flow $\boldsymbol{u}_b$ satisfying stationary boundary conditions is a common simplification of the Galerkin projected equations\cite{Holmes,schlegel2015long}. Thus, $\boldsymbol{\tilde{u}}$ is
\begin{equation}
\boldsymbol{\tilde{u}} = U^\top (\boldsymbol{u} - \boldsymbol{u}_b).
\end{equation}
In this representation, the $k$th-component of nonlinear term is
\begin{equation}
\boldsymbol{\mathcal{N}}_k(\boldsymbol{\tilde{u}}) = \boldsymbol{\tilde{u}}^\top  F^k \boldsymbol{\tilde{u}},
\end{equation}
where $ij$-component of matrix $F^k \in \mathbb{R}^{r \times r}$ is
\begin{equation}
F^k_{ij}={-\langle(\boldsymbol{\phi}_i\cdot\nabla)\boldsymbol{\phi}_j,{\boldsymbol{\phi}}_k\rangle}.
\end{equation}
The linear term is
\begin{equation}
\boldsymbol{\mathcal{L}}(\boldsymbol{\tilde{u}}) = G \boldsymbol{\tilde{u}},
   \label{linearG}
\end{equation}
where $ki$-component of matrix $G \in \mathbb{R}^{r \times r}$ is
\begin{equation}
G_{ki}={\frac{1}{Re}\langle{\nabla^2}\boldsymbol{\phi}_i,{\boldsymbol{\phi}}_k\rangle}-\langle(\boldsymbol{u}_b\cdot\nabla)\boldsymbol{\phi}_j,{\boldsymbol{\phi}}_k\rangle-\langle(\boldsymbol{\phi}_i\cdot\nabla)\boldsymbol{u}_b,{\boldsymbol{\phi}}_k\rangle.
 \label{Galerkin_linearfluid}
\end{equation}
This derivation is based on a vast amount of previous studies\cite{Noack_1994, cylinder7, GP_deane, mypaper_6, Satomanifold_2024, mypaper2}, and the effect of the pressure term is neglected in the linear term\cite{pressure2}.


\subsection{DMD}
DMD was proposed by Schmid~\cite{DMD} to extract coherent structures from time-series data of flow fields. DMD assumes a linear time evolution from a snapshot matrix $X  \in \mathbb{R}^{N \times M}$ to $X' \in \mathbb{R}^{N \times M}$, where superscript $'$ represents variables after time advancement. 
 This relationship is expressed as
\begin{equation}
X' = AX,
\label{Aevolution}
\end{equation}
where $A$ is a linear operator that approximates the dynamics underlying the data. DMD provides a way to compute the eigenvalues and eigenvectors of this operator, which characterize the temporal growth rates and spatial structures of the flow. In this sense, DMD approximates the linear evolution of the system. Resulting in the approximation, the matrix $A$ is computed from $X$ and $X'$ as
 \begin{equation}
{A}=X'{X}^{\dagger},
  \label{Aapproximation}
\end{equation}
where the superscript $\dagger$ denotes the Moore–Penrose pseudoinverse. Because the eigenstructure of $A$ reflects key properties of the underlying dynamical system, the goal of DMD is to determine both its eigenvalues and eigenvectors.

Using the definition of the Moore–Penrose pseudoinverse, $A$ can be computed as
 \begin{equation}
{A}=X'V_{\text{full}}S_{\text{full}}^{-1}U_{\text{full}}^\top ,
  \label{AapproximationA}
\end{equation}
where $V_{\text{full}} \in \mathbb{R}^{M \times M}$ is the matrix of right singular vectors, $S_{\text{full}} \in \mathbb{R}^{M \times M}$ is the diagonal matrix of singular values of $X$, and $U_{\text{full}} \in \mathbb{R}^{N \times M}$ contains the left singular vectors, which correspond to the POD basis vectors when $r = M$. 
In practice, however, the rank of $X$ is generally less than $M$, and $A$ is often too large to handle directly in numerical computations. Therefore, DMD uses a low-dimensional approximation of $A$ defined as 
 \begin{equation}
\tilde{A}= U^{T}AU \approx U^{T}X'VS^{-1}U^\top U = U^{T}X'VS^{-1},
  \label{AapproximationA}
\end{equation}
where $\tilde{A} \in \mathbb{R}^{r \times r}$, $U \in \mathbb{R}^{N \times r}$, $V \in \mathbb{R}^{r \times M}$, and $S \in \mathbb{R}^{r \times r}$ are the truncated SVD components of $X$.

The eigenvalues of the matrix $\tilde{A}$ characterize the temporal evolution of the corresponding eigenvectors. The growth rate and frequency of the $k$th mode are computed as follows:
\begin{equation}
\sigma_k = \frac{\text{Real}\{\log({\lambda}_k)\}}{\Delta T},
  \label{DMDgrowth}
\end{equation}
\begin{equation}
f_k = \frac{\text{Imag}\{\log({\lambda}_k)\}}{2\pi \Delta T},
  \label{DMDfreq}
\end{equation}
where $\text{Real}(\cdot)$ and $\text{Imag}(\cdot)$ denote the real and imaginary parts, respectively, and $\lambda_k$ is the $k$th eigenvalue of $\tilde{A}$. The $\log(\cdot)$ denotes the complex logarithm, with the argument (phase angle) defined in the range $-\pi$ to $\pi$.
In projected DMD, the full-state eigenvectors $\boldsymbol{\varphi}_{k}$ is obtained from the reduced-order eigenmode $\boldsymbol{\tilde{\varphi}}_{k}$ of $\tilde{A}$ via
 \begin{equation}
\boldsymbol{\varphi}_{k}=U \boldsymbol{\tilde{\varphi}}_{k}.
  \label{fulleigenmode}
\end{equation}

\subsection{Grassmann manifold interpolation}
We introduce a method to estimate subspaces under varying conditions using two different subspaces spanned by POD basis vectors. To formalize the treatment of subspaces, we employ the Grassmann manifold, defined as
\begin{equation}
  \mathrm{Gr}(N,r) := \{ \mathcal{S}\subset\mathbb{R}^{N}~|~ \mathrm{dim}(\mathcal{S}) =r \}.  
\end{equation}
which is the set of all $r$-dimensional subspaces of an $N$-dimensional vector space. In the context of POD for fluid flow, $N$ corresponds to the dimension of the vector space and $r$ to the number of retained POD basis vectors. Thus, estimating subspaces under different conditions reduces to estimating elements on the Grassmann manifold. It should be noted, however, that the Grassmann manifold is not a vector space, as simple linear operations between subspaces are not generally defined.

Here, we introduce the tangent space in one element (point) on the Grassmann manifold $\mathrm{Gr}(N,r)$, whose subspace is spanned by $U_o$. The tangent space is the space orthogonal to the subspace $[U_o]$, defined as 
\begin{equation}
  \mathrm{T}_{[U_o]} = \{ \Delta\in\mathbb{R}^{N\times r}~|~U_o^\top \Delta = 0\},
\end{equation}
where $\Delta$ represents a tangent vector. Strictly speaking, $\Delta$ is a matrix of vectors.
In contrast to Grassmann manifolds, which are not vector spaces, tangent spaces are vector spaces. Hence, the elements on the Grassmann manifold projected onto the tangent space, i.e. tangent vectors are operable, and interpolation on the tangent space is possible. 

The mapping from an element on the Grassmann manifold $U$ to a tangent vector $\Delta$ referred to as logarithmic maps is defined to be
\begin{equation}
  \mathrm{Log}_{[U_o]}^\mathrm{Gr} \rightarrow [U] \mapsto \Delta, \label{eq:GrLog}
\end{equation}
and is computed from\cite{edelman1998geometry}:
\begin{equation}
  \Delta = V_U\mathrm{arctan}(\Sigma_U)W_U^\top ,
  \label{eq:GrLog_mat}
\end{equation}
where $V_U  \in  \mathbb{R}^{N \times r}$, $\Sigma_U  \in  \mathbb{R}^{r \times r}$, and $W_U  \in  \mathbb{R}^{r \times r}$ are obtained using $U$ presented below
\begin{equation}
  (I-U_oU_o^\top )U(U_o^\top U)^{-1} \stackrel{\rm{SVD}}{=} V_U \Sigma_U W_U^\top .
\end{equation}
Note that thin SVD is used here. For the numerical computation process, the left-hand side is treated by transforming it to 
\begin{equation}
  (I-U_oU_o^\top )U(U_o^\top U)^{-1} = U(U_o^\top U)^{-1} - U_o,
\end{equation}
to avoid the treatment of large matrices.
That is, a logarithmic mapping of an element on a Grassmann manifold yields a vector on the tangent space, i.e., a tangent vector. Note that when $U=U_o$, the tangent vector is $\boldsymbol{0}$, which indicates that the tangent space is a vector space whose origin corresponds to $U_o$, and the tangent vector from the subspace spanned by $U$ identifies the subspace of other conditions in the tangent space. 
To obtain the subspace on the Grassmann manifold pointed to by the tangent vector $\Delta$, we define the inverse mapping of the logarithmic map referred to as the exponential map presented below
\begin{equation}
  \mathrm{Exp}^{\mathrm{Gr}}_{[U_o]}: \Delta \mapsto [U].\label{eq:GrGeodesic}
\end{equation}
The exponential map on the Grassmann manifold can be computed by\cite{edelman1998geometry}:
\begin{equation}
  U = [U_o W_{\Delta}\mathrm{cos}(\Sigma_{\Delta})W_{\Delta}^\top +V_{\Delta}\mathrm{sin}(\Sigma_{\Delta})W_{\Delta}^\top ],
  \label{eq:GrExp_mat}
\end{equation}
where $V_{\Delta}  \in  \mathbb{R}^{N \times r}$, $\Sigma_{\Delta}  \in  \mathbb{R}^{r \times r}$, and $W_{\Delta}  \in  \mathbb{R}^{r \times r}$ are obtained by thin SVD of $\Delta$ presented below
\begin{equation}
  \Delta \stackrel{\rm{SVD}}{=} V_{\Delta} \Sigma_{\Delta} W_{\Delta}^\top .
\end{equation}

On the tangent space, we now consider interpolating the subspace $U^{\text{int}}$ in condition $\eta$, where superscript ${\text{int}}$ represents interpolated variables, from the subspaces $U_0$ and $U_1$ in conditions $\eta_0$ and $\eta_1$, respectively. Interpolation in the tangent space $\mathrm{T}_{[U_0]}$ can be performed using the tangent vector of $U_1$. In the case of linear interpolation, the interpolated tangent vector $\Delta^{\text{int}}$ is computed from  
\begin{equation}
  \Delta^{\text{int}} = \frac{\eta - \eta_0}{\eta_1 - \eta_0}\Delta,
    \label{eq:tan_inter}
\end{equation}
where $\Delta$ is the tangent vector of $U_1$ in $\mathrm{T}_{[U_0]}$. Here, when $\eta=\eta_1$, the subspace pointed to by the tangent vector coincides with the subspace spanned by $U_1$. However, the matrix $U^{\text{int}}_1$ obtained by the exponential mapping of tangent vector $\Delta^{\text{int}}$ does not necessarily coincide with $U_1$ because the choice of a set of orthonormal bases for constructing a subspace is not unique. The $U_{\text{out}}$ obtained by mapping $U_1$ to the exponential and then to the logarithmic map is
\begin{equation}
  U_{\text{out}} = UR^\top ,
  \label{eq:GrExp_mat_procrastes}
\end{equation}
where $R=(U^\top _0U)$ is a regular matrix. This is a Procrustes transformation of $U$ with respect to $U_0$. The derivation is provided in \ref{apena}.
Strictly speaking, in order to distinguish between $U_\text{out}$ and $U_1$, which have the same subspace but are constructed on different bases, we need to define logarithmic and exponential mappings on the Stiefel manifold. However, because logarithmic mapping on the Stiefel manifold requires convergence computation\cite{zimmermann2017matrix, zimmermann2019manifold, ELOMARI2025113564}, it is practical to treat $U_{\text{out}}$ by considering to be obtained after the Procrustes transformation in interpolation on a Grassmann manifold.

\begin{figure}[htbp]
  \centering\includegraphics[width=8cm,keepaspectratio]{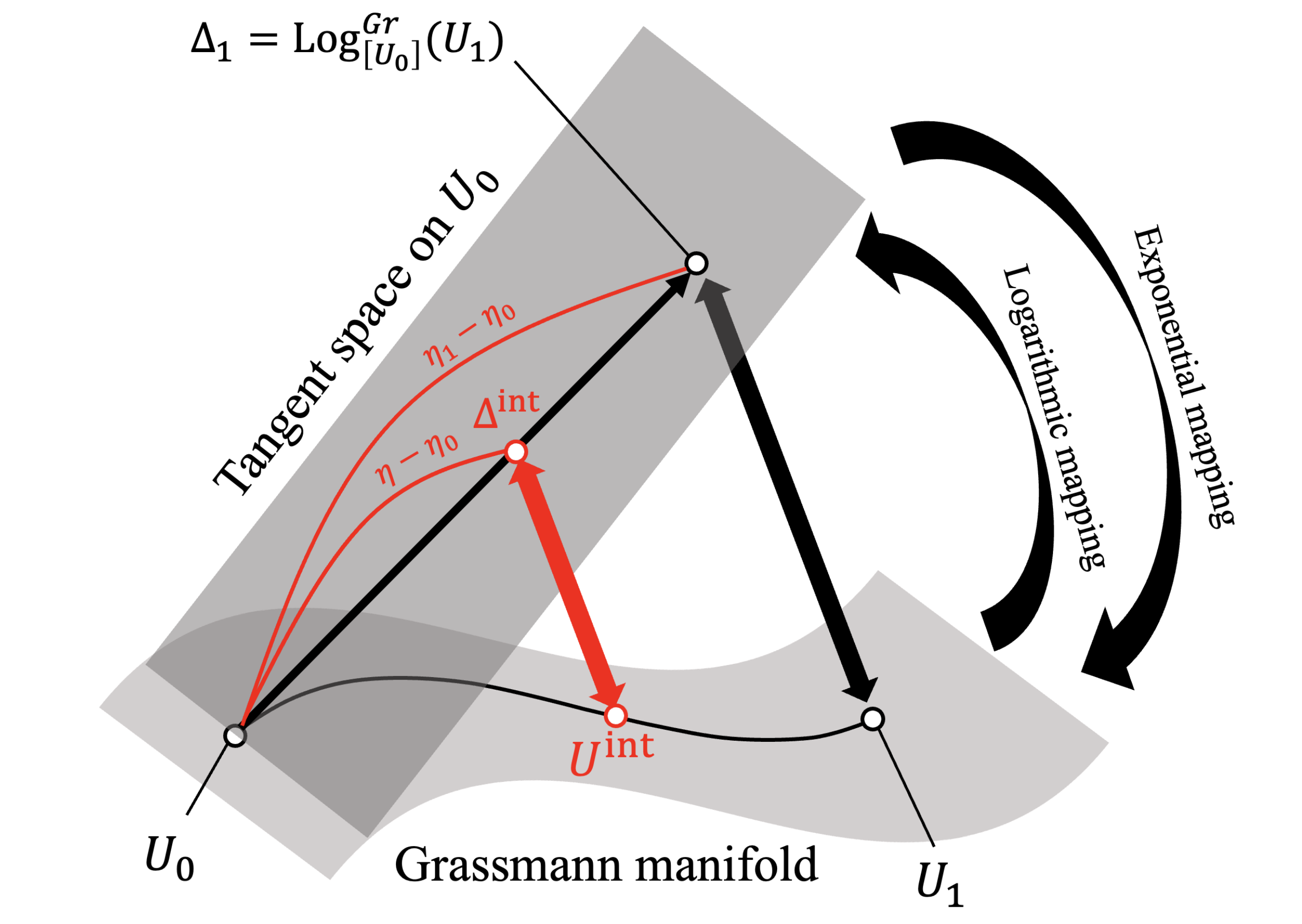}
\caption{Conceptual diagram of interpolation on the Grassmann manifold.}
 \label{fig:figure_Grassman}
\end{figure}

Figure \ref{fig:figure_Grassman} shows the conceptual diagram of interpolation process of obtaining the other subspace from the two subspaces $U_0$ and $U_1$, and that process is summarized by the following 
\begin{enumerate}
	\item In the tangent space in $U_0$, compute the tangent vector $\Delta_1$ pointing to $U_1$ using a logarithmic map from Eq. (\ref{eq:GrLog_mat}).
	\item The interpolation is conducted in tangent space, and the tangent vector $\Delta^{\text{int}}$ is obtained. In the case of linear interpolation, it is given by Eq. (\ref{eq:tan_inter}). 
	\item The interpolated subspace (basis) $U^{\text{int}}$ is obtained from exponential map to $\Delta^{\text{int}}$, represented by  Eq. (\ref{eq:GrExp_mat}).
\end{enumerate}

\section{Parameterized ROM with operator interpolation}\label{ROM}
In this section, we present two types of methods for estimating the linear operator, as well as its eigenvalues and eigenvectors, at an arbitrary parameter value $\eta$. The estimation procedure is formulated using snapshot matrices $X$ and $X'$ obtained at two known parameter values, $\eta_0$ and $\eta_1$, with $\eta_0 < \eta_1$. The snapshot matrices corresponding to a given parameter $\eta$ are denoted by $X(\eta)$ and $X'(\eta)$.
\subsection{Galerkin-projection-based interpolation ROM (GapiROM)}
As described in the DMD algorithm, the dimension of the linear operator is typically too large to handle directly in numerical computations. To address this, we reduce the dimensionality of the operator by projecting the governing equations onto a subspace spanned by POD basis vectors via Galerkin projection.
The bases for constructing this subspace are obtained by applying POD to the dataset matrices $X(\eta_0)$ and $X(\eta_1)$. 
Let $U(\eta_0)$ and $U(\eta_1)$ represent the matrices containing the $r$ POD basis vectors corresponding to the largest eigenvalues, extracted from the two datasets:
\begin{equation}
  \begin{split}
{X}(\eta_0) \stackrel{\rm{POD}}{ \rightarrow } U(\eta_0) = [\boldsymbol{\phi}_1(\eta_0), \, \boldsymbol{\phi}_2(\eta_0), \, \cdots, \, \boldsymbol{\phi}_r(\eta_0)] \in  \mathbb{R}^{n \times r},\\
{X}(\eta_1) \stackrel{\rm{POD}}{ \rightarrow } U(\eta_1) = [\boldsymbol{\phi}_1(\eta_1), \, \boldsymbol{\phi}_2(\eta_1), \, \cdots, \, \boldsymbol{\phi}_r(\eta_1)]  \in  \mathbb{R}^{n \times r}.
  \label{AapproximationA}
   \end{split}
\end{equation}
Given a general parameter condition $\eta$ the basis matrix $U(\eta) \in  \mathbb{R}^{n \times r}$  is approximated by interpolating  $U(\eta_0)$ and $U(\eta_1)$ in the tangent space of $U(\eta_0)$ on Grassmann manifold.
The reduced linear operator in the subspace spanned by $U(\eta)$ is obtained by Galerkin projection of the governing equations. In particular, the reduced linear operator is derived from the operator $G$ in Eq. (\ref{linearG}).
Its elements are computed from the POD basis vectors and the base flow using Eq. (\ref{Galerkin_linearfluid}). 
When the parameter $\eta$ includes the Reynolds number, the variation of the operator with respect to $\eta$ reflects the Reynolds-number dependence on the coefficient of the viscous term. We denote by $G^{\text{int}(\eta)}$ the reduced operator constructed using the interpolated POD basis vectors $U(\eta)$.

The eigenvalues and eigenvectors of the full-dimensional operator under the condition $\eta$ are approximated by solving the eigenvalue problem of the $r \times r$ matrix $G^{\text{int}(\eta)}$. Letting $\boldsymbol{\tilde{\varphi}}_{k}(\eta)$  denote the $k$th eigenvector of $G^{\text{int}(\eta)}$, the corresponding approximate eigenvector in the full space is given by
\begin{equation}
\boldsymbol{\varphi}_k(\eta) = U(\eta) \tilde{\boldsymbol{\varphi}}_k(\eta).
\label{invpro0}
\end{equation}

\begin{figure}[htbp]
  \centering\includegraphics[width=8cm,keepaspectratio]{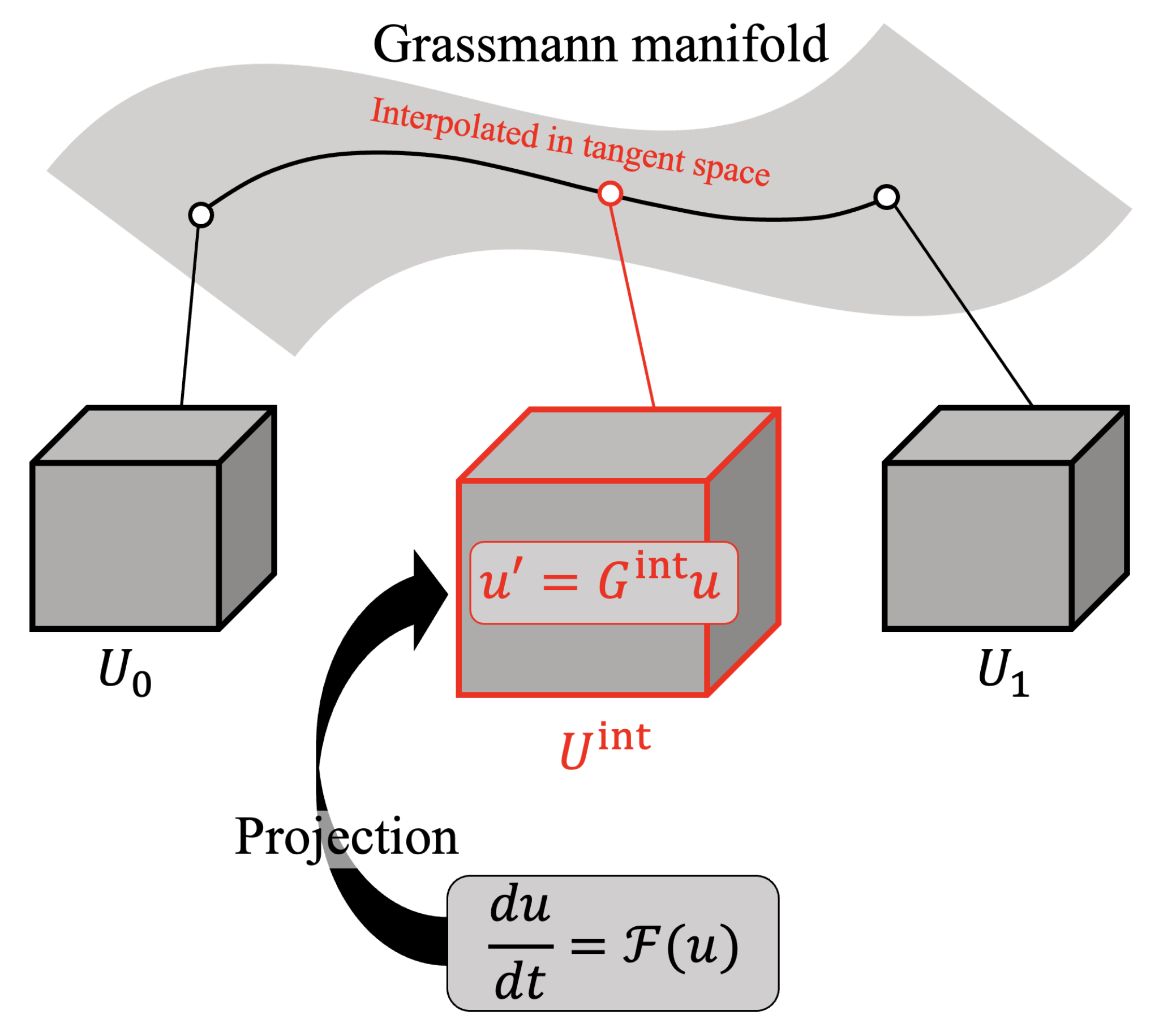}
  \caption{Conceptual diagram of the GapiROM.}
 \label{fig:figure_Gapi}
\end{figure}

The ROM that constructs low-dimensional operators in this manner and obtains their spectral properties is referred to as Galerkin-projection-based interpolation ROM (GapiROM) in this study. As illustrated conceptually in Fig.~\ref{fig:figure_Gapi}, GapiROM constructs a low-dimensional operator by projecting the governing equations onto a subspace obtained via Grassmann manifold interpolation.

\subsection{DMD-based operator interpolation ROM  (DoiROM)}
To illustrate the prediction of the parameter $\eta$ based on two known conditions, $\eta_0$ and $\eta_1$, we begin by formulating the reduced operator within the framework of DMD. The discrete-time reduced linear operator, denoted by $\tilde{A}(\eta)$, advances the reduced-state (velocity) vectors according to
\begin{equation}
\tilde{\boldsymbol{u}}(t + \Delta T) \left( = \tilde{\boldsymbol{u}}' \right) = \tilde{A}(\eta) \tilde{\boldsymbol{u}}(t),
\end{equation}
where $\Delta T$ is the time interval between snapshots. Since $\tilde{A}(\eta)$ depends on the sampling interval, it is more appropriate to convert it to a continuous-time operator, which is independent of $\Delta T$, before performing interpolation.

To derive the continuous-time operator, we begin with a dynamical system parameterized by $\eta$:
\begin{equation}
\frac{\partial \boldsymbol{u}}{\partial t}= \boldsymbol{\mathcal{F}}(\boldsymbol{u}; \eta).
\label{general_eq_para}
\end{equation}
By approximating this system by linear operator yields
\begin{equation}
\frac{\partial \boldsymbol{u}}{\partial t}= F(\eta)\boldsymbol{u},
\label{general_eq_para_linear}
\end{equation}
where $F(\eta)$ is the linear operator corresponding to the parameter $\eta$. Introducing a projection matrix $P$, the system is represented in reduced form as
\begin{equation}
\frac{\partial (P\boldsymbol{u})}{\partial t}= FP\boldsymbol{u}.
\label{general_eq_projection_DMD}
\end{equation}
The projection matrix $P$ is defined as the orthogonal projector
\begin{equation}
P = UU^\top ,
\label{projection_operator}
\end{equation}
where the columns of $U$ form an orthonormal basis for the reduced subspace.

Since the basis is time-independent, the reduced-order system can be expressed as
\begin{equation}
\frac{\partial (U^\top \boldsymbol{u})}{\partial t} = U^\top  F U U^\top  \boldsymbol{u}.
\label{general_eq_projection_DMD_tilde_0}
\end{equation}
By defining the reduced state as $\tilde{\boldsymbol{u}} = U^\top  \boldsymbol{u}$ and the reduced operator as $\tilde{F} = U^\top  F U$, we obtain
\begin{equation}
\frac{\partial \tilde{\boldsymbol{u}}}{\partial t} = \tilde{F} \tilde{\boldsymbol{u}}.
\label{general_eq_projection_DMD_tilde}
\end{equation}
This equation describes the continuous-time evolution of the reduced-order state. Mathematically, the continuous-time operator $\tilde{F}$ and the discrete-time operator $\tilde{A}$ are related through the matrix exponential \cite{Shieh_1980,shieh1986determining}:
\begin{equation}
\tilde{A} =  \mathrm{EXP}(\tilde{F} \Delta T),
\label{FA_relation}
\end{equation}
where $ \mathrm{EXP}(\cdot)$ represents matrix exponential.

To interpolate the system behavior at an arbitrary parameter value $\eta$, the continuous-time operator is linearly approximated using the operators obtained at the known conditions $\eta_0$ and $\eta_1$:
\begin{equation}
\tilde{F}^{\text{int}}(\eta) = \frac{\eta - \eta_0}{\eta_1 - \eta_0} \tilde{F}(\eta_1) - \frac{\eta - \eta_1}{\eta_1 - \eta_0} \tilde{F}(\eta_0).
\label{interpolated_F}
\end{equation}
This interpolation can also be expressed in terms of the continuous-time operator derived from the matrix logarithm of the discrete-time operator:
\begin{equation}
\tilde{F}^{\text{int}}(\eta) = \mathrm{LOG}(\tilde{A}^{\text{int}})= \frac{\eta - \eta_0}{\eta_1 - \eta_0} \frac{\mathrm{LOG}\left\{\tilde{A}(\eta_1)\right\}}{\Delta T_1} - \frac{\eta - \eta_1}{\eta_1 - \eta_0} \frac{\mathrm{LOG}\left\{\tilde{A}(\eta_0)\right\}}{\Delta T_0},
\end{equation}
where the logarithm $\mathrm{LOG}(\cdot)$ refers to the principal value of the matrix logarithm, and $\Delta T_0$ and $\Delta T_1$ denote the time intervals associated with $\tilde{A}(\eta_0)$ and $\tilde{A}(\eta_1)$, respectively. The matrix logarithm can be computed using the geometric series method \cite{Shieh_1980}.
The interpolated continuous-time operator $\tilde{F}^{\text{int}}(\eta)$ is therefore given by
\begin{equation}
\tilde{F}^{\text{int}}(\eta) = \left\{ \frac{\eta - \eta_0}{\eta_1 - \eta_0} \frac{\mathrm{LOG}\left\{\tilde{A}(\eta_1)\right\}}{\Delta T_1} - \frac{\eta - \eta_1}{\eta_1 - \eta_0} \frac{\mathrm{LOG}\left\{\tilde{A}(\eta_0)\right\}}{\Delta T_0} \right\}.
\label{general_eq_interpolate_DMD}
\end{equation}

The reduced basis $U$ corresponding to an intermediate parameter value $\eta$ is constructed via interpolation on the Grassmann manifold. As shown in \ref{apena}, this interpolation yields a smooth transition between $U_0$ and $U_{\text{out}} = U_1 R^\top $, where $R$ is the regular matrix obtained from the Procrustes problem \cite{golub2013matrix} applied to $U_0$ and $U_1$. However, this approach does not exactly reproduce $U_1$, leading to a discrepancy when reconstructing the full-order operator from reduced-order quantities.
Specifically, reconstructing the full-order operator at $\eta_1$ using the original basis $U_1$ yields
\begin{equation}
A_1 = U_1 \tilde{A}_1 U_1^\top,
\end{equation}
while reconstructing the full-order operator $A_{\text{out}}$ using the interpolated basis $U_{\text{out}}$ results in
\begin{equation}
A_{\text{out}} = U_{\text{out}} \tilde{A}_1 U_{\text{out}}^\top  = U_1 R^\top  \tilde{A}_1 R U_1^\top  \neq A_1.
\end{equation}
Here, the interpolated basis $U_{\text{out}}$ is obtained as
\begin{equation}
[U_{\text{out}}] = \mathrm{Exp}^{\mathrm{Gr}}_{[U_0]} \circ \mathrm{Log}_{[U_0]}^\mathrm{Gr} ([U_1]).
\end{equation}
To resolve this inconsistency, we define a rotated reduced operator
\begin{equation}
\tilde{A}_1^R = R \tilde{A}_1 R^\top,
\label{procrastes_U_A'}
\end{equation}
which compensates for the rotational misalignment between $U_1$ and $U_{\text{out}}$, as discussed in \ref{apena}. The interpolation is then performed between $\tilde{A}_0$ and the rotated operator $\tilde{A}_1^R$:
\begin{equation}
\tilde{F}^{\text{int}}(\eta) = \frac{\eta - \eta_0}{\eta_1 - \eta_0} \frac{\mathrm{LOG}(\tilde{A}_1^R)}{\Delta T_1} - \frac{\eta - \eta_1}{\eta_1 - \eta_0} \frac{\mathrm{LOG}(\tilde{A}_0)}{\Delta T_0}.
\label{general_eq_interpolate_DMD}
\end{equation}

\begin{figure}[htbp]
\centering
\includegraphics[width=12cm,keepaspectratio]{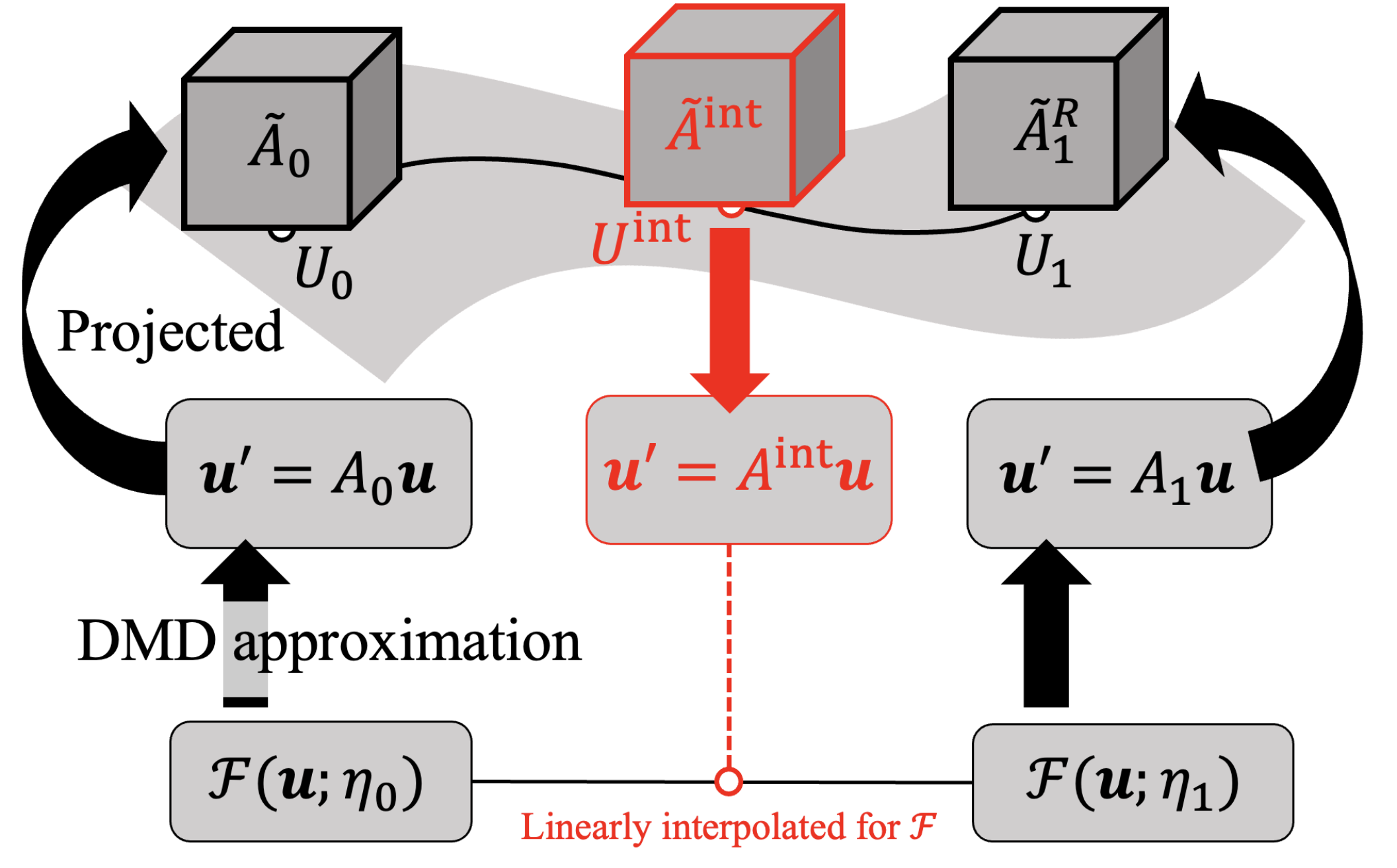}
\caption{Conceptual diagram of the DoiROM.}
\label{fig:figure_Doi}
\end{figure}

We refer to the ROM constructed through this procedure as the DMD-based operator interpolation ROM (DoiROM). DoiROM is designed to interpolate between linear operators identified via DMD at different parameter values. A central challenge stems from the fact that these operators are defined in distinct reduced subspaces. To address this, DoiROM performs interpolation at two levels: first, the subspaces are interpolated using Grassmann manifold techniques; second, the operators are interpolated within the aligned subspace, as illustrated in Fig.~\ref{fig:figure_Doi}.

\section{Result and discussion}\label{result}
We demonstrate the applicability of the operator-based ROM using three test cases. Although all involve flow around a bluff body, the characteristics of the governing operators differ among them. In the first and second cases, interpolation is performed across different Reynolds numbers, whereas in the third case, it is performed with respect to the aspect ratio of elliptical cylinders.

\subsection{Growth of oscillations in flow around a circular cylinder at varying Reynolds numbers}
The first case examines the onset or decay of unsteady vortex shedding from a steady flow around a two-dimensional circular cylinder, which is a canonical scenario in stability analysis and operator-based numerical methods. In this setting, we employ the linearized operator about the steady flow (base flow), for which the existence of a discrete set of eigenmodes has been established in previous studies\cite{Kumar_2006,GSA1}. The dataset for the ROM is constructed via global stability analysis using a time-stepping approach. Details of the numerical method and dataset preparation for global stability analysis are provided in \ref{apenb}.

\begin{figure}[htbp]
\begin{tabular}{cc}
\multicolumn{1}{l}{(a)}  &  \multicolumn{1}{l}{(b)}\\
  \begin{minipage}[b]{0.48\linewidth}
          \centering\includegraphics[width=6cm,keepaspectratio]{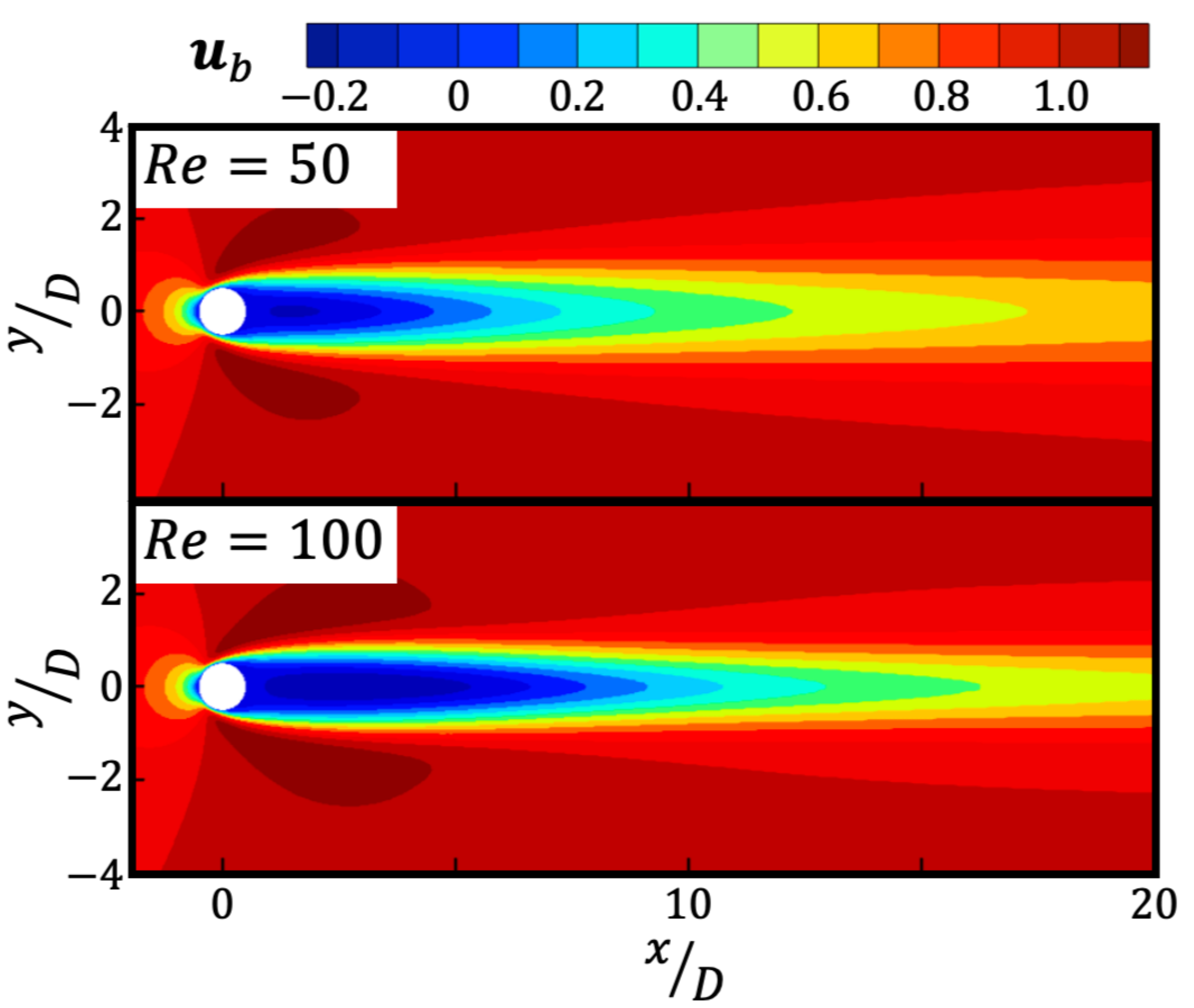}
  \end{minipage}
  &
  \begin{minipage}[b]{0.48\linewidth}
          \centering\includegraphics[width=6cm,keepaspectratio]{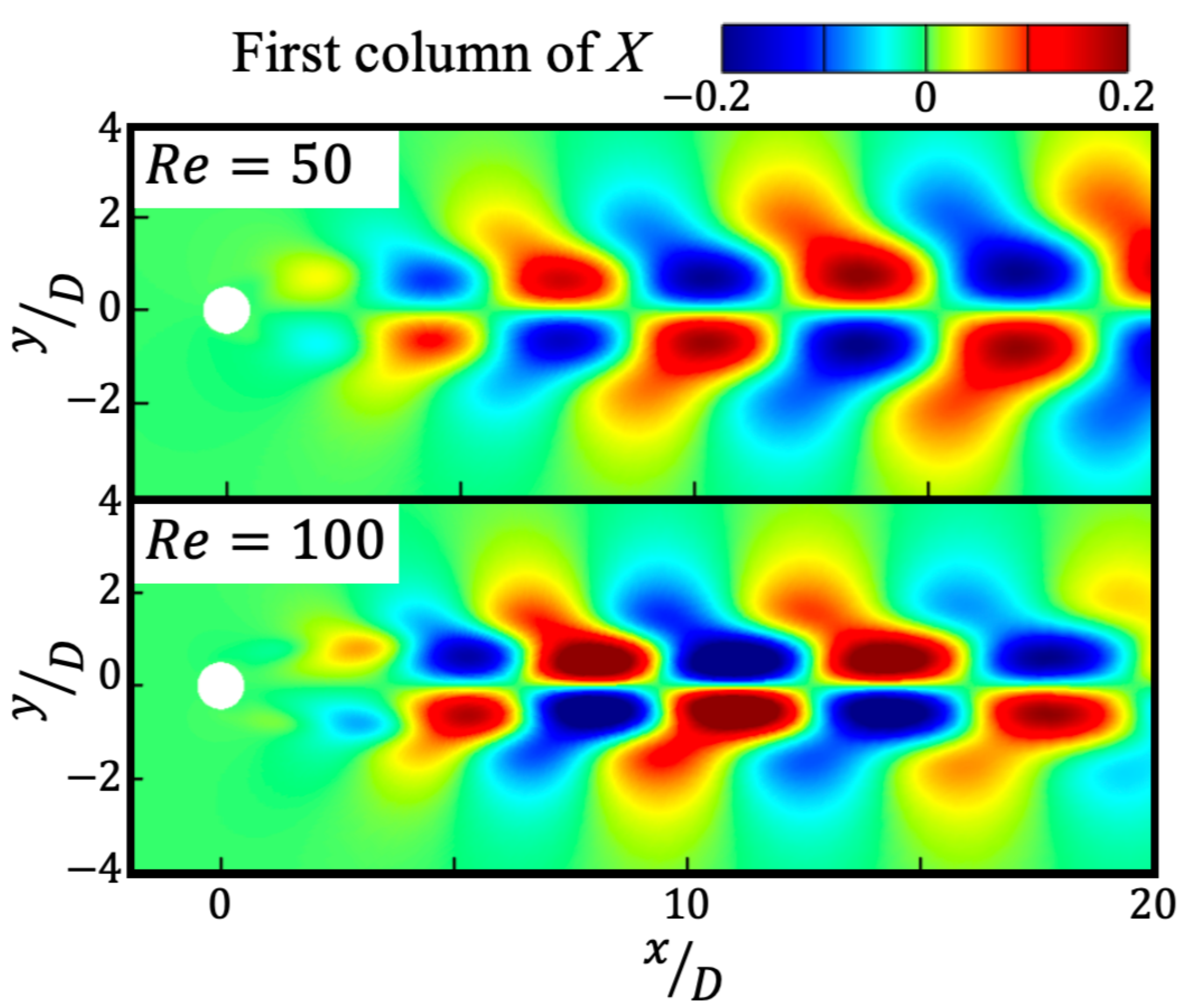}
  \end{minipage}
  \end{tabular}
  \captionsetup{justification=raggedright,singlelinecheck=false}
\caption{Spatial distribution of (a) the steady flow field (base flow) and (b) a representative snapshot from the dataset $X$.}
 \label{fig:figure_GSA_data}
\end{figure}

We briefly present representative snapshots of the flow field and the corresponding POD basis vectors obtained from the snapshot matrix $X$. Figure~\ref{fig:figure_GSA_data} shows the spatial distribution of the base flow and a typical snapshot from the dataset matrix $X$, which serves as the foundation for deriving the linearized operators and their eigenvectors at Reynolds numbers $50$ and $100$. The Reynolds number $Re$ is defined as
\begin{equation}
\begin{split}
Re=\frac{U_\infty D}{\nu},
\end{split}
\end{equation}
where $D$ is the cylinder diameter, $U_\infty$ is the freestream velocity, and $\nu$ is the kinematic viscosity.
The base flow features a symmetric twin-vortex structure aligned along the wake centerline ($y = 0$). As the Reynolds number increases, the streamwise extent of these vortices also increases. In contrast, the snapshots forming matrix $X$ exhibit asymmetry with respect to $y = 0$, reflecting perturbations that grow or decay from the base flow. These asymmetric patterns are characteristic of the evolution of unstable modes extracted through global stability analysis.

\begin{figure}[tbhp]
\begin{center}
\begin{tabular}{cc}
\multicolumn{1}{l}{(a)}  &  \multicolumn{1}{l}{(b)}\\
  \begin{minipage}[b]{0.45\linewidth}
          \centering\includegraphics[width=4.5cm,keepaspectratio]{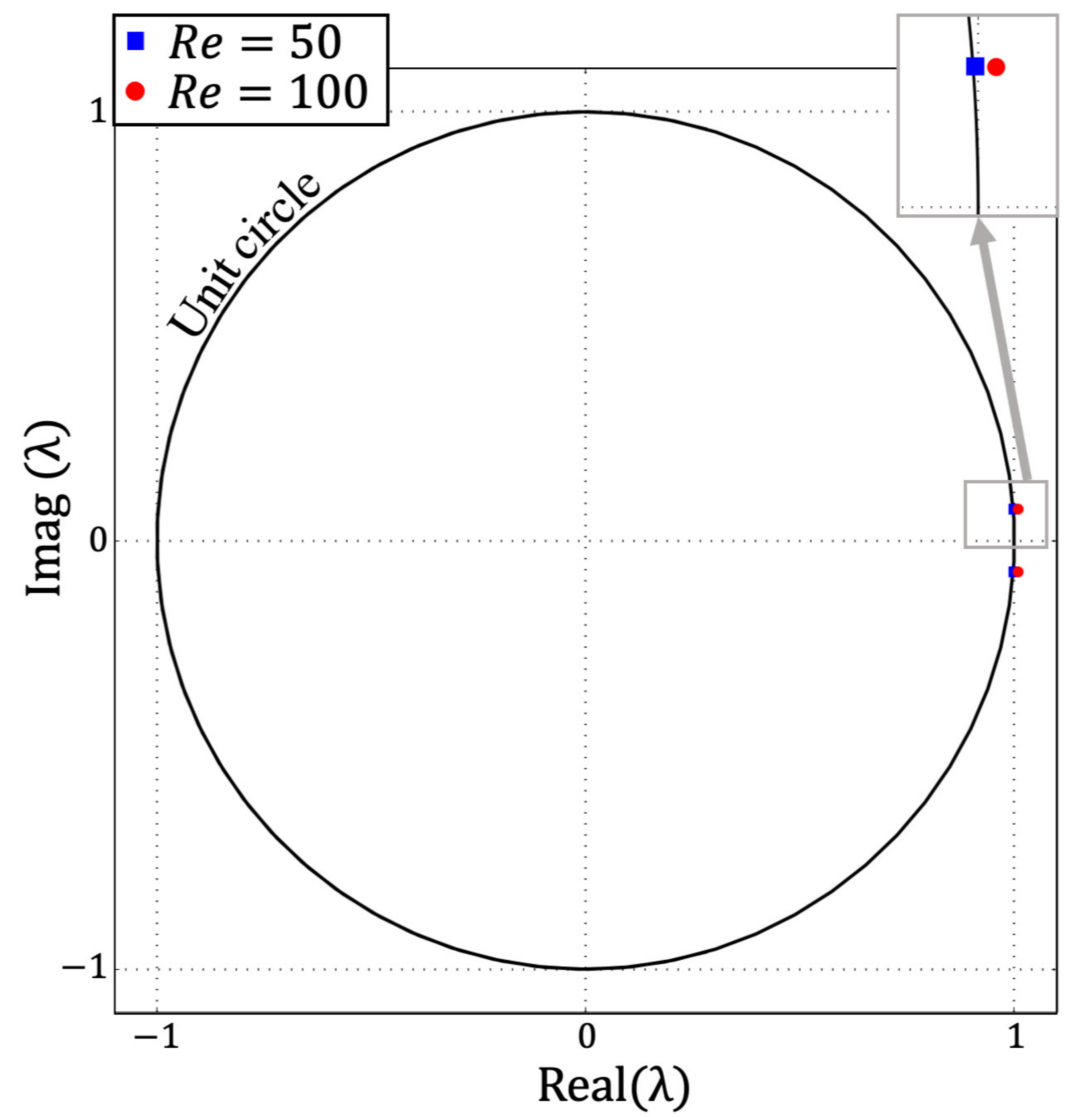}
  \end{minipage}
  &
  \begin{minipage}[b]{0.45\linewidth}
          \centering\includegraphics[width=4.5cm,keepaspectratio]{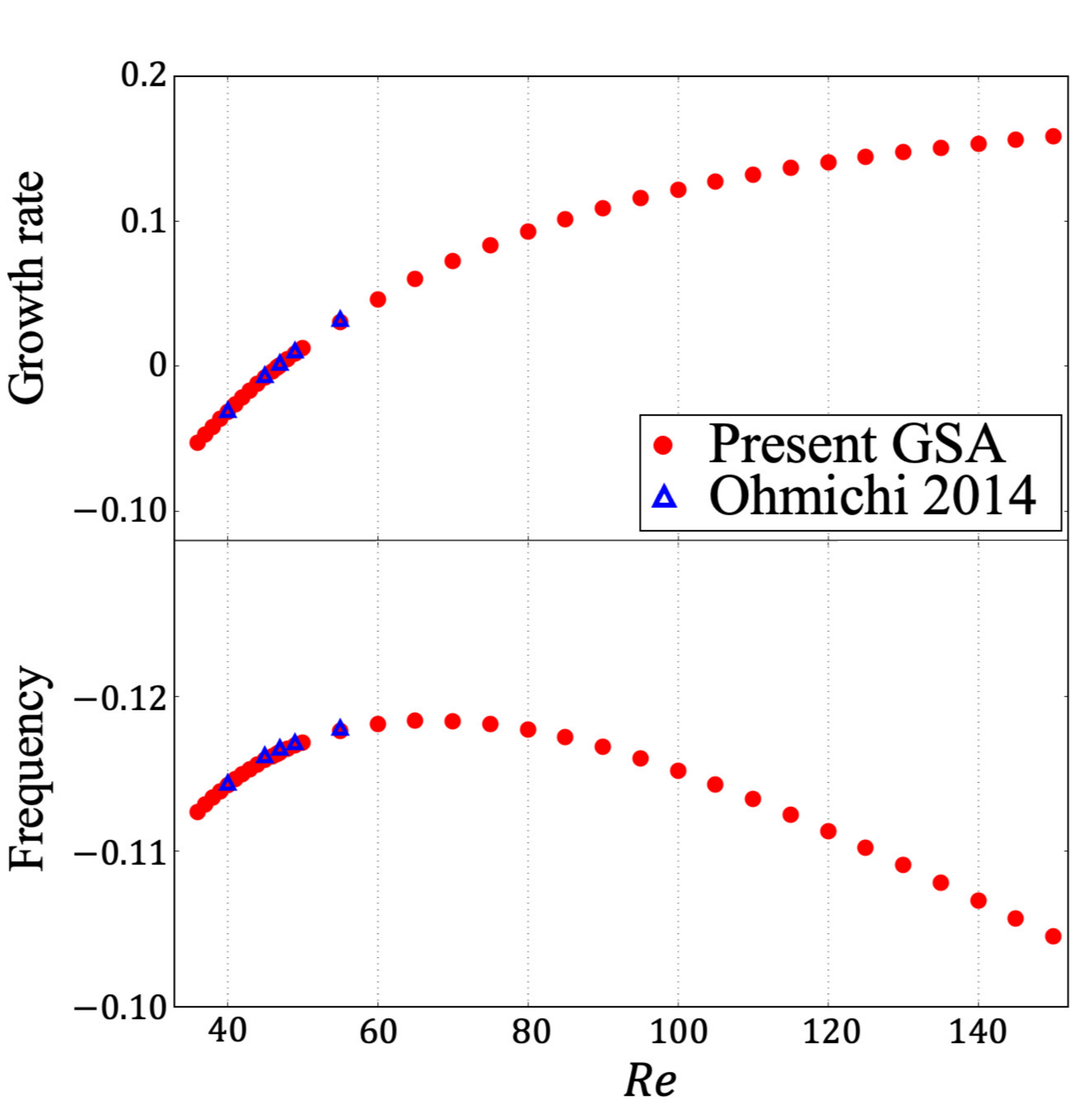}
  \end{minipage}\\
  \multicolumn{1}{l}{(c)}  &  \multicolumn{1}{l}{}\\
  \multicolumn{2}{c}{
    \begin{minipage}[b]{0.9\linewidth}
        \centering\includegraphics[width=10cm,keepaspectratio]{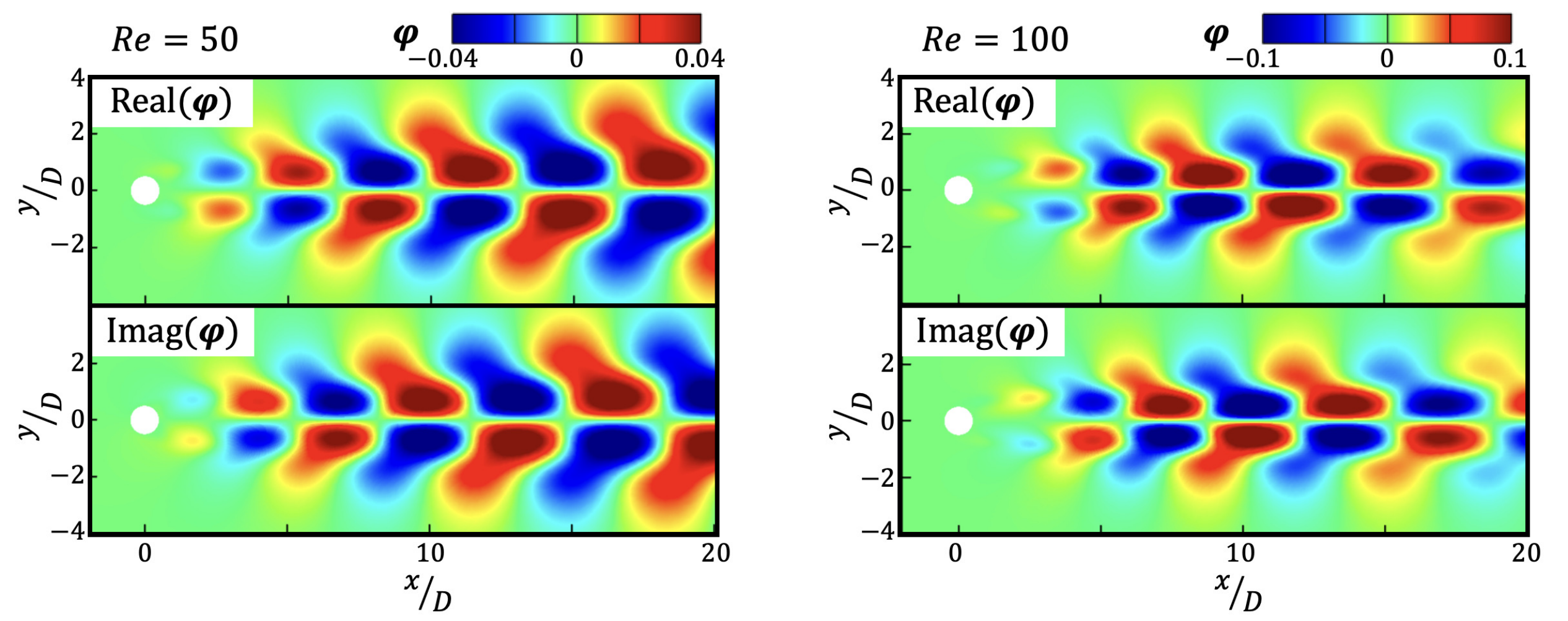}
    \end{minipage}
  }\\
  \end{tabular}
  \end{center}
  \captionsetup{justification=raggedright,singlelinecheck=false}
\caption{(a) Eigenvalue distributions of the linearized operator at $Re = 50$ and $100$. (b) Reynolds number dependence of the growth rate and frequency. (c) Spatial distributions of the eigenmodes at $Re = 50$ (left) and $Re = 100$ (right).}
 \label{fig:figure_GSA_eigenvalue_Re}
\end{figure}

Figure~\ref{fig:figure_GSA_eigenvalue_Re} (a) presents the eigenvalue distributions of the linearized operators about the base flow at $Re=50$ and $100$, obtained via global stability analysis. In both cases, a pair of complex conjugate eigenmodes is observed. Remarkably, this pair alone captures over 99\% of the total kinetic energy contained in the flow snapshots forming the matrix $X$. The emergence of such eigenmodes in the linear stability analysis of steady flow past a cylinder is well documented in previous studies\cite{Ohmichi_D,variousshapes1, GIANNETTI_2007, Leontini_2014}. Although similar complex-conjugate eigenmodes were also identified at other Reynolds numbers, their eigenvalue spectra are omitted here for conciseness.

Figure~\ref{fig:figure_GSA_eigenvalue_Re} (b) presents the growth rates and frequencies derived from the eigenvalues at various Reynolds numbers. These results are consistent with those reported in a previous study\cite{Ohmichi_D}. The growth rate crosses zero at a Reynolds number of approximately $46.6$, indicating a Hopf bifurcation point\cite{shift,Kumar_2006}. Beyond this critical point, the steady flow transitions into an unsteady flow, signifying the onset of unsteadiness for $Re > 46.6$. The oscillation frequency remains nearly constant at approximately $0.11$, with a slight decrease observed beyond $Re = 70$. Notably, for $Re < 46.6$, although the flow remains physically steady, the eigenmodes of the linearized operator still exhibit oscillatory behavior. These modes correspond to decaying oscillations with associated frequencies, reflecting latent dynamic features of the system.

Figure~\ref{fig:figure_GSA_eigenvalue_Re} (c) shows the spatial distributions of the eigenmodes. Although only the real value is physically observable, both the real and imaginary parts of the complex-conjugate pair are needed to reconstruct the full instantaneous velocity field. Since this eigenmode pair captures more than 99\% of the kinetic energy of the flow, the spatial patterns of the snapshots and the modes shown in Fig.~\ref{fig:figure_GSA_data} (b) are nearly identical, differing only by a phase shift.

Before discussing the operator-based ROM with interpolation across flow conditions, we first examine the prediction performance without subspace interpolation. For simplicity, we focus on the case where $\eta_0: (Re=40)$ and $\eta_1: (Re=60)$ in the GapiROM framework.
In the absence of subspace interpolation on the Grassmann manifold, the orthonormal basis at a given Reynolds number is approximated by
\begin{equation}
U(Re) \approx U(Re = 40).
\label{noint}
\end{equation}

\begin{figure}[htbp]
\begin{tabular}{cc}
\multicolumn{1}{l}{(a)}  &  \multicolumn{1}{l}{(b)}\\
  \begin{minipage}[b]{0.48\linewidth}
          \centering\includegraphics[height=5cm,keepaspectratio]{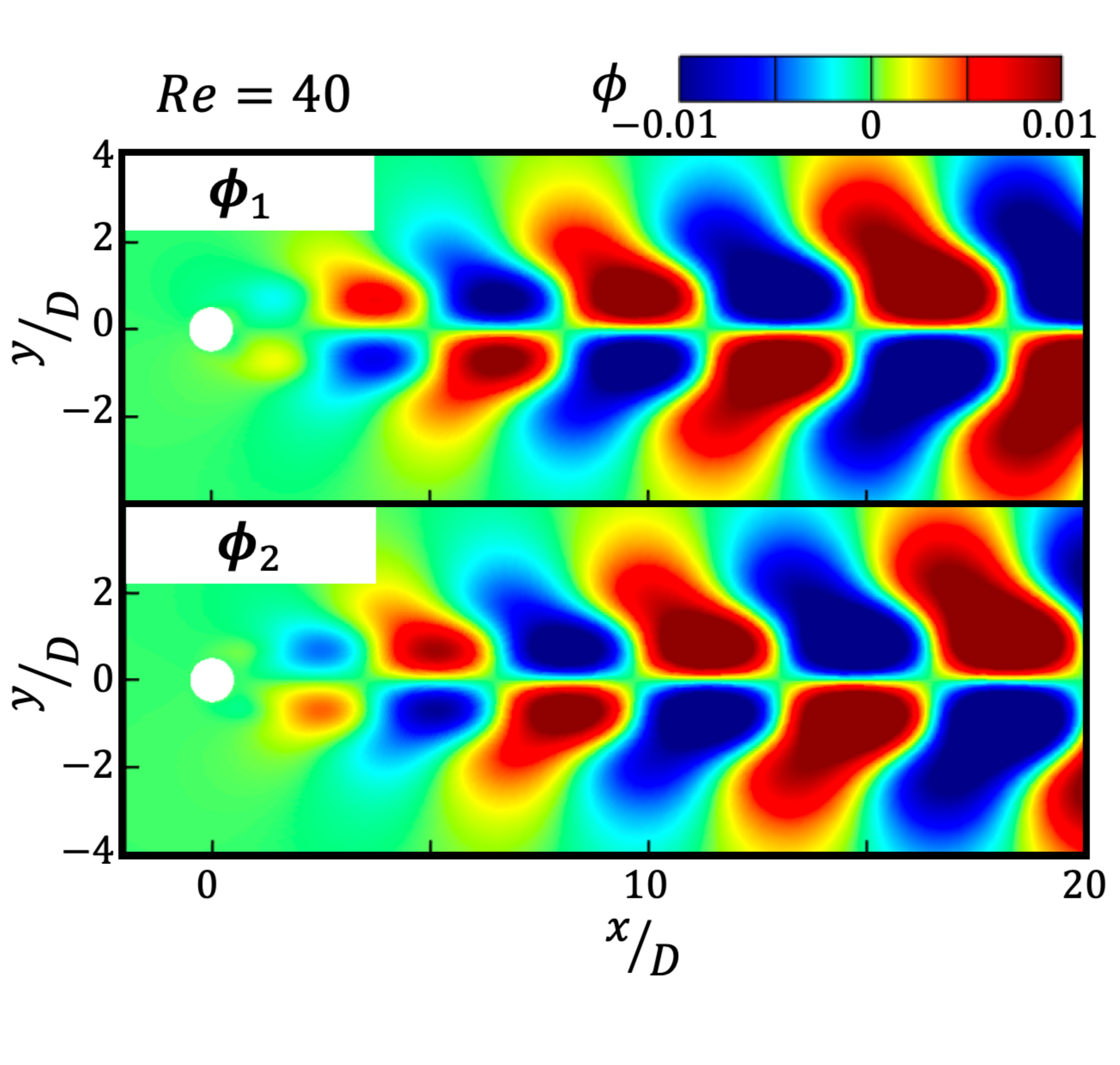}
  \end{minipage}
  &
  \begin{minipage}[b]{0.48\linewidth}
          \centering\includegraphics[height=5cm,keepaspectratio]{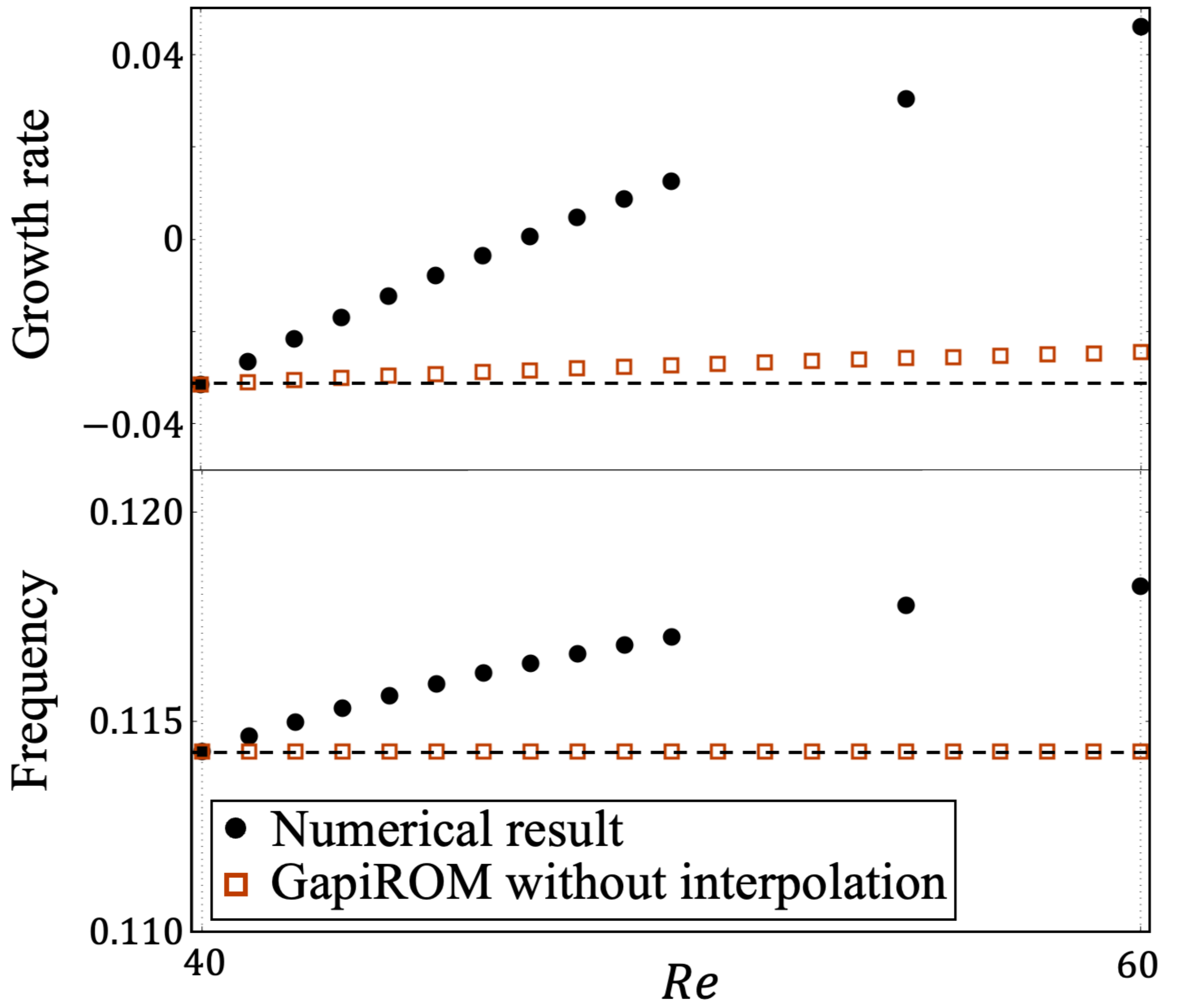}
  \end{minipage}\\
  \multicolumn{1}{l}{(c)}  &  \multicolumn{1}{l}{}\\
  \multicolumn{2}{c}{
    \begin{minipage}[b]{1\linewidth}
        \centering\includegraphics[width=12cm,keepaspectratio]{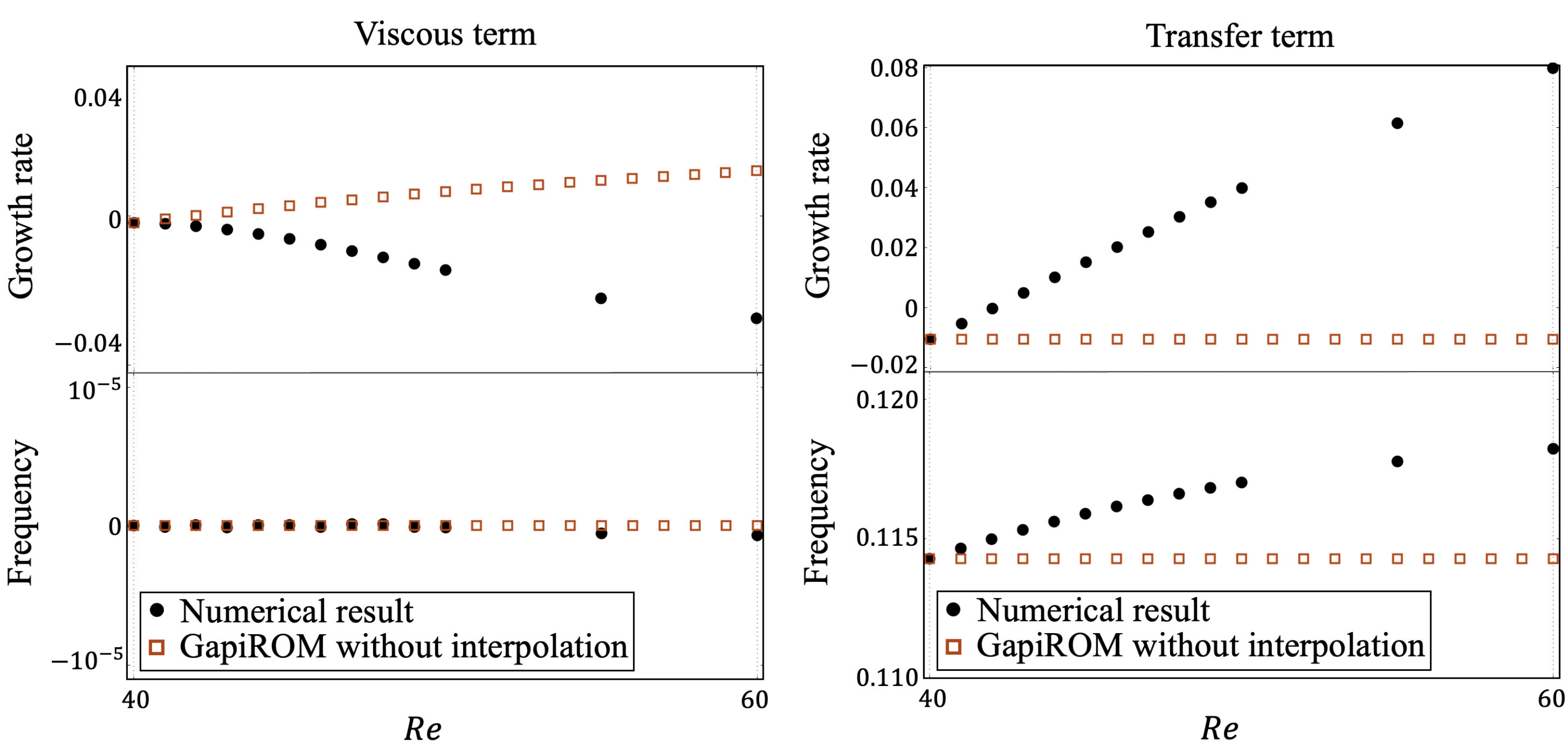}
    \end{minipage}
  }\\
  \end{tabular}
  \captionsetup{justification=raggedright,singlelinecheck=false}
\caption{(a) Spatial distribution of the POD basis vectors at $Re = 40$. (b) Reynolds number dependence of the growth rate and frequency in GapiROM without interpolation, where the basis vectors are directly estimated from those at $Re = 40$ using Eq. (\ref{noint}). Both growth rate and frequency exhibit minimal variation with Reynolds number. (c) The eigenvalue is analytically decomposed into two components, as shown in Eq. (\ref{eigenseparate3}). The variation in frequency and growth rate observed in the full-order model is primarily attributed to changes in the spatial distribution of the basis vectors, rather than to viscous effects. This underscores the importance of accurately capturing basis variation across parameters.}
 \label{fig:figure_GSA_eigenvalue_noint}
\end{figure}

Figure~\ref{fig:figure_GSA_eigenvalue_noint} (a) displays the first and second POD basis vectors at $Re = 40$, which are used for prediction. These basis vectors resemble the real and imaginary parts of the eigenmodes obtained from the global stability analysis at $Re = 50$, shown in Fig.~\ref{fig:figure_GSA_eigenvalue_Re} (c). Using the POD basis vectors at $Re = 40$ as a fixed basis, we predict the eigenvalues at other Reynolds numbers. Although the basis remains constant, the operator varies with Reynolds number due to its dependence on the viscous term in the governing equations, as described in Eq.(\ref{Galerkin_linearfluid}).
Figure~\ref{fig:figure_GSA_eigenvalue_noint} (b) presents the Reynolds number dependence of the growth rate and frequency, computed from the eigenvalues of the operator obtained from GapiROM without interpolation. While the predicted growth rate increases slightly with Reynolds number, the rate of increase is lower than that obtained from global stability analysis (i.e., the full-order model). In contrast, the predicted frequency remains nearly constant across all Reynolds numbers in the GapiROM without interpolation.

For further insight, we analytically decompose the eigenvalues obtained from GapiROM into two components: a viscous term that explicitly depends on the Reynolds number, and other terms that do not. 
In this GapiROM setting, the reduced linear operator $G$ is a $2 \times 2$ matrix, as only two POD basis vectors are used. The eigenvalues of this operator can be expressed analytically as
\begin{equation}
\begin{split}
\lambda=\frac{\text{tr}(G) \pm \sqrt{-1}\sqrt{-\text{tr}(G)^2+4\text{det}(G)}}{2}.
    \label{eigenseparate1}
\end{split}
\end{equation}
To simplify the expression and gain physical insight, we apply a common approximation inspired by the symmetry of dynamical systems near a Hopf bifurcation point~\cite{mypaper_6,schlegel2015long,shift}:
\begin{equation}
\begin{split}
G_{22} \approx G_{11},\,\,\,\,\,
G_{21} \approx -G_{12}.
    \label{eigenseparate2}
\end{split}
\end{equation}
Under this approximation, the growth rate $\sigma$ and oscillation frequency $f$ can be written as
\begin{equation}
\begin{split}
\sigma &= \frac{\text{tr}(G)}{2} \approx  G_{11},\\
f &=\frac{1}{4\pi}\sqrt{-\text{tr}(G)^2+4\text{det}(G)}\\
 &= \frac{1}{4\pi}\sqrt{-\text{tr}(G)^2+4(G_{11}G_{22}-G_{12}G_{21})}\\
  &\approx  \frac{1}{4\pi}\sqrt{-4G_{11}^2+4(G_{11}^2+G_{12}^2)}\\
  &=\frac{1}{2\pi}G_{12}.
    \label{eigenseparate3}
\end{split}
\end{equation}
This formulation clearly shows that the growth rate and frequency can be analytically separated into contributions from the viscous term, represented by the inner product ${\frac{1}{Re} \langle \nabla^2 \boldsymbol{\phi}_i, \boldsymbol{\phi}_k \rangle}$, and other components, such as convection and linearized nonlinear effects. These include $-\langle(\boldsymbol{u}_b\cdot\nabla)\boldsymbol{\phi}_j,{\boldsymbol{\phi}}_k\rangle-\langle(\boldsymbol{\phi}_i\cdot\nabla)\boldsymbol{u}_b,{\boldsymbol{\phi}}_k\rangle$.

Figure~\ref{fig:figure_GSA_eigenvalue_noint} (c) shows the Reynolds number dependence of the viscous and non-viscous terms in both GapiROM and the full-order model. For the full-order model, the values are computed using the POD basis vectors (from matrix $X$) obtained through global stability analysis at each Reynolds number.
The viscous contribution to the growth rate exhibits completely opposite trends in GapiROM and the full-order model as the Reynolds number varies. In contrast, the viscous contribution to the frequency remains nearly zero across all Reynolds numbers in both models. This behavior is consistent with the result in Fig.~\ref{fig:figure_GSA_eigenvalue_noint} (b), where the frequency predicted by GapiROM without interpolation remains nearly constant as the Reynolds number changes. This trend arises because, in this ROM, only the viscous term depends on the Reynolds number. Theoretical analysis also confirms that the viscous term does not explicitly affect the frequency\cite{freeman_2024}.

Focusing on the non-viscous contributions, the full-order model exhibits a pronounced dependence on Reynolds number in both growth rate and frequency. Since these quantities are determined solely by the POD basis vectors and the base flow, the observed variations reflect changes in the spatial structure of the POD basis vectors with respect to Reynolds number.
As expected, GapiROM without interpolation fails to capture these variations because it uses a fixed basis across all Reynolds numbers. This limitation underscores the importance of incorporating subspace interpolation to account for the dependence of basis vectors on Reynolds number in operator-based ROMs.

\begin{figure}[htbp]
\begin{tabular}{cc}
\multicolumn{1}{l}{(a)}  &  \multicolumn{1}{l}{(b)}\\
  \begin{minipage}[b]{0.48\linewidth}
          \centering\includegraphics[height=4.5cm,keepaspectratio]{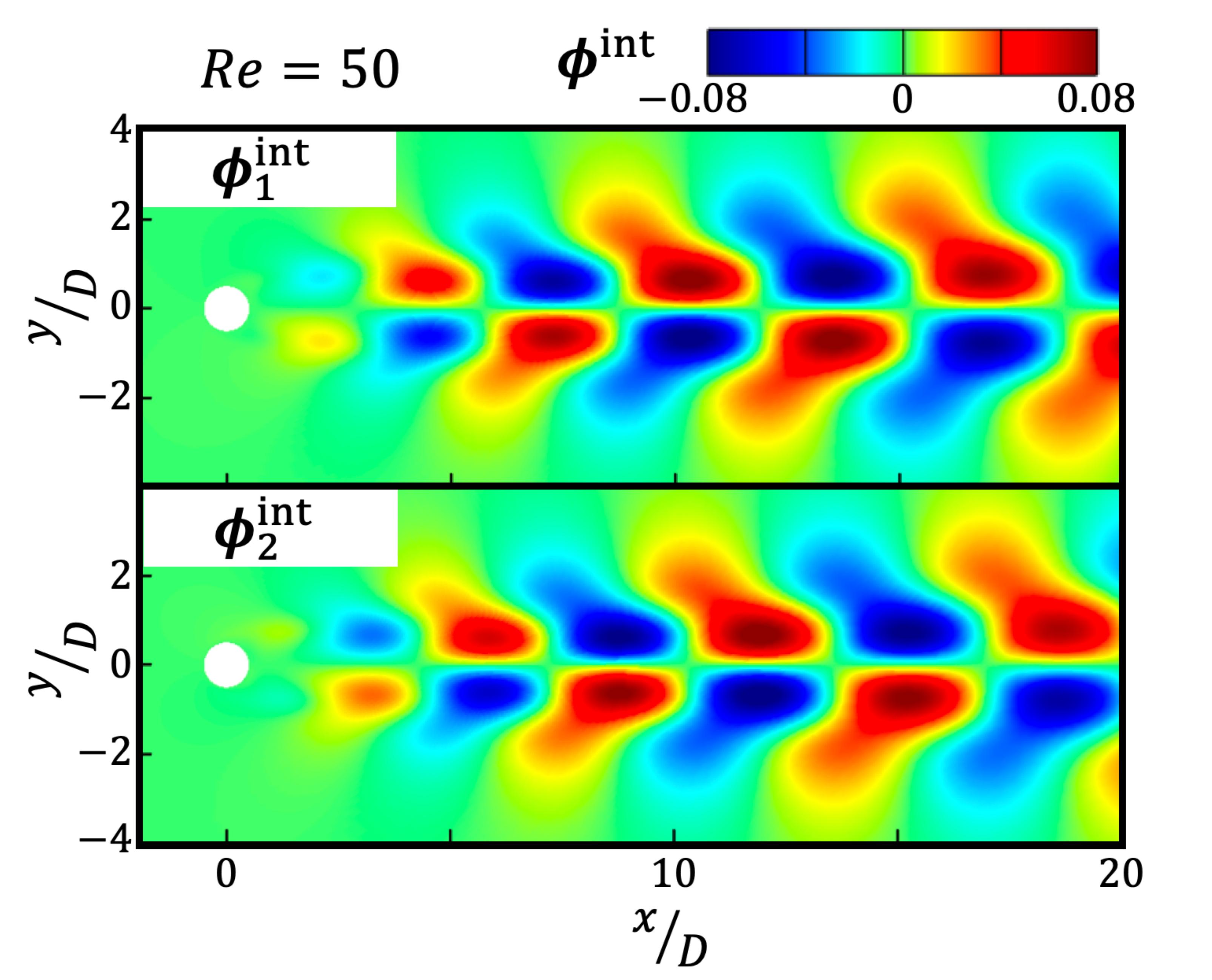}
  \end{minipage}
  &
  \begin{minipage}[b]{0.48\linewidth}
          \centering\includegraphics[height=4.5cm,keepaspectratio]{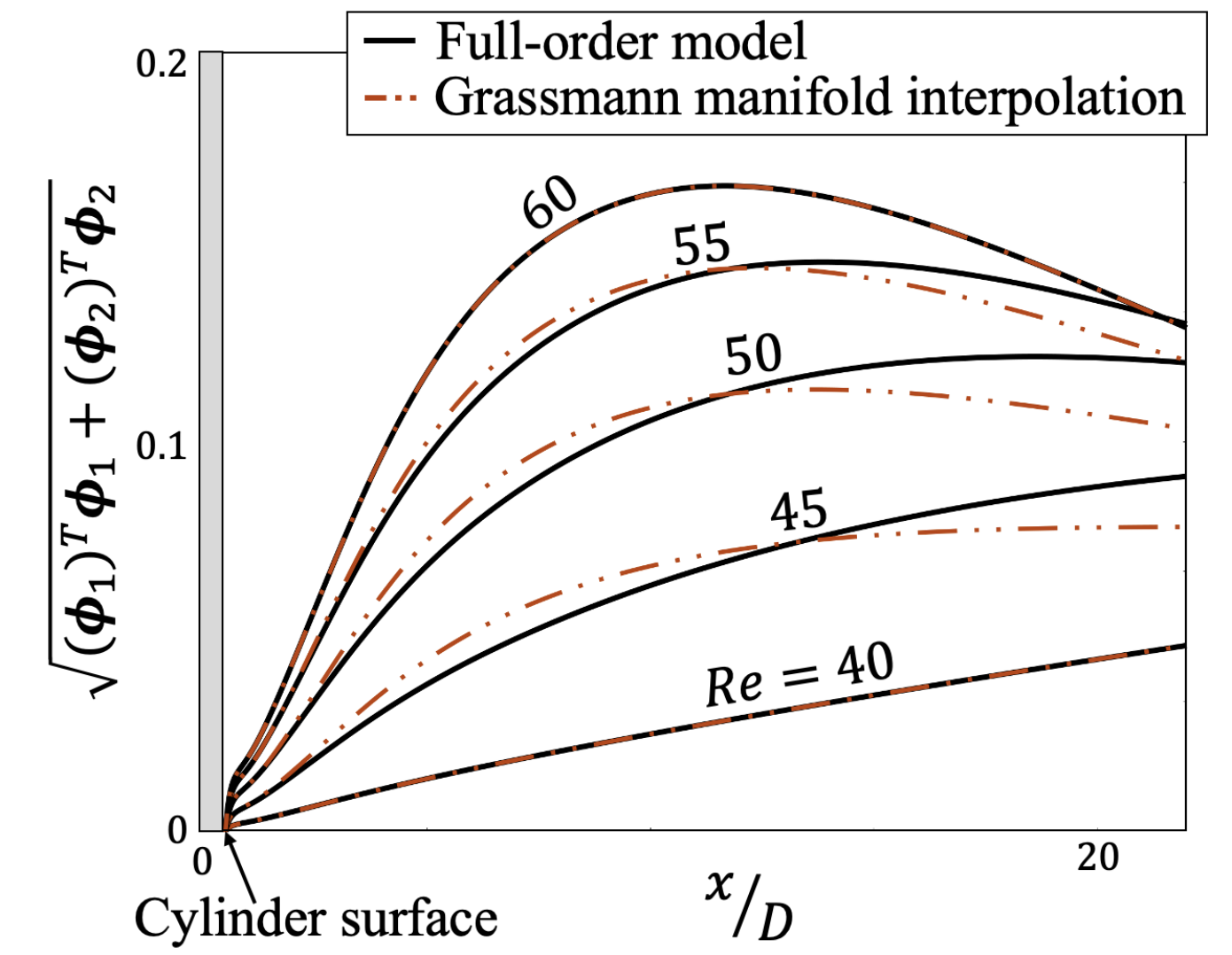}
  \end{minipage}\\
  \multicolumn{1}{l}{(c)}  &  \multicolumn{1}{l}{(d)}\\
  \begin{minipage}[b]{0.48\linewidth}
          \centering\includegraphics[height=5cm,keepaspectratio]{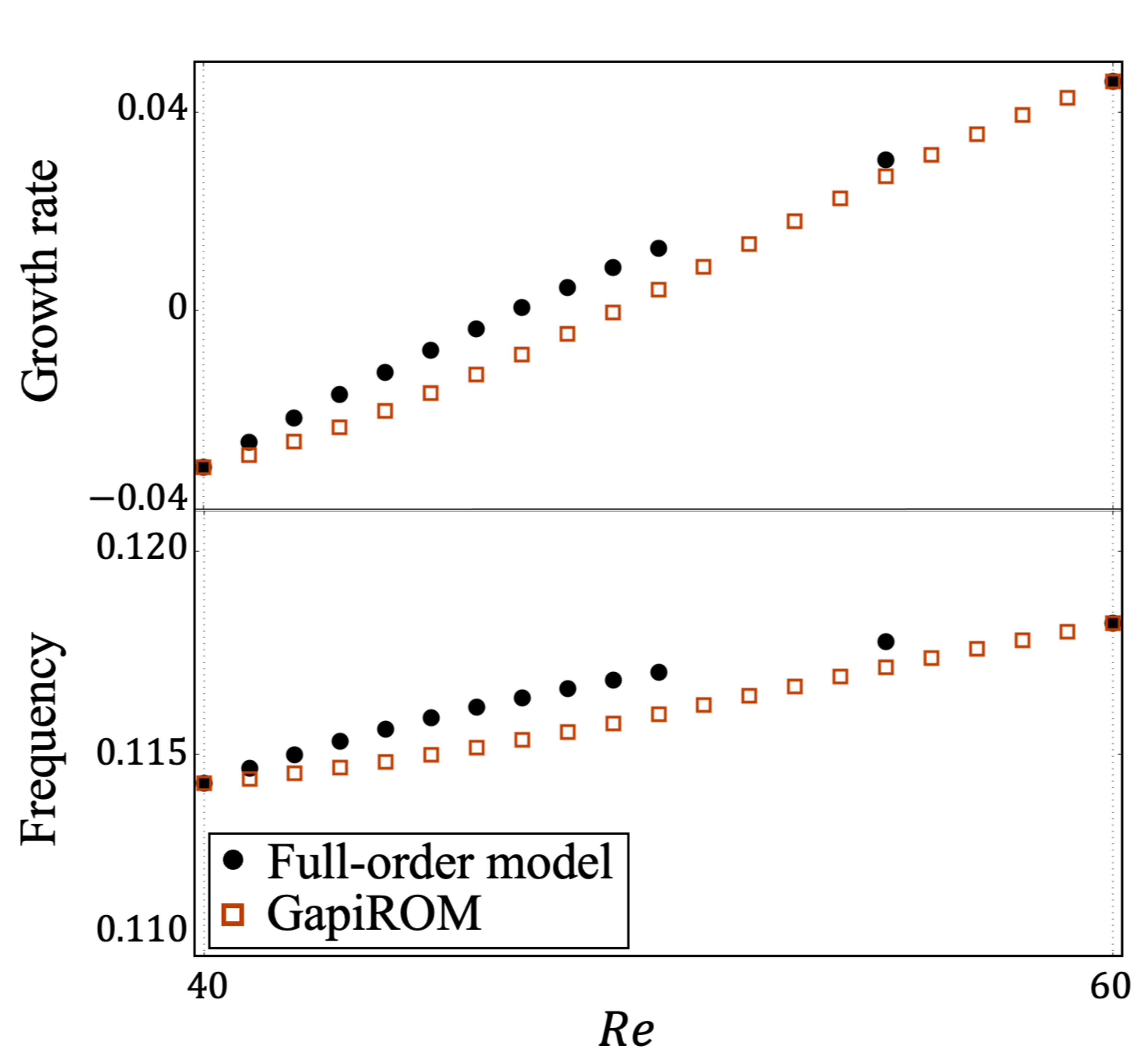}
  \end{minipage}
  &
  \begin{minipage}[b]{0.48\linewidth}
          \centering\includegraphics[height=5cm,keepaspectratio]{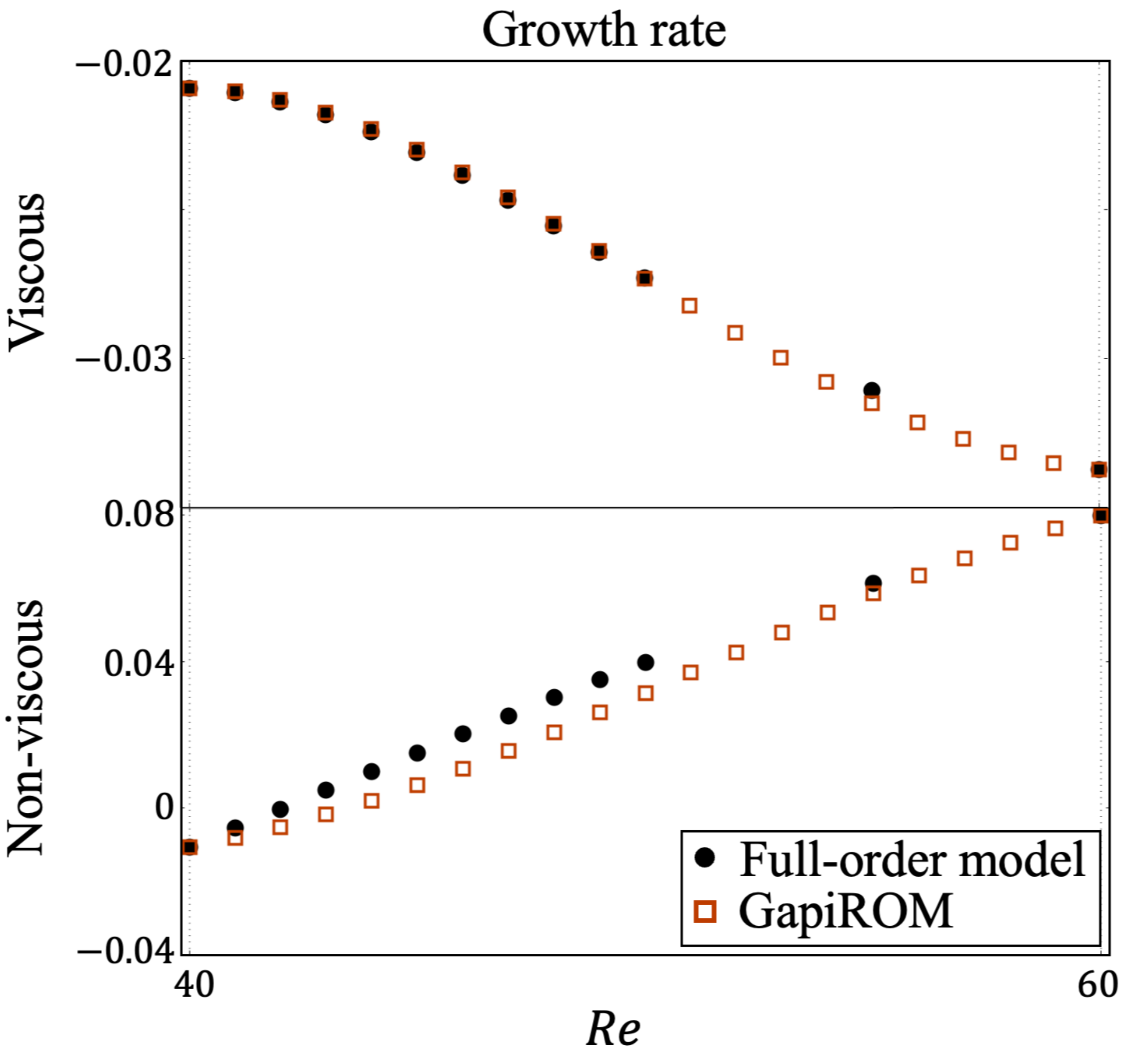}
  \end{minipage}
  \end{tabular}
  \captionsetup{justification=raggedright,singlelinecheck=false}
\caption{(a) Spatial distribution of the basis at $Re = 50$, obtained via Grassmann manifold interpolation using $\eta_0 = Re_0 = 40$ and $\eta_1 = Re_1 = 60$. (b) Comparison of mode distributions at $Re = 40,\ 45,\ 50,\ 55,$ and $60$. (c) Reynolds number dependence of the growth rate and frequency. (d) Reynolds number dependence of the viscous and non-viscous contributions to the growth rate.}
 \label{fig:figure_GSA_GapiROM}
\end{figure}

We predict the Reynolds-number dependence of the basis vectors using GapiROM, which interpolates reduced subspaces on the Grassmann manifold. The interpolation is performed between two reference Reynolds numbers, $\eta_0: (Re=40)$ and $\eta_1: (Re=60)$. Figure~\ref{fig:figure_GSA_GapiROM} (a) shows the two basis vectors, $\boldsymbol{\phi}^{\text{int}}_1$ and $\boldsymbol{\phi}^{\text{int}}_2$, at $Re = 50$, obtained via Grassmann manifold interpolation. These basis vectors exhibit spatial structures that capture physically meaningful features, particularly the asymmetric wake pattern characteristic of a Kármán vortex street.

Figure~\ref{fig:figure_GSA_GapiROM} (b) compares the spatial distributions of the basis vectors at $y = 0$ interpolated on the Grassmann manifold for $Re = 40,\ 45,\ 50,\ 55,\ 60$ with the corresponding POD basis vectors obtained from the matrix $X$ (full-order model). It is important to emphasize that interpolation on the Grassmann manifold does not directly interpolate the basis vectors themselves, but rather the subspaces they span. As a result, the interpolated basis vectors are not expected to exactly match the original POD basis vectors. However, in cases where a single dominant oscillatory eigenmode is present, the magnitudes of the spatial distributions remain identical across all basis vectors spanning the same subspace, regardless of the specific basis vector used. This behavior is evident at $Re = 60$, where the spatial distributions of the interpolated and full-order basis vectors coincide, up to a phase shift.

The interpolated basis vectors at $Re = 45,\ 50,\ 55$ closely resemble those obtained from the full-order model at the corresponding Reynolds numbers, indicating successful interpolation. In addition, the interpolated basis vectors vary smoothly with Reynolds number, likely due to the use of linear interpolation in the tangent space of the Grassmann manifold. In contrast, the POD basis vectors obtained from the full-order model do not exhibit strictly linear behavior. For example, a noticeable deviation appears between $Re = 55$ and $60$ near $x/D \approx 20$. These nonlinearities in the full-order model likely contribute to the discrepancies between the full-order and interpolated results.

Figure~\ref{fig:figure_GSA_GapiROM} (c) presents the growth rates and frequencies across various Reynolds numbers predicted by GapiROM using the interpolated basis. Since the reference Reynolds numbers for interpolation, $Re=40$ and $60$, yield growth rates and frequencies that closely match those of the full-order model, any discrepancies at intermediate values are likely attributable to differences in the basis, as shown in Fig.~\ref{fig:figure_GSA_GapiROM} (b). Nonetheless, the GapiROM results with interpolation show better agreement with the full-order model than that obtained without interpolation, as seen in Fig.\ref{fig:figure_GSA_eigenvalue_noint} (b).

As in the analysis without interpolation, the growth rate and frequency are decomposed into viscous and non-viscous components. Figure~\ref{fig:figure_GSA_GapiROM} (d) illustrates the Reynolds-number dependence of these components for the growth rate. The frequency components are omitted, as the viscous contribution remains nearly zero across all Reynolds numbers, consistent with previous GapiROM results without interpolation. The effect of the viscous term on the growth rate is accurately captured through subspace interpolation. Most of the prediction error originates from the non-viscous component, which may reflect nonlinear effects or limitations in representing the dynamics within a linearly interpolated subspace. Nonetheless, accounting for condition-dependent variations in the basis vectors clearly improves prediction accuracy.

\begin{figure}[htbp]
  \centering\includegraphics[width=12cm,keepaspectratio]{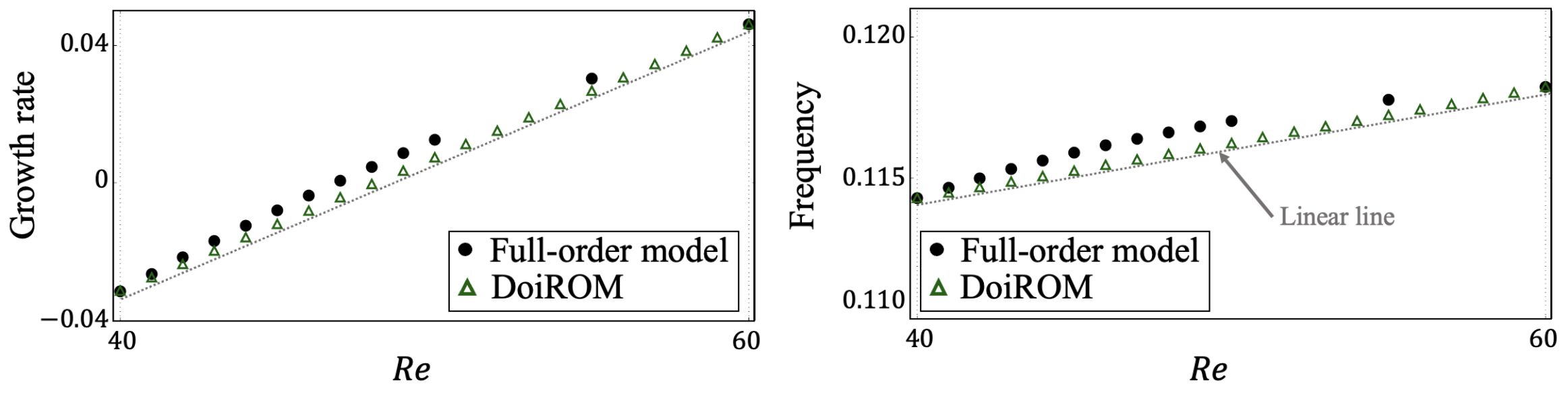}
\caption{Reynolds number dependence of the growth rate and frequency predicted by DoiROM.}
 \label{fig:figure_GSA_Doi}
\end{figure}

As the second operator-based ROM incorporating interpolation, predictions were made using DoiROM. The reference conditions are the same as those used in Fig.~\ref{fig:figure_GSA_GapiROM}; in particular, the interpolated subspace in DoiROM is identical to that employed in GapiROM. Figure \ref{fig:figure_GSA_Doi} shows the Reynolds-number dependence of the growth rate and frequency predicted by DoiROM. A linear relationship is observed between the growth rate and frequency at the two reference points, $\eta_0: (Re = 40)$ and $\eta_1: (Re = 60)$, indicating that the interpolation scheme in DoiROM operates linearly on the operator $\mathcal{F}$. This linear trend between the endpoints is a natural consequence of the interpolation method. In this sense, DoiROM provides a simple and interpretable model, where the accuracy of interpolation is directly reflected in the predicted results.

\subsection{Periodic flow around a circular cylinder}
\begin{figure}[htbp]
\begin{tabular}{cc}
\multicolumn{1}{l}{(a)}  &  \multicolumn{1}{l}{(b)}\\
  \begin{minipage}[b]{0.48\linewidth}
          \centering\includegraphics[height=3cm,keepaspectratio]{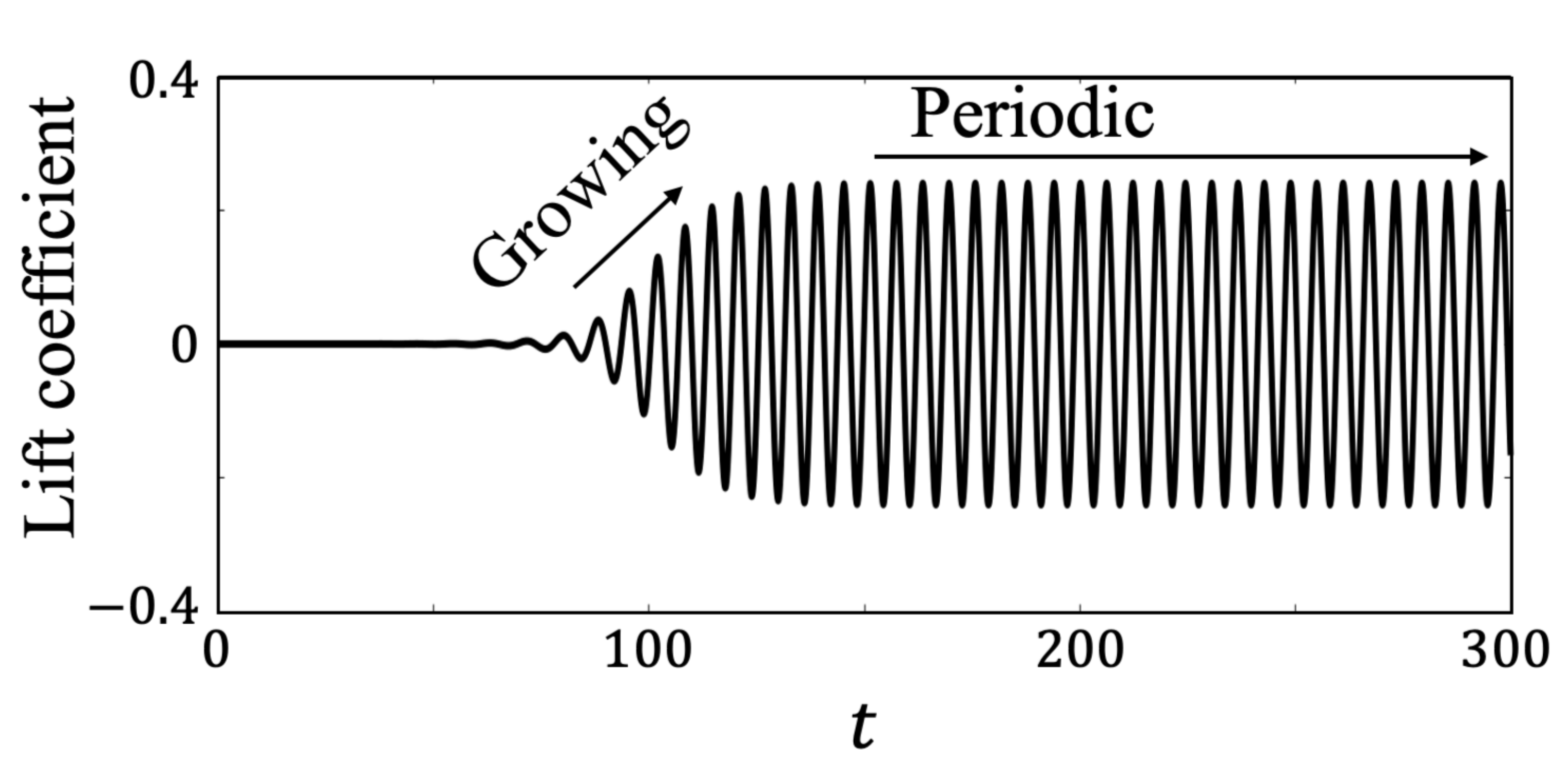}
  \end{minipage}
  &
  \begin{minipage}[b]{0.48\linewidth}
          \centering\includegraphics[height=3cm,keepaspectratio]{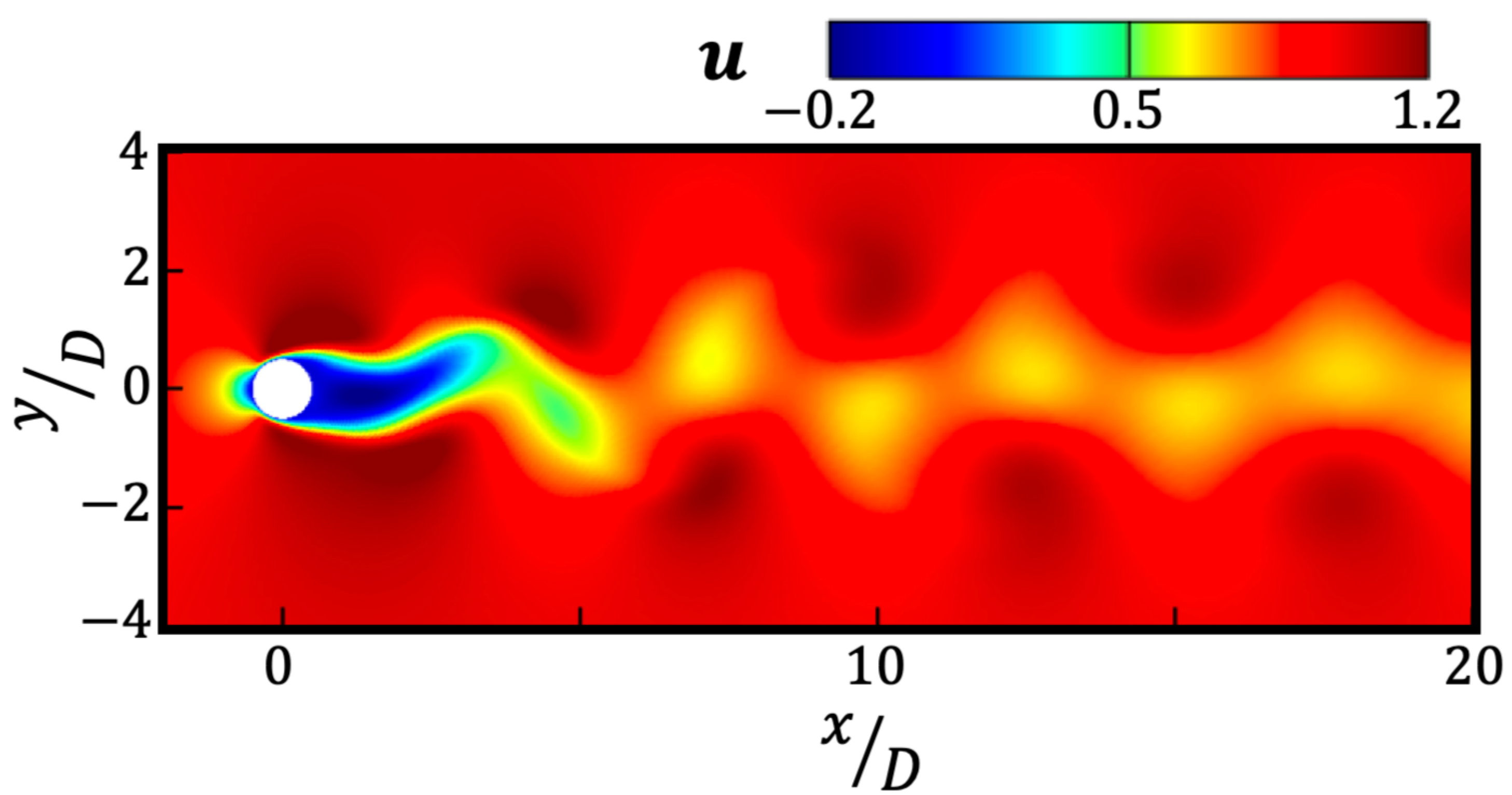}
  \end{minipage}\\
  \multicolumn{1}{l}{(c)}  &  \multicolumn{1}{l}{(d)}\\
  \begin{minipage}[b]{0.48\linewidth}
          \centering\includegraphics[height=3cm,keepaspectratio]{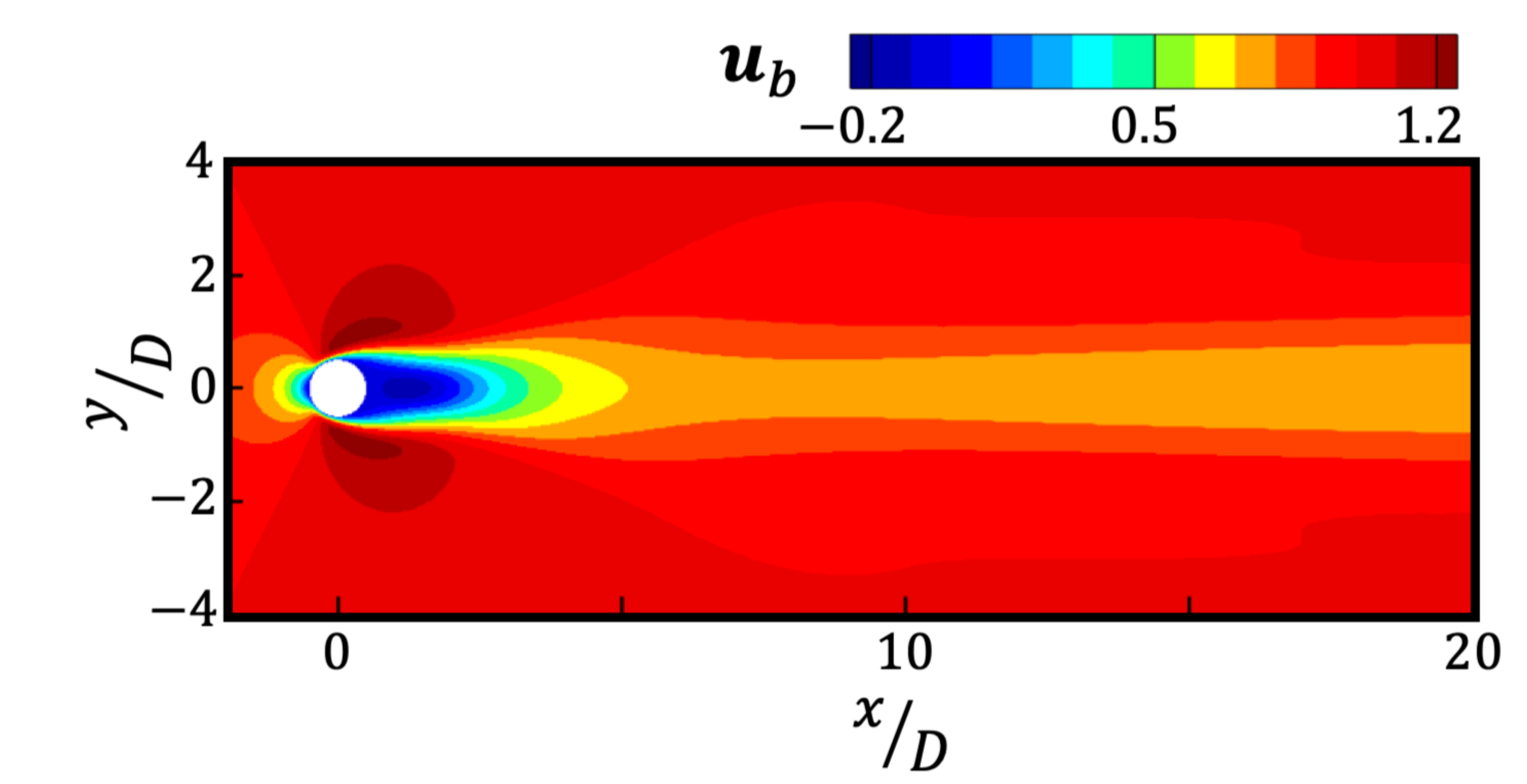}
  \end{minipage}
  &
  \begin{minipage}[b]{0.48\linewidth}
          \centering\includegraphics[height=3cm,keepaspectratio]{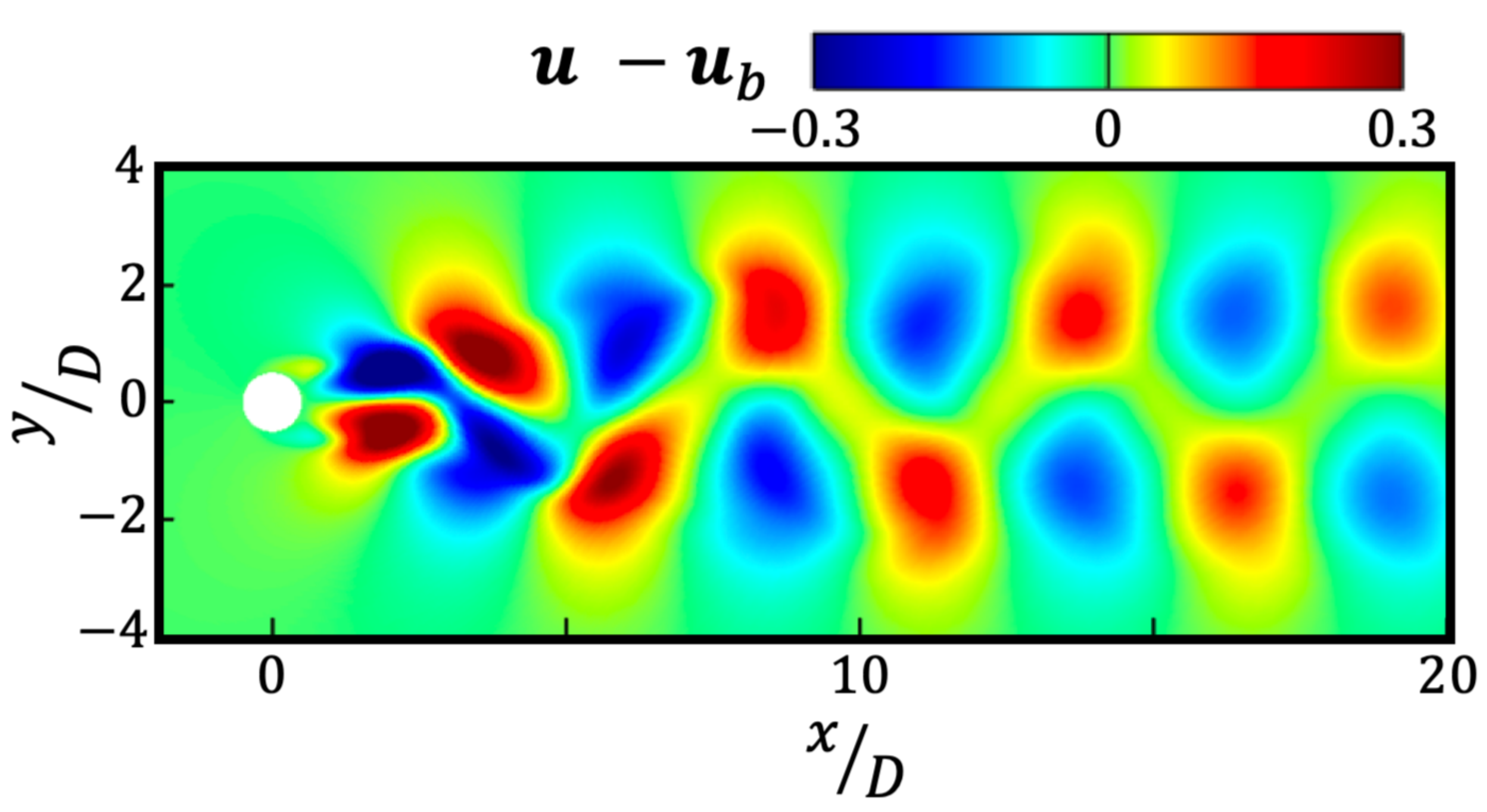}
  \end{minipage}
  \end{tabular}
  \captionsetup{justification=raggedright,singlelinecheck=false}
\caption{(a) Time history of the lift coefficient acting on the cylinder. (b) Instantaneous flow field at $t = 2250$, corresponding to a fully developed periodic state. (c) Time-averaged velocity field of the periodic flow, used as the base flow for GapiROM. (d) Velocity fluctuation field at $t = 2250$.}
 \label{fig:figure_peri_base}
\end{figure}

The second case considers a fully developed periodic flow past a two-dimensional circular cylinder. We first describe the flow characteristics and the procedure used to generate the dataset. The dataset is obtained from time-dependent simulations of the incompressible Navier–Stokes equations.
Figure~\ref{fig:figure_peri_base} (a) shows the time history of the lift coefficient on the cylinder at Reynolds number $100$. Starting from a uniform flow, the simulation is advanced in time until the flow reaches a fully periodic state. The oscillatory growth phase preceding the periodic regime is driven by unstable modes, as identified by global stability analysis.

Figure~\ref{fig:figure_peri_base} (b) shows an instantaneous velocity field in the periodic regime, illustrating unsteady vortex shedding in the wake of the cylinder.
The base flow required for the POD–Galerkin model is typically the time-averaged velocity field. Figure~\ref{fig:figure_peri_base} (c) presents this time-averaged field, denoted $\boldsymbol{u}_b$. Although this base flow resembles that used in the global stability analysis (see Fig.~\ref{fig:figure_GSA_data} (a)), its recirculation region is slightly smaller than that in the GSA base flow.

To construct the POD–Galerkin model, the snapshot matrix $X$ must represent fluctuations about the base flow. Each snapshot is therefore obtained by subtracting the time-averaged velocity field from the instantaneous velocity field. An example of such a fluctuation field is shown in Fig.~\ref{fig:figure_peri_base} (d). The fluctuation exhibits an asymmetric structure relative to the base flow. It is important to note that in the POD–Galerkin model, the fluctuation field is used as the snapshot data, whereas in DMD, the velocity field $\boldsymbol{u}$ is used directly.

\begin{figure}[htbp]
\begin{tabular}{cc}
  \multicolumn{1}{l}{(a)}  &  \multicolumn{1}{l}{}\\
  \multicolumn{2}{c}{
    \begin{minipage}[b]{1\linewidth}
        \centering\includegraphics[width=12cm,keepaspectratio]{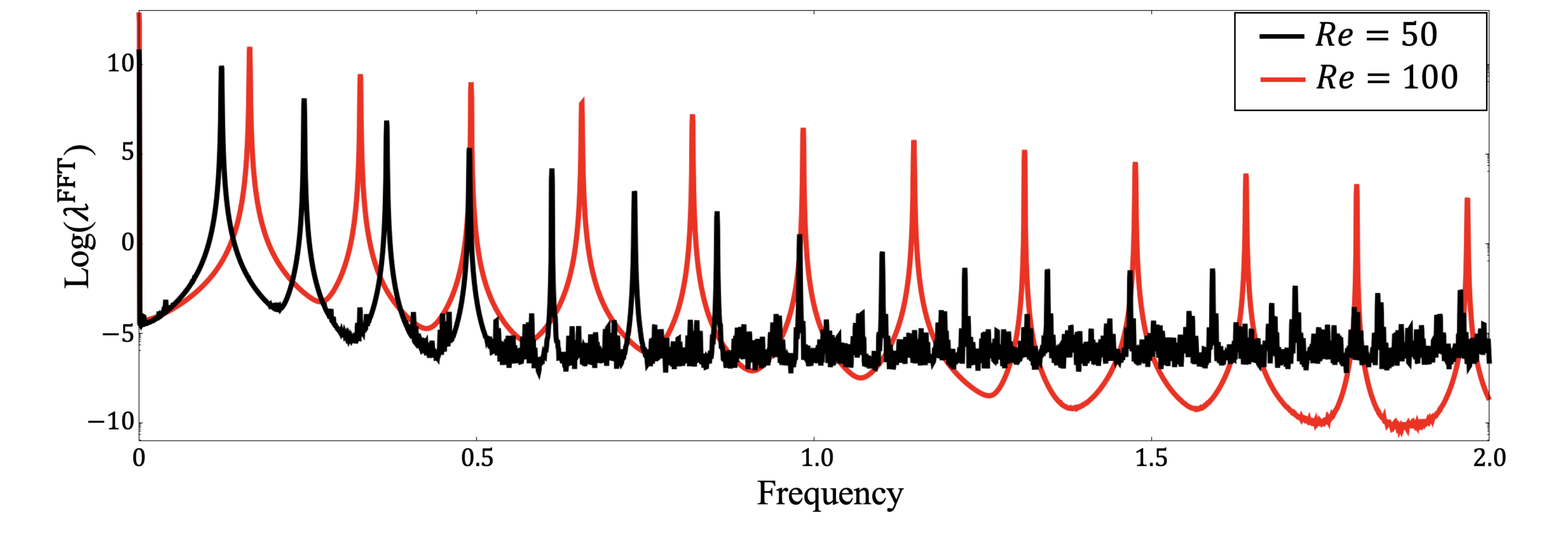}
    \end{minipage}
  }\\
\multicolumn{1}{l}{(b)}  &  \multicolumn{1}{l}{(c)}\\
  \begin{minipage}[b]{0.48\linewidth}
          \centering\includegraphics[height=6cm,keepaspectratio]{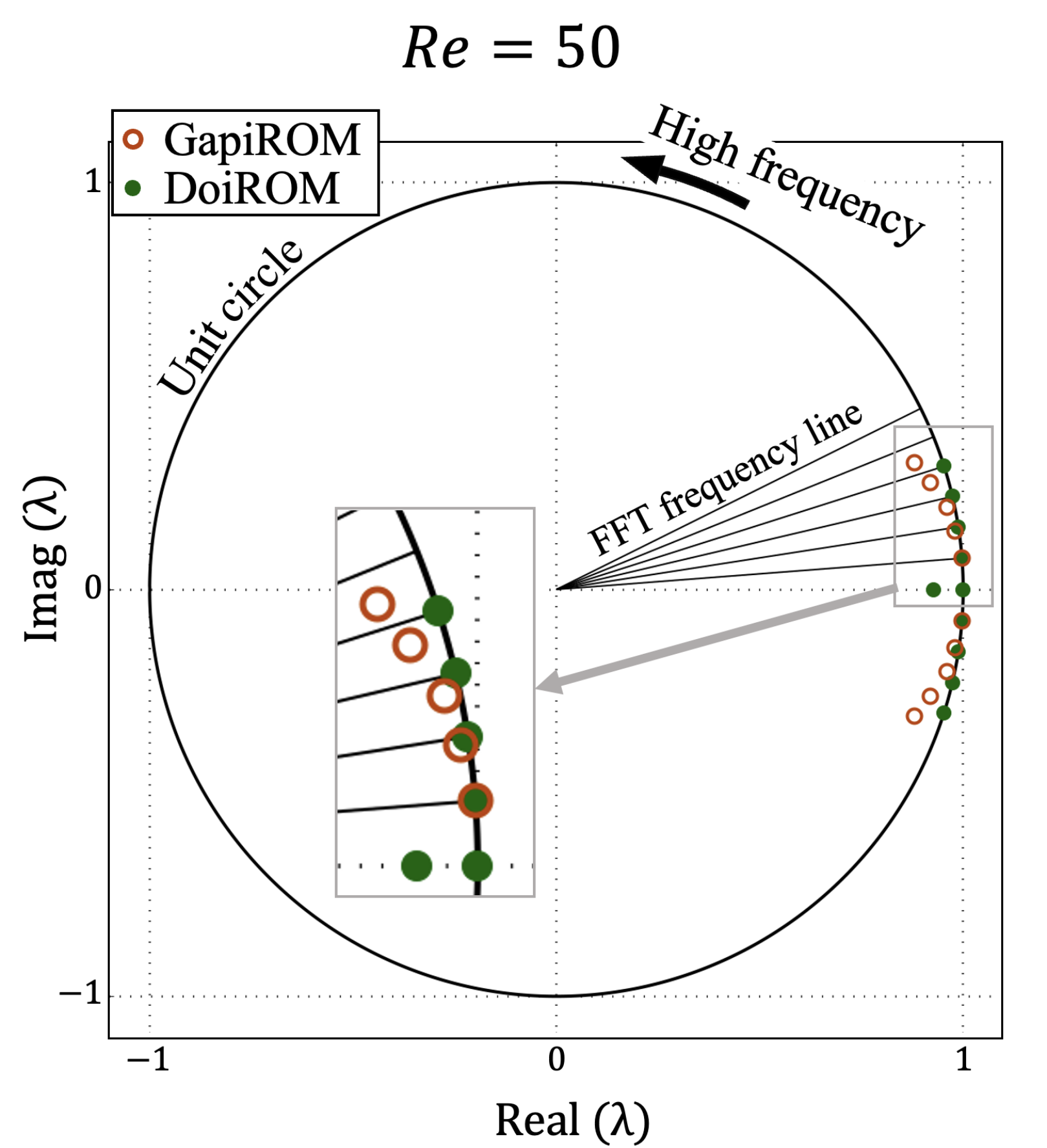}
  \end{minipage}
  &
  \begin{minipage}[b]{0.48\linewidth}
          \centering\includegraphics[height=6cm,keepaspectratio]{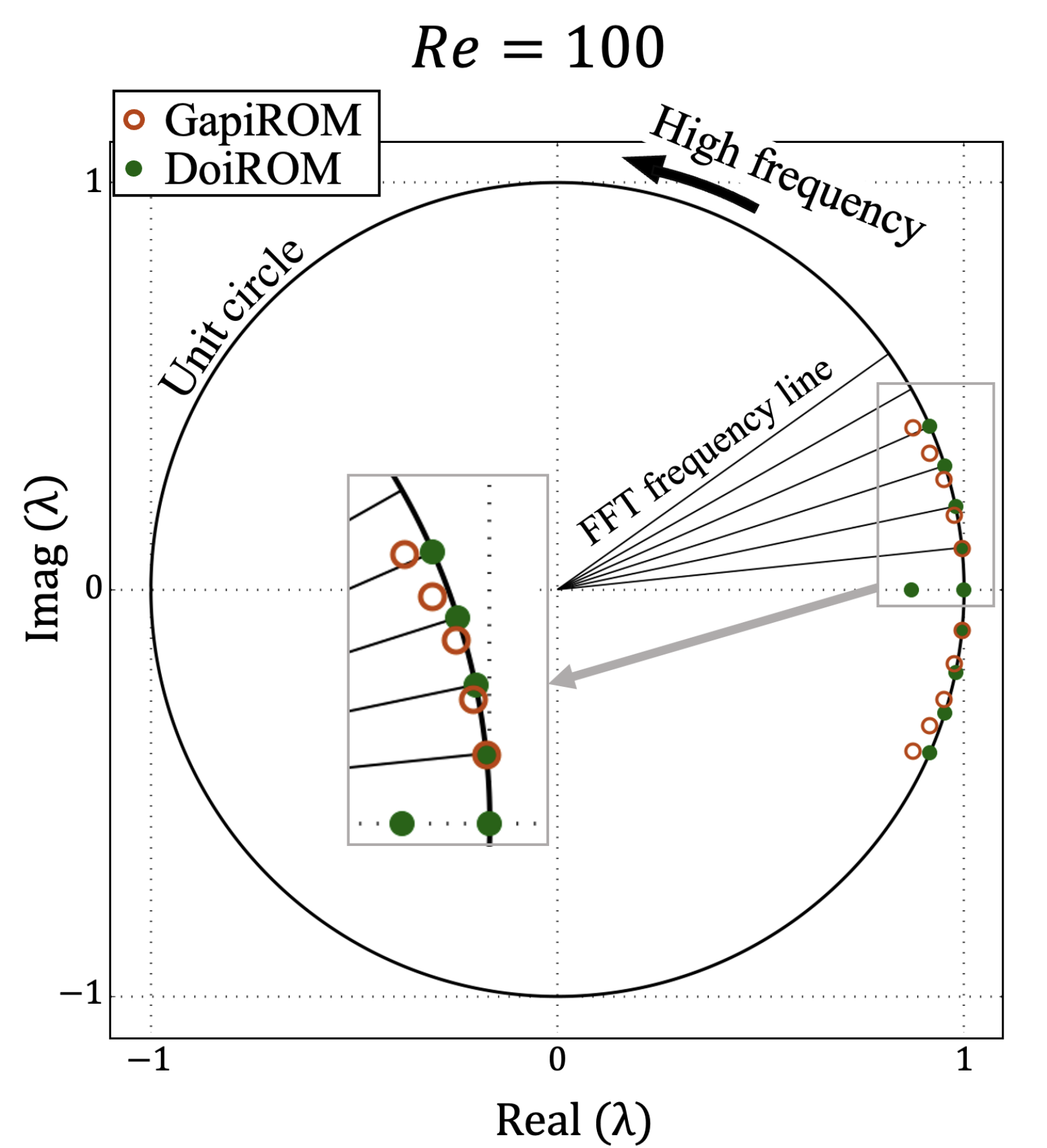}
  \end{minipage}\\
  \end{tabular}
  \captionsetup{justification=raggedright,singlelinecheck=false}
\caption{Comparison of frequencies obtained from FFT, GapiROM, and DoiROM using bases at $Re = 50$ and $100$. (a) Frequency spectrum from FFT analysis of the numerical simulation. (b) Eigenvalue distributions of GapiROM and DoiROM at $Re = 50$. (c) Eigenvalue distributions at $Re = 100$. In all cases, the ROM predictions are performed at the same Reynolds numbers used to construct the respective bases.}
 \label{fig:figure_peri_FFT}
\end{figure}

We performed an FFT analysis of numerically computed time-series velocity data at all spatial grid points for $Re=50$ and $100$.
The summed power spectral density is shown in Fig.~\ref{fig:figure_peri_FFT} (a). Spectral peaks appear at uniformly spaced frequencies, indicating that the flow consists of harmonic components of the fundamental frequency $f_1$. At higher frequencies, the noise level near each peak increases, particularly for the lower Reynolds number ($Re=50$). This increase is attributed to two main factors: numerical errors in the FFT computation and numerical errors in the simulation. In both GapiROM and DoiROM, the dimensionality reduction via POD suppresses the effects of these noisy components because POD basis vectors with low energy, which are often associated with high-frequency noise, are truncated.

The eigenvalue spectra of GapiROM and DoiROM at $Re=50$ and $100$ are computed using POD basis vectors obtained at the corresponding Reynolds numbers. Figures~\ref{fig:figure_peri_FFT} (b) and (c) show the eigenvalue distributions at $Re=50$ and $100$, respectively. The number of retained POD basis vectors is set to $10$, following previous studies on POD–Galerkin models of cylinder flow~\cite{GP_deane}. To ensure a consistent comparison with DoiROM, the eigenvalues of GapiROM are transformed as $e^{\lambda^{\text{GapiROM}}\Delta T}$, where $\Delta T = 0.1$, matching the value used in DoiROM. The eigenvalues of GapiROM lie within the unit circle, except for those corresponding to the fundamental frequency. These components exhibit a slightly positive growth rate on the order of $\mathcal{O}(10^{-3})$ for both Reynolds numbers.
Because the GapiROM operator is constructed by linearizing the governing equations around the time-averaged flow (that is, the base flow), it is particularly effective at capturing dynamics near this flow. As shown in previous linear stability analyses of the time-averaged flow~\cite{Mittal_2007,barkley2006linear,mypaper_7}, eigenmodes of the fundamental frequency typically emerge in the vicinity of the base flow. The observation that only the fundamental frequency eigenmodes exhibit growth in GapiROM therefore aligns with its theoretical foundation.

In contrast, all nonzero-frequency eigenvalues in DoiROM lie precisely on the unit circle, reflecting the neutrally stable nature of the periodic flow. Although both DoiROM and GapiROM employ linearized representations of flow dynamics, it is important to note that DMD, and therefore DoiROM, identifies a best-fit linear operator directly from the dataset~\cite{brunton2021modern}. This approach is conceptually distinct from the linearization method used in GapiROM. The frequencies extracted by DoiROM closely match those obtained via FFT, as their angular positions align with the peaks in the FFT spectrum.
Quantitative comparisons of the fundamental frequency are summarized in Table~\ref{table_freq}. The frequency values predicted by both ROMs show close agreement with the FFT results and with previously reported results~\cite{Mittal_2005,cylinder4}. In modeling modal eigenvalues for fully developed periodic flows, DoiROM proves effective, particularly since GapiROM does not capture the eigenvalues of higher-order harmonic frequencies.


\begin{table}[hbpt]
 \caption{The lowest frequency of $Re=50$ and $100$.}
 \label{table2_2}
 \centering
  \begin{tabular}{|c|cc|}
   \hline
   Re & 50 & 100 \\
   \hline \hline
   Present FFT ($\Delta f = 6.1\times 10^{-4} $) & $0.1221$   & $0.1642$ \\
    \hline
   Present GapiROM & $0.1222$   & $0.1642$ \\
    \hline
   Present  DoiROM & $0.1223$   & $0.1640$ \\
    \hline
   Mittal \cite{Mittal_2005}  & $0.1227$  & $0.1644$ \\
     \hline
   Jiang and Cheng \cite{cylinder4}  & $0.1238$  & $0.1652$ \\
   \hline
  \end{tabular}
  \label{table_freq}
\end{table}

\begin{figure}[htbp]
\begin{tabular}{cc}
\multicolumn{1}{l}{(a)}  &  \multicolumn{1}{l}{(b)}\\
  \begin{minipage}[b]{0.48\linewidth}
          \centering\includegraphics[height=6cm,keepaspectratio]{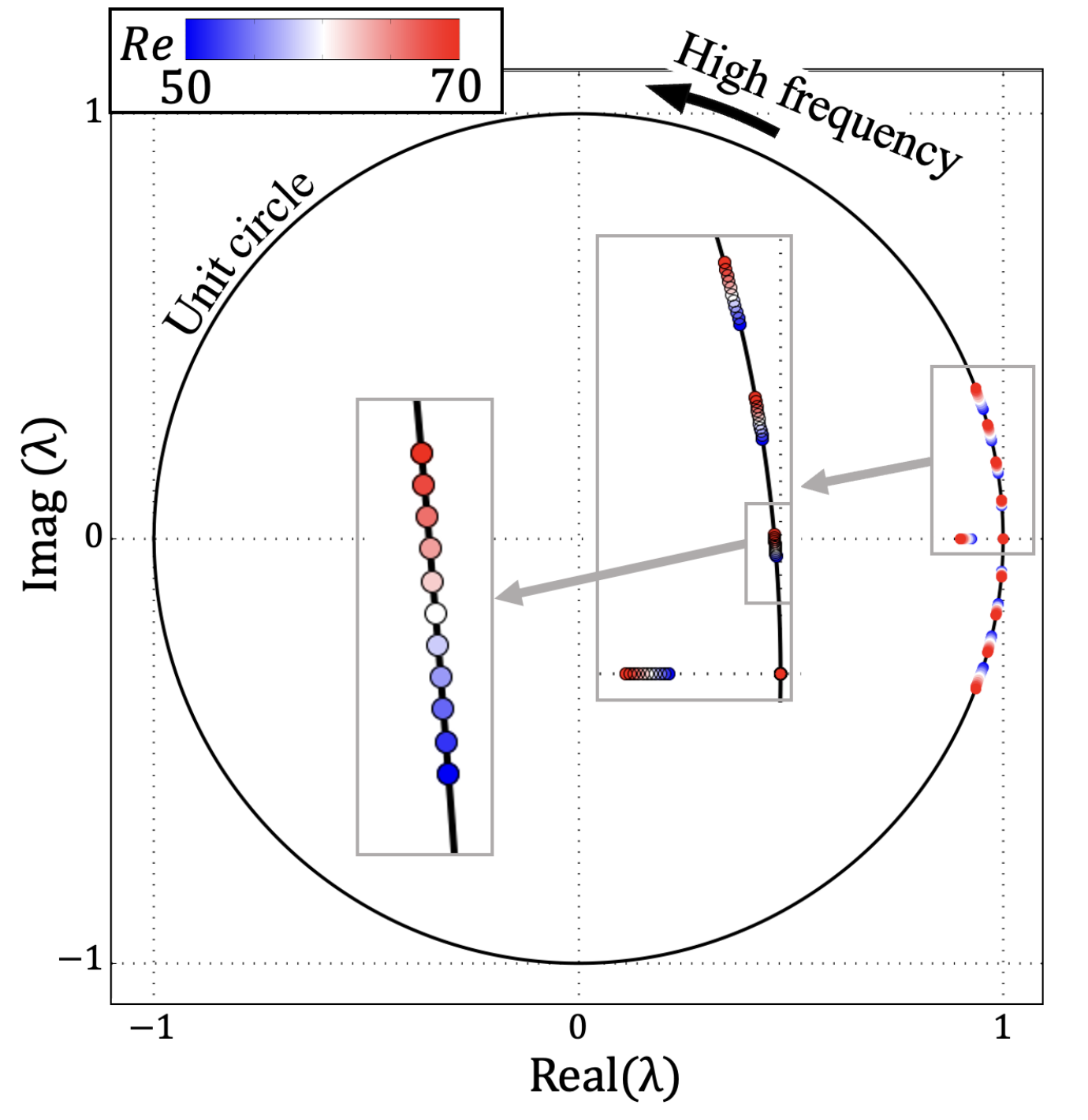}
  \end{minipage}
  &
  \begin{minipage}[b]{0.48\linewidth}
          \centering\includegraphics[height=6cm,keepaspectratio]{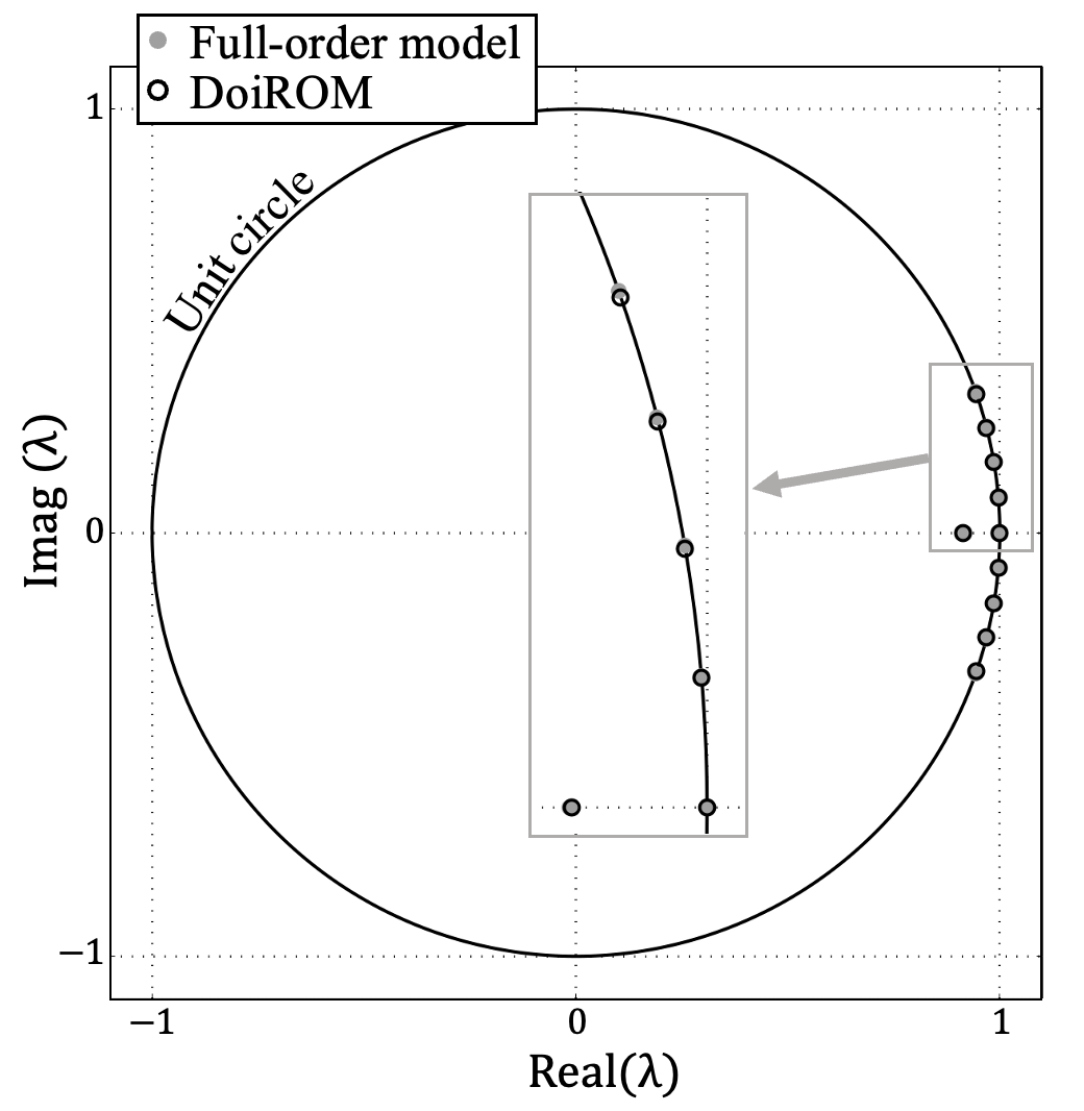}
  \end{minipage}\\
  \multicolumn{1}{l}{(c)}  &  \multicolumn{1}{l}{(d)}\\
  \begin{minipage}[b]{0.48\linewidth}
          \centering\includegraphics[height=6cm,keepaspectratio]{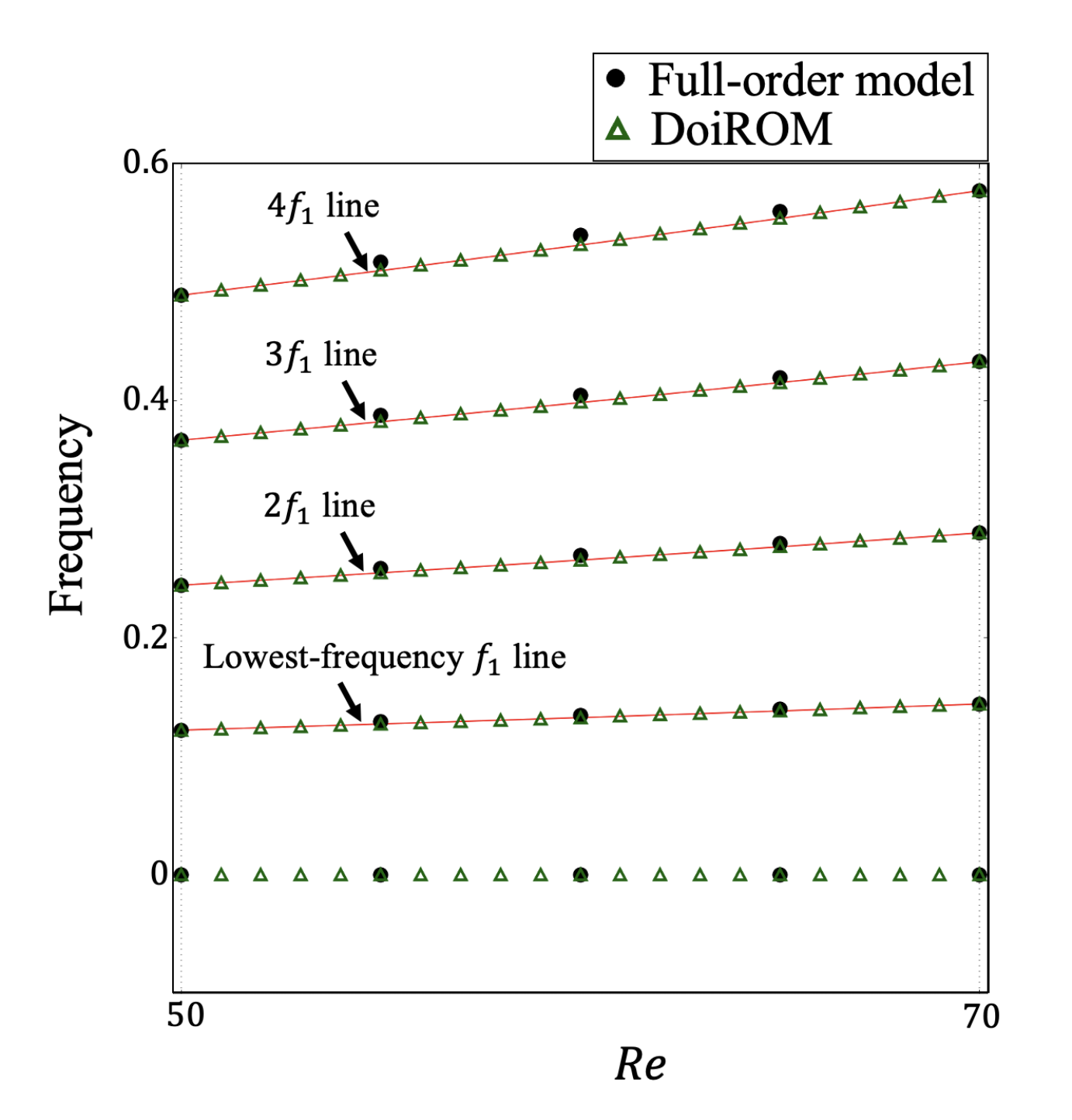}
  \end{minipage}
  &
  \begin{minipage}[b]{0.48\linewidth}
          \centering\includegraphics[height=6cm,keepaspectratio]{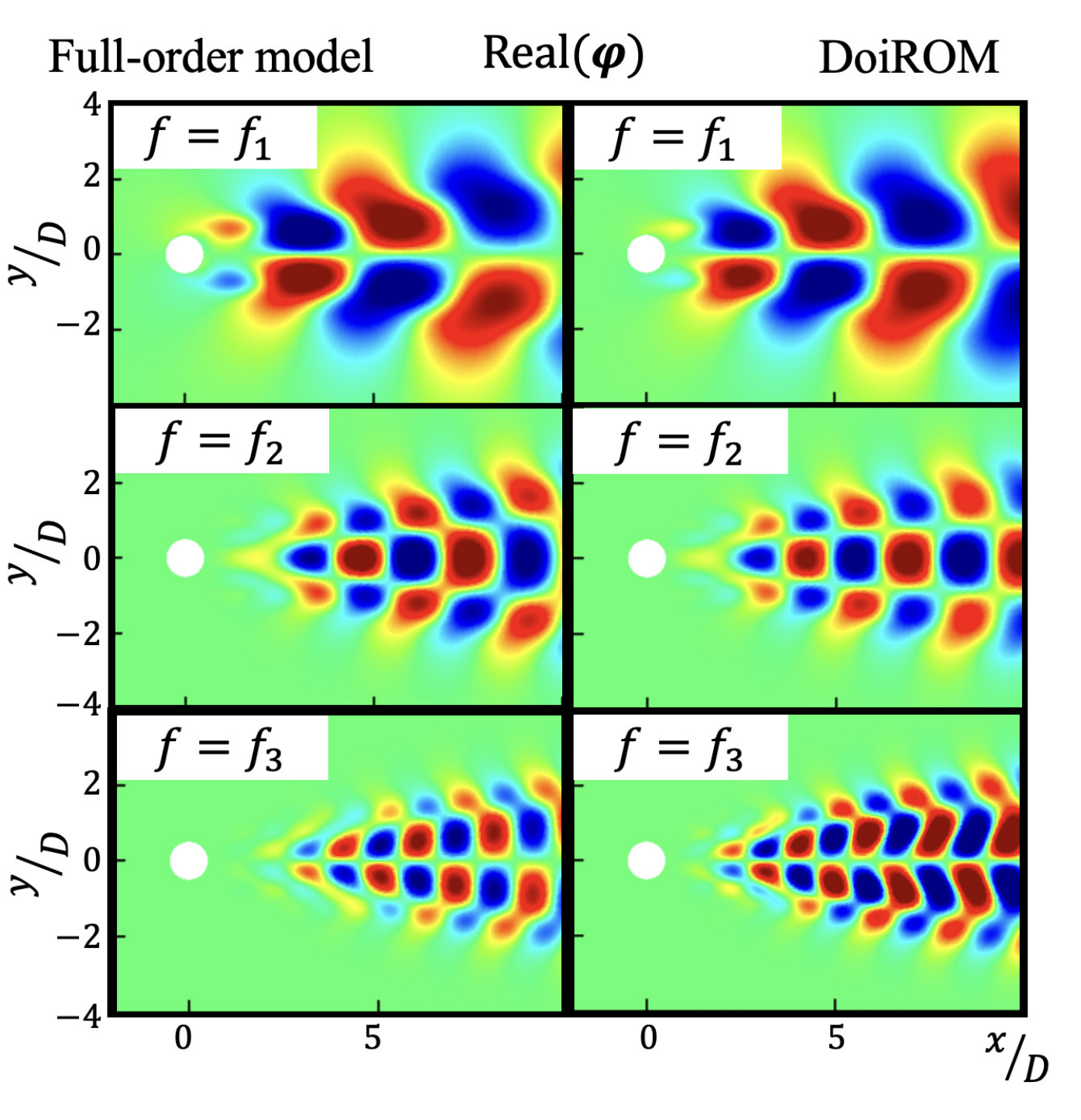}
  \end{minipage}
  \end{tabular}
  \captionsetup{justification=raggedright,singlelinecheck=false}
\caption{(a) Predicted eigenvalue distributions at $Re = 50$–$70$ using DoiROM, with reference parameters $\eta_0 = 50$ and $\eta_1 = 70$. (b) Comparison of eigenvalue distributions at $Re = 60$ between the full-order model and DoiROM. (c) Reynolds number dependence of the frequencies computed from the eigenvalues of both the full-order model and DoiROM. The predicted frequencies correspond to harmonics of the fundamental frequency $f_1$, indicated by red lines. (d) Comparison of representative eigenmodes from the full-order model and DoiROM at $Re = 60$, showing the real parts of the eigenmodes.}
 \label{fig:figure_peri_DoiROM}
\end{figure}

We investigated the predictive performance of DoiROM in terms of eigenvalues and eigenmodes at Reynolds numbers different from those used to construct the subspace. GapiROM is omitted from the remainder of this paper because it cannot capture the harmonic characteristics of periodic flows. Figure~\ref{fig:figure_peri_DoiROM} (a) shows the predicted eigenvalue distributions at Reynolds numbers ranging from $50$ to $70$, using $\eta_0: (Re = 50)$ and $\eta_1: (Re = 70)$ as the reference conditions. A smooth variation in the eigenvalue distribution is observed across this range. At $Re = 60$, Fig.~\ref{fig:figure_peri_DoiROM} (b) compares the eigenvalue distribution obtained from the full-order model via DMD with that predicted by DoiROM. The predicted eigenvalues generally agree well with those of the full-order model, particularly for the lower-frequency eigenmodes that dominate the flow dynamics around the cylinder.

Figure~\ref{fig:figure_peri_DoiROM} (c) shows the Reynolds-number dependence of the frequencies computed from the eigenvalues of both the full-order model and DoiROM. The frequencies predicted by DoiROM vary linearly between the values at the reference Reynolds numbers. All nonzero frequencies align with the harmonics of the fundamental frequency $f_1$, as indicated by the red line in the figure. This result indicates that the DoiROM prediction preserves the operator structure, including its harmonic content, an intrinsic physical feature of the periodic flow at the reference conditions.

Figure~\ref{fig:figure_peri_DoiROM} (d) compares the eigenmodes obtained from DoiROM with those from the DMD (full-order model). The predicted eigenmodes show qualitative agreement, even for higher-frequency eigenmodes, except for differences in phase. It is important to note that, even for the same linear operator, the phase of eigenmodes is arbitrary and may not align exactly. These results demonstrate that the operator estimation via DoiROM is effective for interpolating periodic flow dynamics.

\begin{figure}[htbp]
\begin{tabular}{c}
\multicolumn{1}{l}{(a)}\\
  \begin{minipage}[b]{1\linewidth}
          \centering\includegraphics[width=10cm,keepaspectratio]{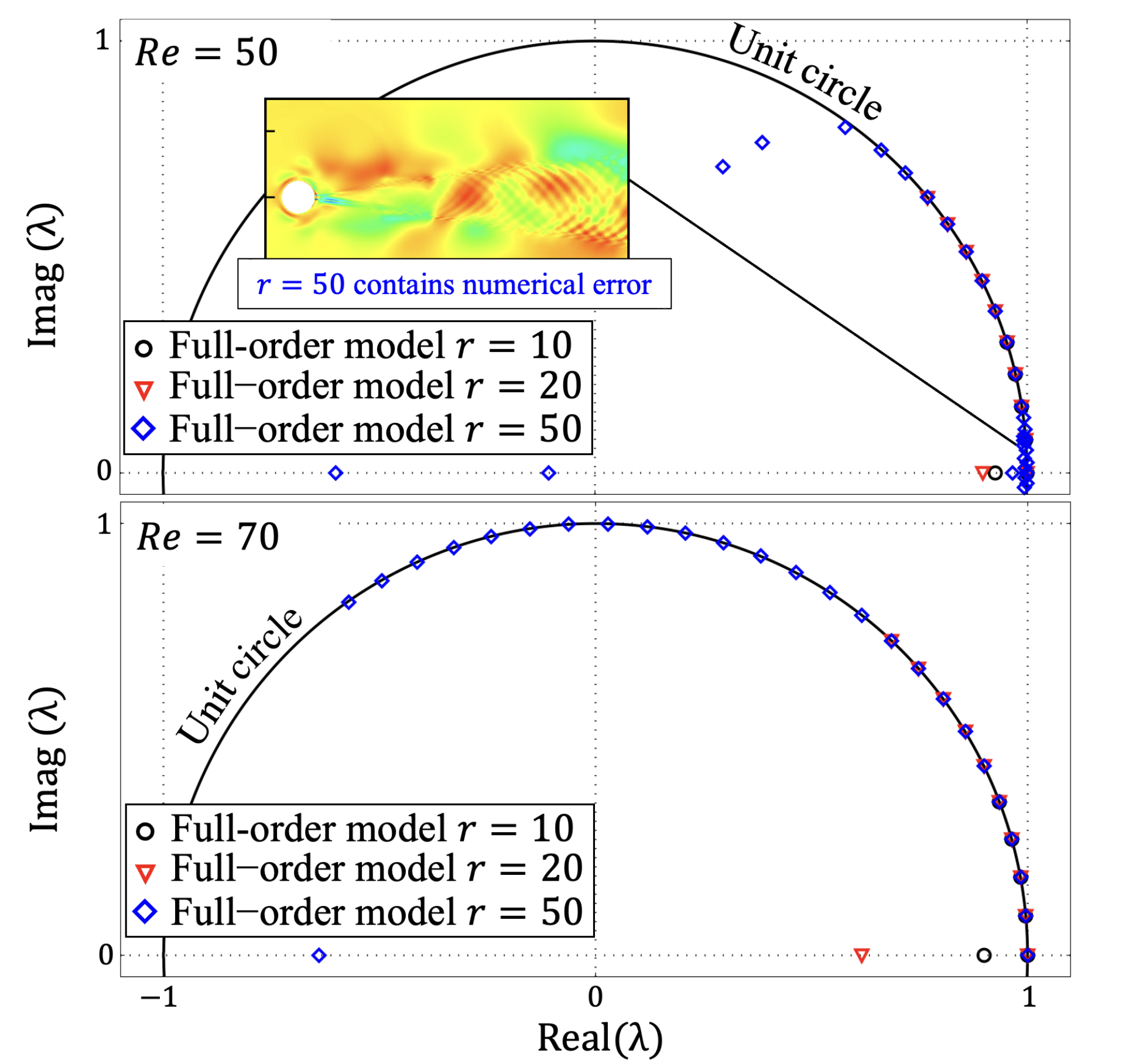}
  \end{minipage}\\
\multicolumn{1}{l}{(b)}\\
  \begin{minipage}[b]{1\linewidth}
          \centering\includegraphics[width=10cm,keepaspectratio]{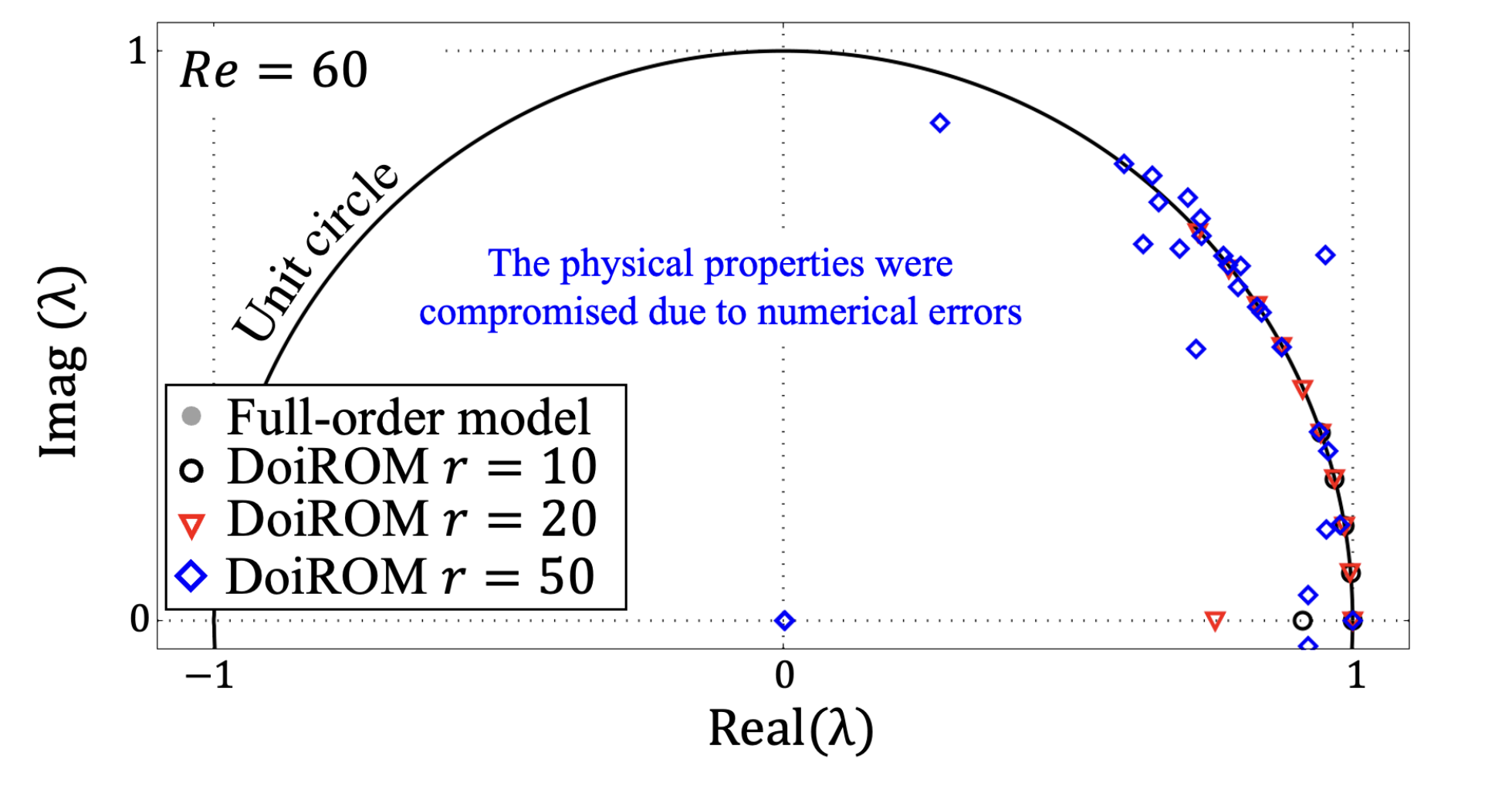}
  \end{minipage}\\
  \end{tabular}
  \captionsetup{justification=raggedright,singlelinecheck=false}
\caption{(a) Eigenvalue distributions of the full-order model at $Re = 50$ (top) and $Re = 70$ (bottom) for different numbers of POD basis vectors, obtained via POD of the snapshot matrix $X$, with $r = 10$, $20$, and $50$. (b) Eigenvalue distributions predicted by DoiROM using $\eta_0: (Re = 50)$ and $\eta_1: (Re = 70)$. At $r = 50$, the presence of non-physical frequencies in the basis at $Re = 50$ leads to reduced accuracy in predicting the harmonics of the fundamental frequency.}
 \label{fig:figure_peri_r}
\end{figure}

We investigated the effect of the subspace dimension $r$, which corresponds to the number of retained POD basis vectors, on the interpolation performance of DoiROM. DMD was performed using $r = 10$, $20$, and $50$ at two reference Reynolds numbers, $Re = 50$ and $70$, to assess how the eigenvalue distribution depends on the choice of subspace dimension. Figure~\ref{fig:figure_peri_r} (a) shows the eigenvalue distributions of DMD at the reference Reynolds numbers. At $Re = 50$, the eigenvalues for $r = 10$ and $20$ appear only at harmonics of the fundamental frequency. In contrast, at $r = 50$, additional eigenvalues emerge at non-harmonic frequencies. These are considered nonphysical based on their spatial structure and likely reflect numerical artifacts associated with low-energy POD basis vectors not representing coherent flow dynamics. At $Re = 70$, however, the eigenvalues remain aligned with the harmonic frequencies even for $r = 50$, which is consistent with the observation from the FFT analysis (Fig.~\ref{fig:figure_peri_FFT} (a)) that noise effects diminish at higher Reynolds numbers.

To evaluate the predictive capability of DoiROM, the eigenvalues at $Re = 60$ are estimated based on the reference conditions at $Re = 50$ and $70$, using $r = 10$, $20$, and $50$. The corresponding eigenvalue distribution from the full-order model is shown in Fig.~\ref{fig:figure_peri_r} (b). For $r = 10$ and $20$, the predicted eigenvalues align well with the harmonics of the fundamental frequency, indicating accurate interpolation of the operators. However, for $r = 50$, while some eigenvalues remain near the harmonic frequencies, many are located inside or outside the unit circle, indicating the presence of decaying or growing modes. These results highlight that using an excessively large number of POD basis vectors can degrade the predictive performance of DoiROM. In particular, the number of POD basis vectors must be selected to ensure that the reduced operators at the reference conditions are not contaminated by numerical errors.

\begin{figure}[htbp]
\begin{tabular}{cc}
\multicolumn{1}{l}{(a)}  &  \multicolumn{1}{l}{(b)}\\
  \begin{minipage}[b]{0.48\linewidth}
          \centering\includegraphics[width=6cm,keepaspectratio]{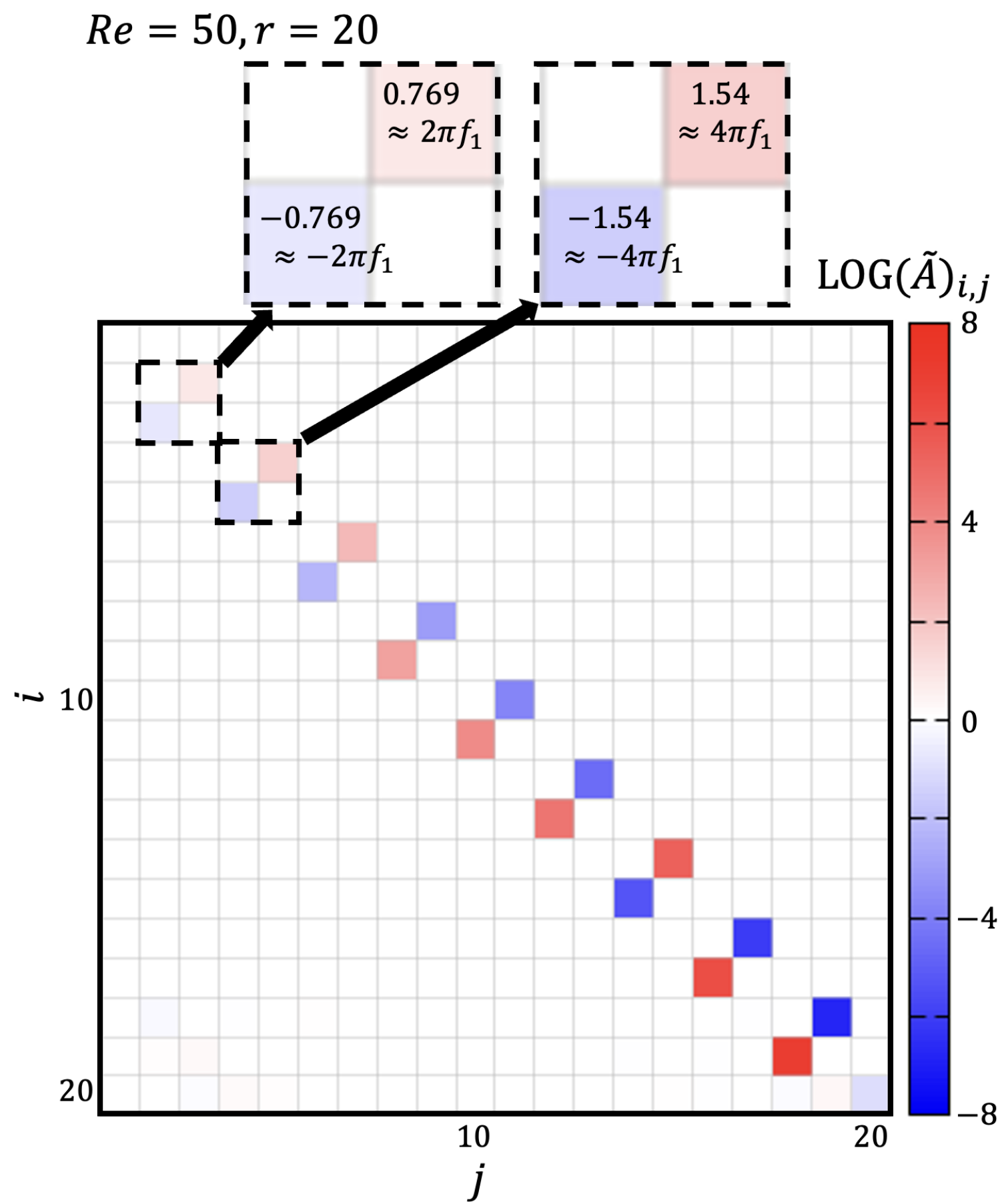}
  \end{minipage}
  &
  \begin{minipage}[b]{0.48\linewidth}
          \centering\includegraphics[width=6cm,keepaspectratio]{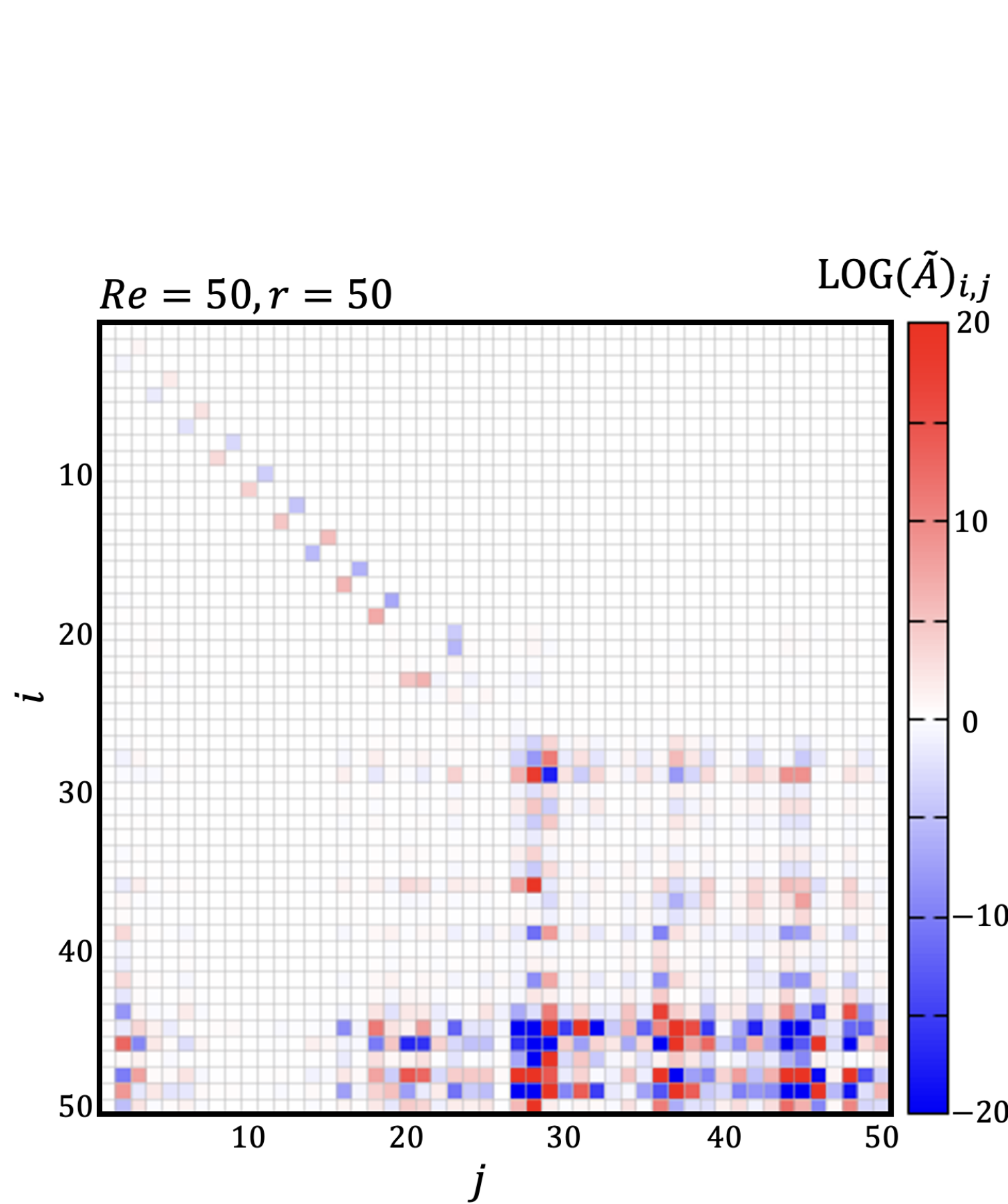}
  \end{minipage}\\
  \multicolumn{1}{l}{(c)}  &  \multicolumn{1}{l}{(d)}\\
  \begin{minipage}[b]{0.48\linewidth}
          \centering\includegraphics[width=6cm,keepaspectratio]{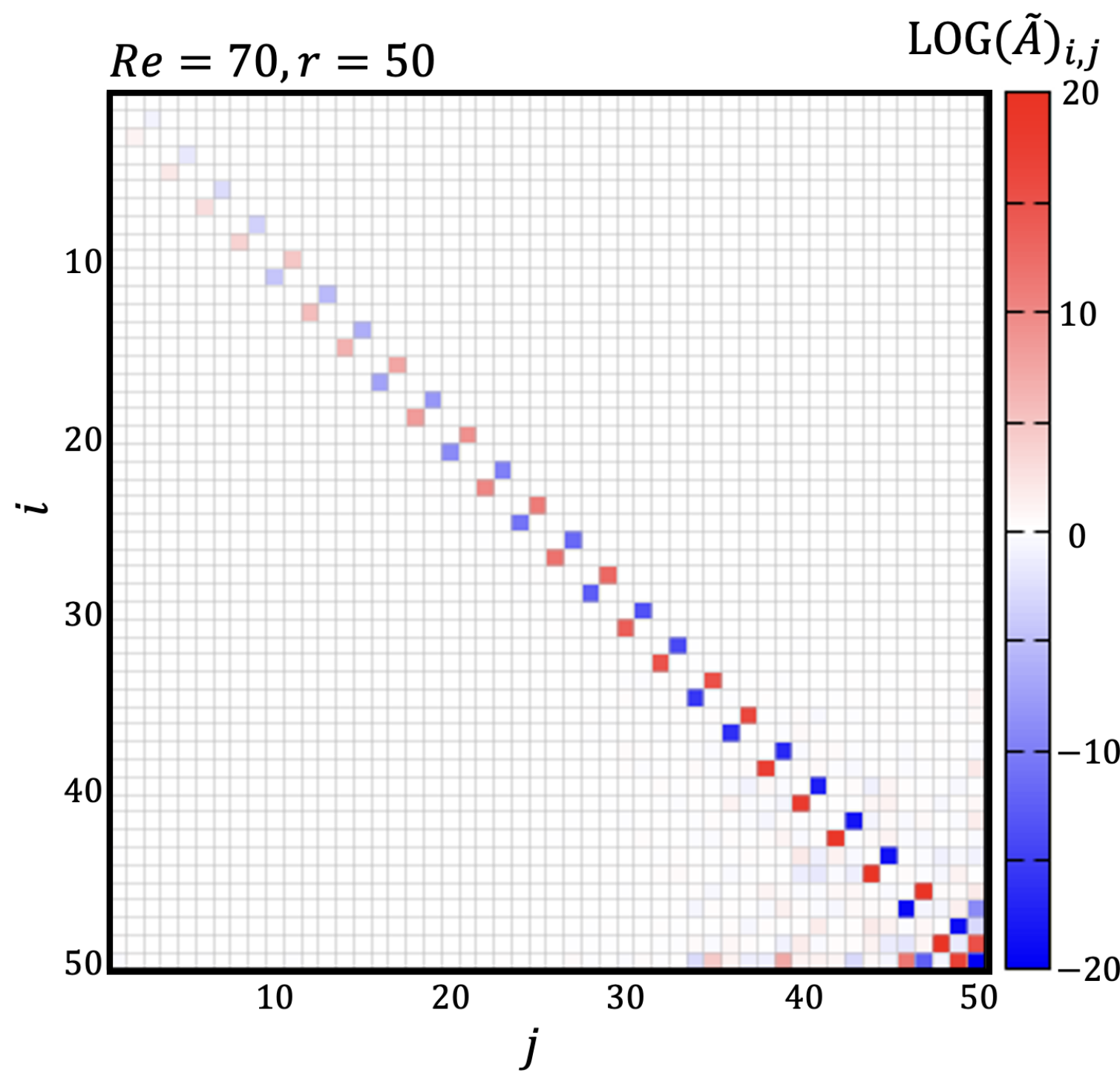}
  \end{minipage}
  &
  \begin{minipage}[b]{0.48\linewidth}
          \centering\includegraphics[width=6cm,keepaspectratio]{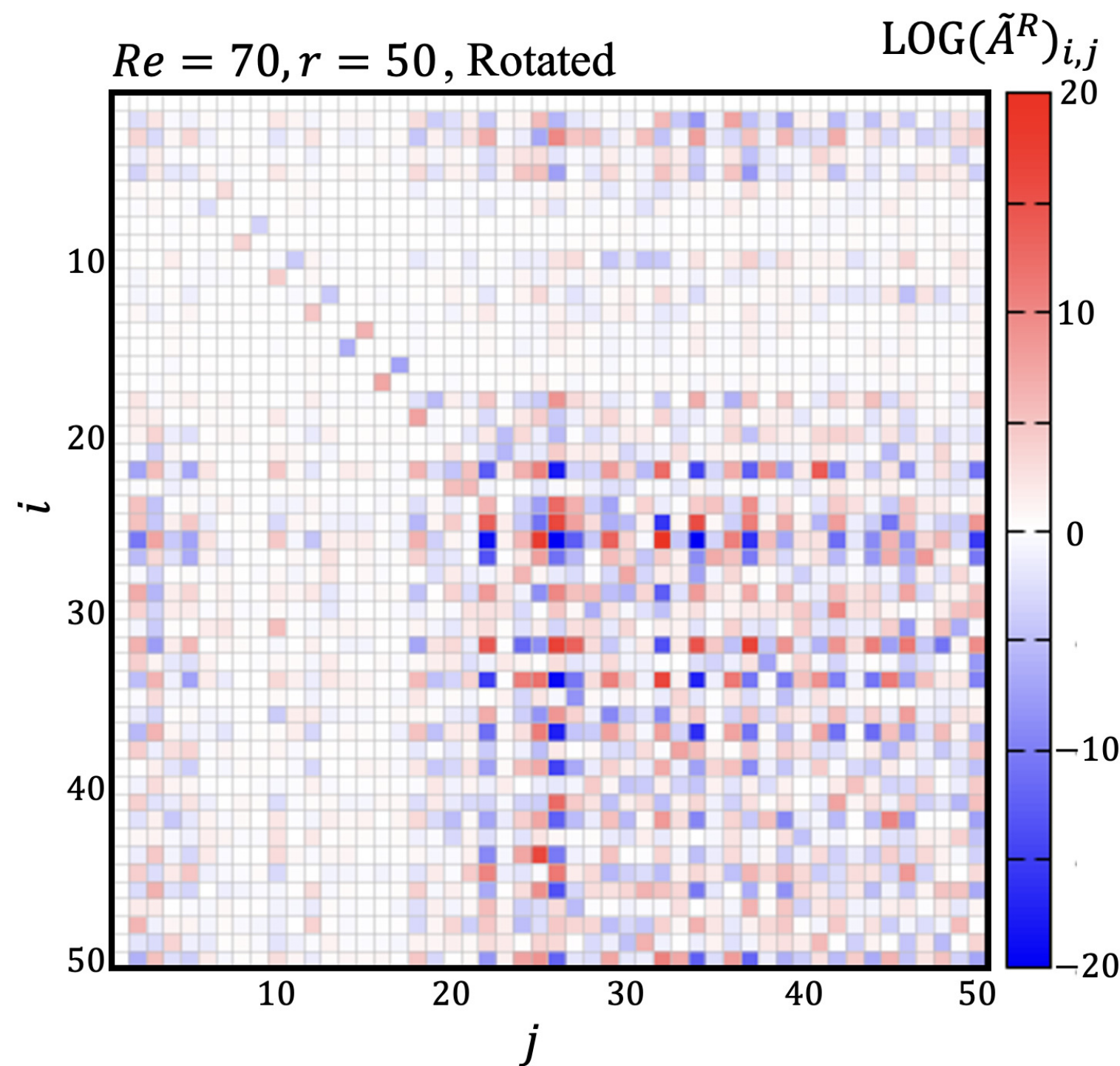}
  \end{minipage}
  \end{tabular}
  \captionsetup{justification=raggedright,singlelinecheck=false}
\caption{Visualization of the matrix elements of the continuous-time linear operator at different Reynolds numbers and subspace dimensions. (a) Operator matrix at $Re = 50$ with subspace dimension $r = 20$, exhibiting a banded structure with dominant elements near the diagonal. (b) Operator matrix at $Re = 50$ with $r = 50$, where nonzero elements are scattered irregularly due to numerical errors in the low-energy POD basis vectors. (c) Operator matrix at $Re = 70$ with $r = 50$, showing a banded structure similar to (a). (d) Transformed operator matrix $\text{Log}(\tilde{A^R})$ at $Re = 70$, obtained using the transformation defined in Eq.~(\ref{procrastes_U_A'}). The banded structure is disrupted, illustrating the adverse impact of fitting to numerical artifacts.}
 \label{fig:figure_peri_last}
\end{figure}

To further investigate why interpolation fails even when physically meaningful eigenvalues and eigenmodes are present alongside numerical errors, we visualized the elements of the linear operator. Figure~\ref{fig:figure_peri_last} shows the distribution of matrix elements for the continuous-time linear operator. In Fig.~\ref{fig:figure_peri_last} (a), the operator matrix at $Re = 50$ with a subspace dimension of $r = 20$ exhibits a banded structure, with significant values concentrated near the diagonal. Notably, the diagonal entries are nearly zero, while the nonzero elements form a regular pattern.

By examining a representative $2 \times 2$ submatrix, we observe that the off-diagonal elements correspond to positive and negative frequencies. This observation is consistent with Eq.~(\ref{eigenseparate3}), which relates the entries of a $2 \times 2$ matrix to the associated frequency and growth rate. Specifically, the growth rate is given by the sum of the diagonal elements, and the fact that these values are close to zero supports the physical consistency of the operator. This relationship suggests that frequency and growth rate are encoded within the submatrix, indicating that the POD basis vectors effectively isolate individual frequencies in periodic flows. Thus, a pair of POD basis vectors represents a single-frequency oscillation.

In Fig.\ref{fig:figure_peri_last} (b), the linear operator matrix at $Re = 50$ with $r = 50$ shows irregularly scattered nonzero entries across the matrix. This irregularity is attributed to numerical errors introduced through the low-energy POD basis vectors. In contrast, Fig.\ref{fig:figure_peri_last} (c) shows the matrix at $Re = 70$ with $r = 50$, where nonzero entries are again concentrated near the diagonal, resembling the structured pattern observed in Fig.\ref{fig:figure_peri_last} (a).
When performing interpolation in DoiROM using the operator matrices at $Re = 50$ and $Re = 70$ with $r = 50$, the matrix at $Re = 70$ is transformed according to Eq.(\ref{procrastes_U_A'}). The element distribution of the transformed matrix, $\mathrm{LOG}(\tilde{A}^R)$, is presented in Fig.~\ref{fig:figure_peri_last} (d). The previously observed banded structure near the diagonal is disrupted by the transformation, indicating a loss of the original near-diagonal structure. 

Interestingly, even within the submatrix defined by $1 \leq i \leq 20$ and $1 \leq j \leq 20$, where a banded structure was evident at $Re = 50$, $r = 50$, the transformed matrix $\mathrm{LOG}(\tilde{A}^R)$ at $Re = 70$ exhibits irregularly scattered nonzero elements. Although the transformation involves an invertible matrix and thus preserves eigenvalues, the eigenvalues of the interpolated operator constructed from these transformed matrices differ significantly from those of the full-order models. This discrepancy indicates a failure to preserve the underlying physical properties through interpolation.

A comparison between the submatrices defined by $1 \leq i, j \leq 20$ in $\mathrm{LOG}(\tilde{A})$ at $Re = 50$ and in $\mathrm{LOG}(\tilde{A}^R)$ at $Re = 70$ reveals differences in the distribution of nonzero elements. Consequently, linear interpolation between these matrices combines regions of zero and nonzero values, resulting in interpolated matrices that lack physical consistency.
As discussed in \ref{apena}, the transformation using a regular matrix $R$ minimizes the Frobenius norm between the bases. However, in this case, the matrix $R$ appears to have aligned the basis at $Re = 70$ with POD basis vectors that contain numerical errors. While such transformations are generally intended to align structurally similar operator matrices, the problem here is that the transformation responds excessively to numerical artifacts. This excessive sensitivity distorts the structure of $\mathrm{LOG}(\tilde{A}^R)$ at $Re = 70$, ultimately undermining the effectiveness of the interpolation.

\subsection{Flow around elliptical cylinders with aspect ratio variation}
Operator interpolation is applied to the flow past cylinders and ellipses with various aspect ratios. The elliptical geometries are generated by applying a conformal mapping to a computational grid originally constructed around a circular cylinder, as detailed in \ref{ellipse}.

\begin{figure}[htbp]
  \centering\includegraphics[width=8cm,keepaspectratio]{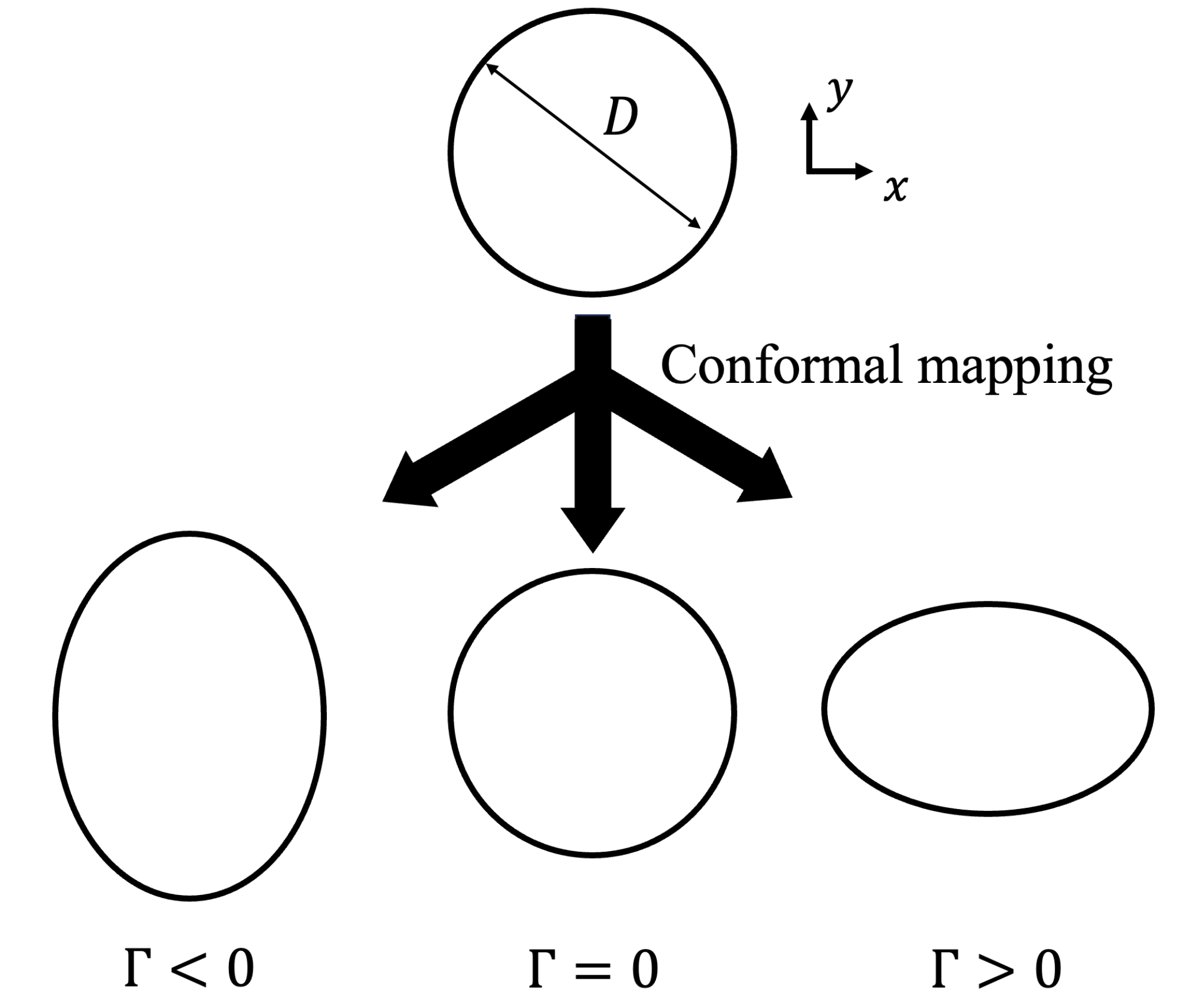}
\caption{Schematic illustration of ellipses generated via conformal mapping, illustrating the signed aspect ratio $\Gamma$.}
 \label{fig:figure_ellipse_conformal}
\end{figure}

The aspect ratio of an ellipse is defined using a signed aspect ratio that characterizes the geometric relationship between the major and minor axes according to the following convention:
\begin{equation}
\Gamma = \left\{1-\frac{\min(D_x, D_y)}{\max(D_x, D_y)}\right\} \operatorname{sign}(D_x - D_y),
   \label{sign_Gamma}
\end{equation}
where $D_x$ and $D_y$ denote the semi-axes in the horizontal ($x$) and vertical ($y$) directions, respectively. The relationship between $\Gamma$ and the shape of the ellipse is illustrated in Fig.~\ref{fig:figure_ellipse_conformal}. A positive value of $\Gamma$ indicates elongation in the $y$-direction, while a negative value indicates elongation in the $x$-direction. The magnitude of $\Gamma$ increases with the degree of elongation. Following our previous study\cite{mypaper2}, the diameter $D$ of the original circular cylinder, i.e., before conformal mapping, is used as the reference length for all geometries. In this section, the Reynolds number is fixed at $100$ for all cases.

Numerical simulations are performed for $\Gamma = -0.7,\ -0.5,\ -0.3,\ 0$, and $0.2$ to investigate the dependence of the flow field on $\Gamma$. The $z$-component of the vorticity, $\omega_z$, is computed from the velocity field following the approach adopted in previous studies on secondary vortex streets in the wake of an ellipse\cite{Jiang_2019,shi2020wakes,variousshapes2}. The nondimensional vorticity is defined as
\begin{equation}
\omega_z = \frac{D}{U_{\infty}}\left(\frac{\partial (u)_y}{\partial x} - \frac{\partial (u)_x}{\partial y}\right),
\label{eqvolz}
\end{equation}
where $(u)_x$ and $(u)_y$ are the velocity components in the $x$- and $y$-directions, respectively. Hereafter, we refer to $\omega_z$ as vorticity.
Note that the vorticity is used solely for flow visualization; all subsequent analyses are based on the velocity field, consistent with the earlier sections.

\begin{figure}[htbp]
  \centering\includegraphics[width=12cm,keepaspectratio]{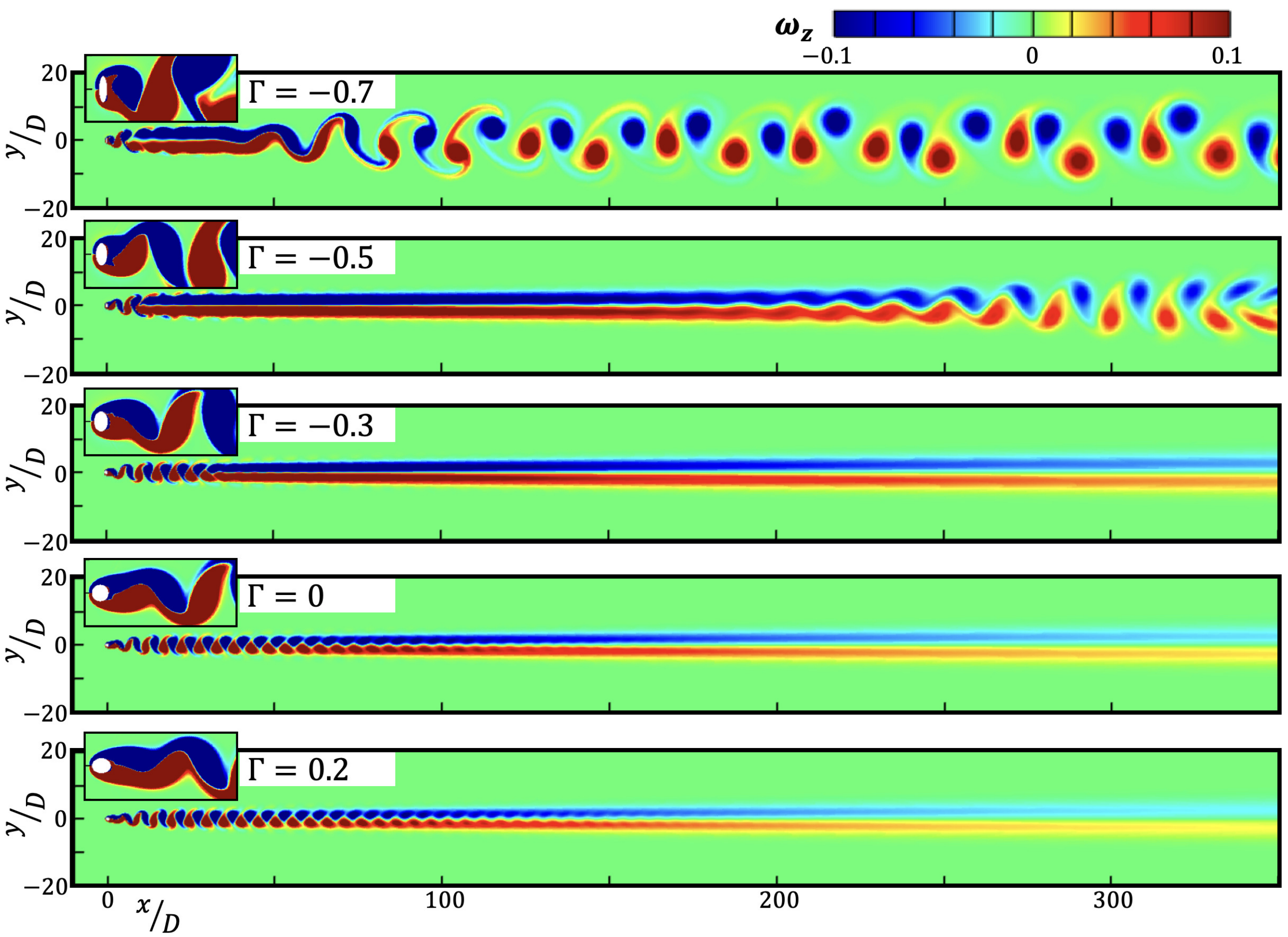}
\caption{Vorticity fields for flows around ellipses with various signed aspect ratios: $\Gamma = -0.7,\ -0.5,\ -0.3,\ 0$, and $0.2$. While the near-field vorticity distributions are similar across all cases, notable differences emerge in the far wake ($x > 100D$). Periodic vortex shedding persists for $\Gamma \geq -0.3$, whereas secondary vortex streets appear for $\Gamma \leq -0.5$.}
 \label{fig:figure_ellipse_vol}
\end{figure}

Figure~\ref{fig:figure_ellipse_vol} presents the vorticity fields for $\Gamma = -0.7,\ -0.5,\ -0.3,\ 0$, and $0.2$. The vorticity distributions near the objects are similar across all cases. In contrast, distinct differences appear in the far wake region ($x > 100D$), particularly when comparing the cases with $\Gamma \geq -0.3$ and those with $\Gamma \leq -0.5$.
For $\Gamma \geq -0.3$, the magnitude of vorticity decays downstream, indicating a periodic flow regime. This behavior is consistent with the flow around a circular cylinder at $Re = 100$. Thus, their spectral characteristics are similar to those shown in Fig.~\ref{fig:figure_peri_FFT} (a). In contrast, the far wake for $\Gamma \leq -0.5$ exhibits a distinct change in vortex pattern, identified as a secondary vortex street. As shown in previous studies\cite{Jiang_2019,variousshapes2}, secondary vortices typically emerge at frequencies lower than the primary Kármán vortex shedding frequency, corresponding to larger spatial scales.

\begin{figure}[htbp]
  \centering\includegraphics[width=12cm,keepaspectratio]{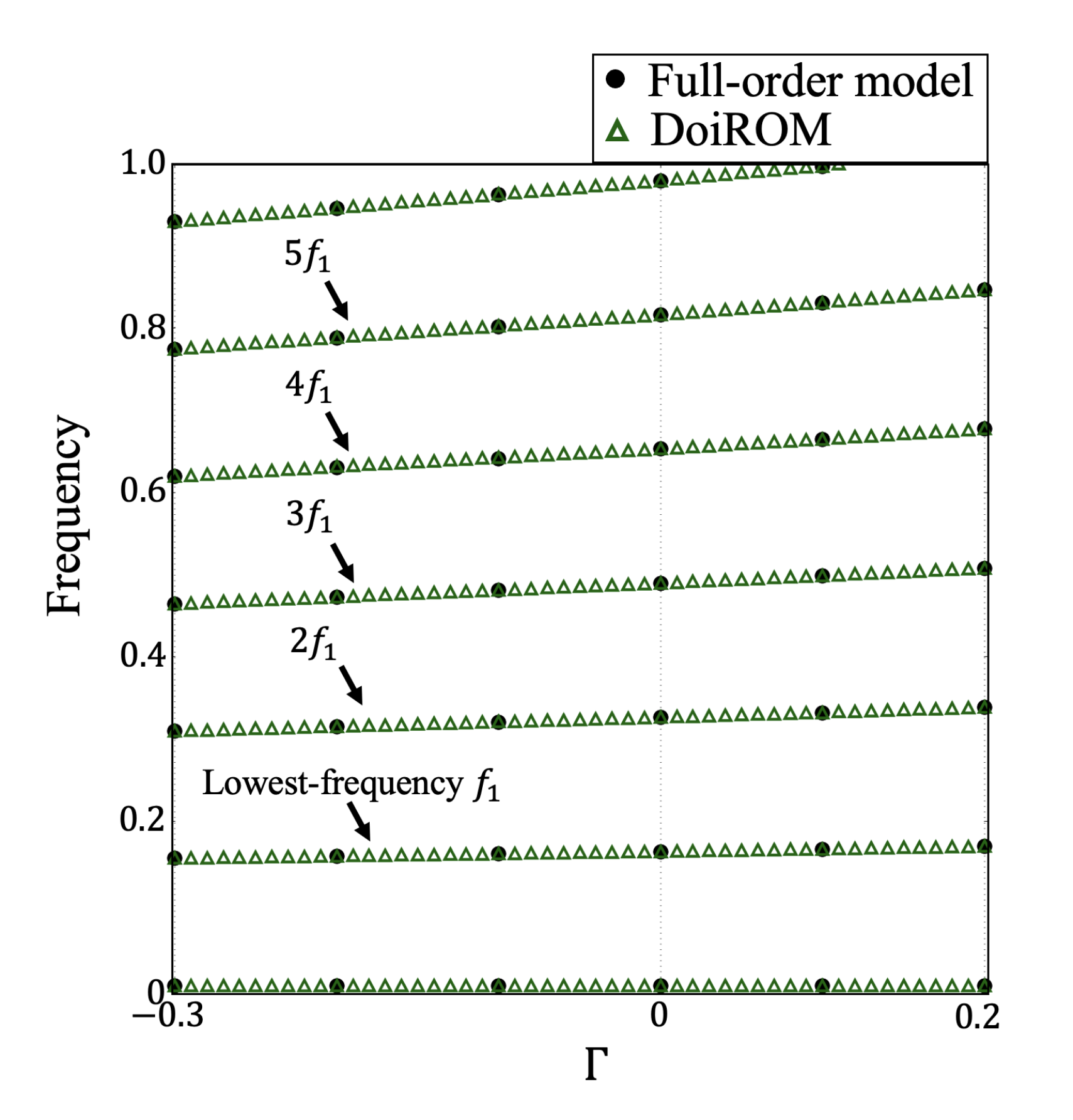}
\caption{Variation of frequencies with respect to the signed aspect ratio $\Gamma$, obtained using DoiROM. The linear trend indicates the effectiveness of operator interpolation across different geometries.}
 \label{fig:figure_ellipse_peri_freq}
\end{figure}

As a test case for applying ROM to interpolate subspace across different geometries, we first consider periodic flows past ellipses with varying aspect ratios in the range $\Gamma \geq -0.3$. Operator interpolation is performed using DoiROM, with reference data corresponding to the three cases shown in Fig.\ref{fig:figure_ellipse_vol}, specifically $\Gamma = -0.3$, $0$, and $0.2$.
Variations in the computational grid due to changes in geometry are accounted for using grid weights, following the methodology described in our previous study\cite{mypaper2}. The number of POD basis vectors $r$ is set to $20$, based on the ROM results presented in Fig.\ref{fig:figure_peri_r}. The DMD results for the reference conditions are omitted here, as they are qualitatively similar to those for periodic flow around a circular cylinder. The corresponding eigenvalues consist of the fundamental frequency and its harmonics, each having zero growth rate.
Figure~\ref{fig:figure_ellipse_peri_freq} shows the variation of frequency depending on $\Gamma$, as obtained from the interpolated operator using the reference conditions. The results indicate that the frequencies vary linearly with changes in signed aspect ratio, demonstrating that operator interpolation remains effective even across different geometries. Furthermore, the harmonic characteristic of periodic flows is well preserved.

\begin{figure}[htbp]
\begin{tabular}{cc}
\multicolumn{1}{l}{(a)}  &  \multicolumn{1}{l}{(b)}\\
  \begin{minipage}[b]{0.48\linewidth}
          \centering\includegraphics[height=6.5cm,keepaspectratio]{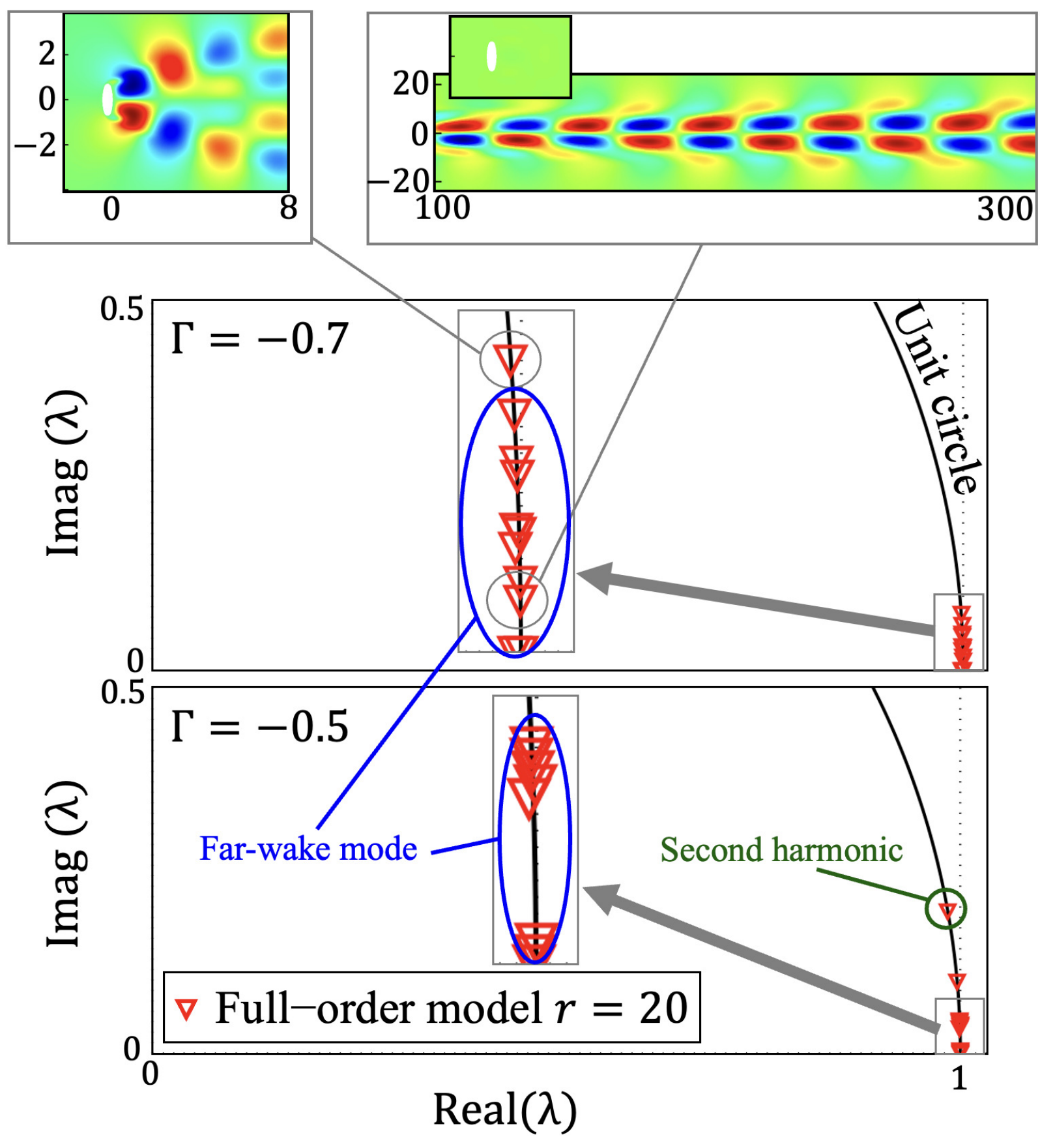}
  \end{minipage}
  &
  \begin{minipage}[b]{0.48\linewidth}
          \centering\includegraphics[height=6.5cm,keepaspectratio]{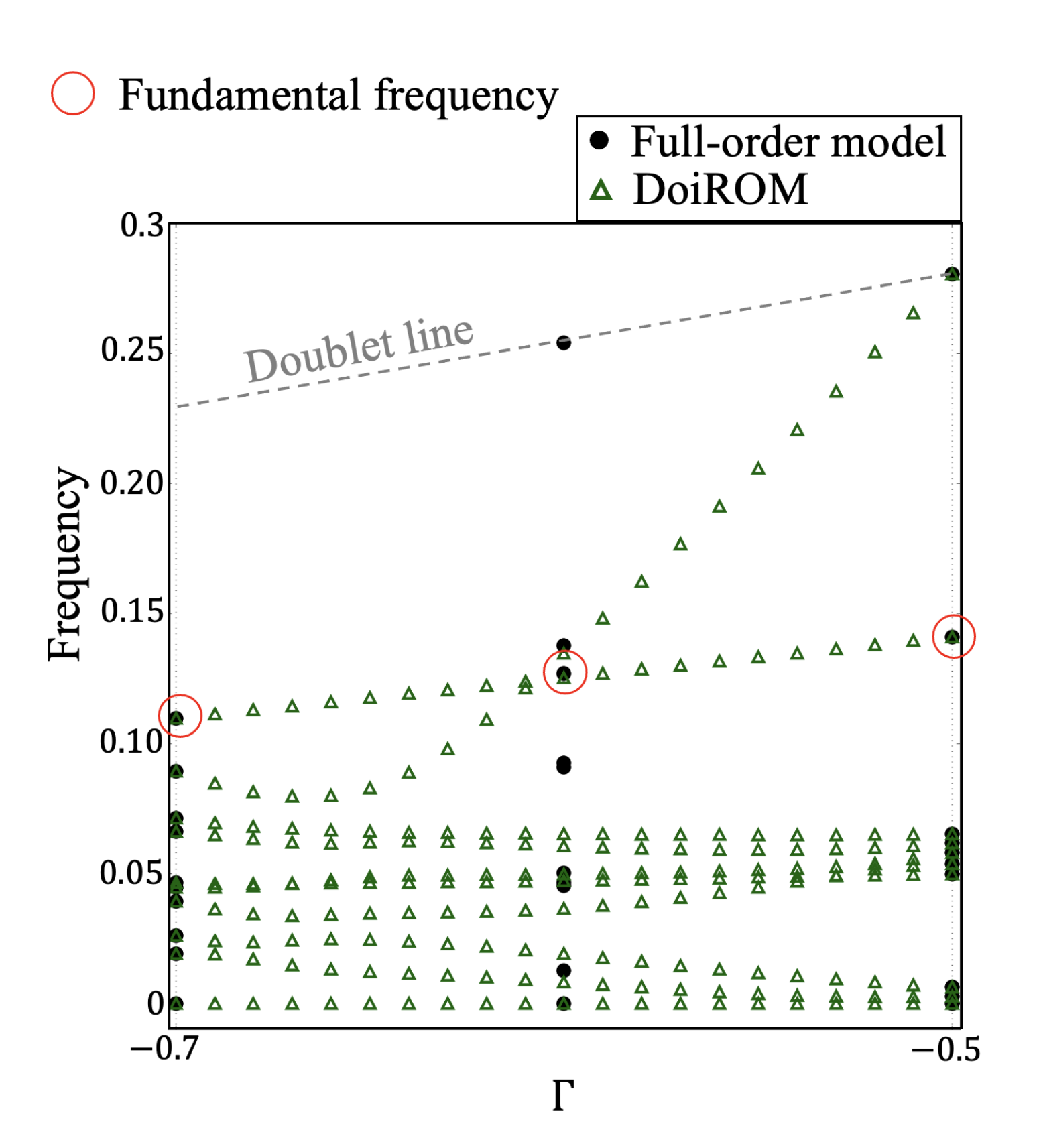}
  \end{minipage}\\
  \end{tabular}
  \captionsetup{justification=raggedright,singlelinecheck=false}
\caption{DMD eigenvalue distributions and frequencies obtained from DoiROM for $-0.7 \leq \Gamma \leq -0.5$.
(a) DMD eigenvalues at $\Gamma = -0.5$ and $-0.7$, with a representative mode at $\Gamma = -0.7$ showing two types: primary Kármán vortex modes near the cylinder (frequency $\approx 0.1$) and secondary vortex street modes in the far wake (blue circles). A harmonic mode is marked with a green circle.
(b) Frequencies predicted by DoiROM. Red circles indicate fundamental frequencies from the full-order DMD results. While the fundamental frequencies are well interpolated, the second harmonics (gray dashed line) are not captured at $\Gamma = -0.5$ due to DMD rank truncation.}
 \label{fig:figure_ellipse_r20_full}
\end{figure}

Operator interpolation using DoiROM is performed for the range $\Gamma = -0.5$ to $-0.7$, where secondary vortex streets emerge. Figure~\ref{fig:figure_ellipse_r20_full} (a) presents the DMD eigenvalue distributions for the velocity fields at $\Gamma = -0.5$ and $-0.7$, along with the spatial distribution of a representative eigenmode at $\Gamma = -0.7$. As shown by the two DMD eigenmode distributions, the eigenmodes can be classified into two types: one associated with the primary Kármán vortex dynamics near the cylinder, and another corresponding to the secondary vortex street in the far wake.
The dimensionless frequency of the eigenmode associated with the primary Kármán vortex is approximately $0.11$ at $\Gamma=-0.7$ and $0.14$ at $\Gamma=-0.5$. For the fundamental frequency, only a single conjugate pair of such eigenmodes is observed. All eigenmodes with lower frequencies, highlighted by blue circles in the figure, were confirmed to be spatially distributed in the far-wake region.

The harmonic eigenmode of the fundamental frequency is indicated by a green circle in the case of $\Gamma = -0.5$. It is worth noting that the number of harmonic eigenmodes shown here is fewer than in the periodic flow cases for $\Gamma = -0.3$ to $0.2$ (see Fig.~\ref{fig:figure_ellipse_peri_freq}) due to truncation of POD basis vectors in the DMD procedure. This reduction does not imply the absence of these harmonic components in the actual flow field. The effect of $r$ selection on the DoiROM results is also discussed later.

Operator interpolation using DoiROM was performed within the range $-0.7 \leq \Gamma \leq -0.5$, with $\eta_0: (\Gamma = -0.7)$ and $\eta_1: (\Gamma = -0.5)$ as the reference conditions. Figure~\ref{fig:figure_ellipse_r20_full} (b) shows how the frequencies derived from the interpolated operators vary with $\Gamma$.
The red circles indicate the fundamental frequencies associated with the primary Kármán vortex shedding, obtained via DMD from the full-order model. The interpolated frequencies show a linear trend connecting the fundamental frequencies at the two reference conditions. Frequencies lower than the fundamental frequency are also interpolated smoothly between the corresponding low-frequency components.
The gray dotted line in the figure marks twice the fundamental frequency. Along this line, second-harmonic frequencies are evident in the DMD frequencies at $\Gamma = -0.7$ and $-0.6$, but are not captured by DoiROM. This discrepancy may stem from the absence of second-harmonic frequencies in the DMD results at $\Gamma = -0.5$, one of the reference conditions. If a higher DMD rank had been used, enabling the extraction of second-harmonic frequencies at $\Gamma = -0.5$, it is likely that DoiROM could have interpolated these frequencies as well.

\begin{figure}[htbp]
\begin{tabular}{cc}
\multicolumn{1}{l}{(a)}  &  \multicolumn{1}{l}{(b)}\\
  \begin{minipage}[b]{0.48\linewidth}
          \centering\includegraphics[height=6.5cm,keepaspectratio]{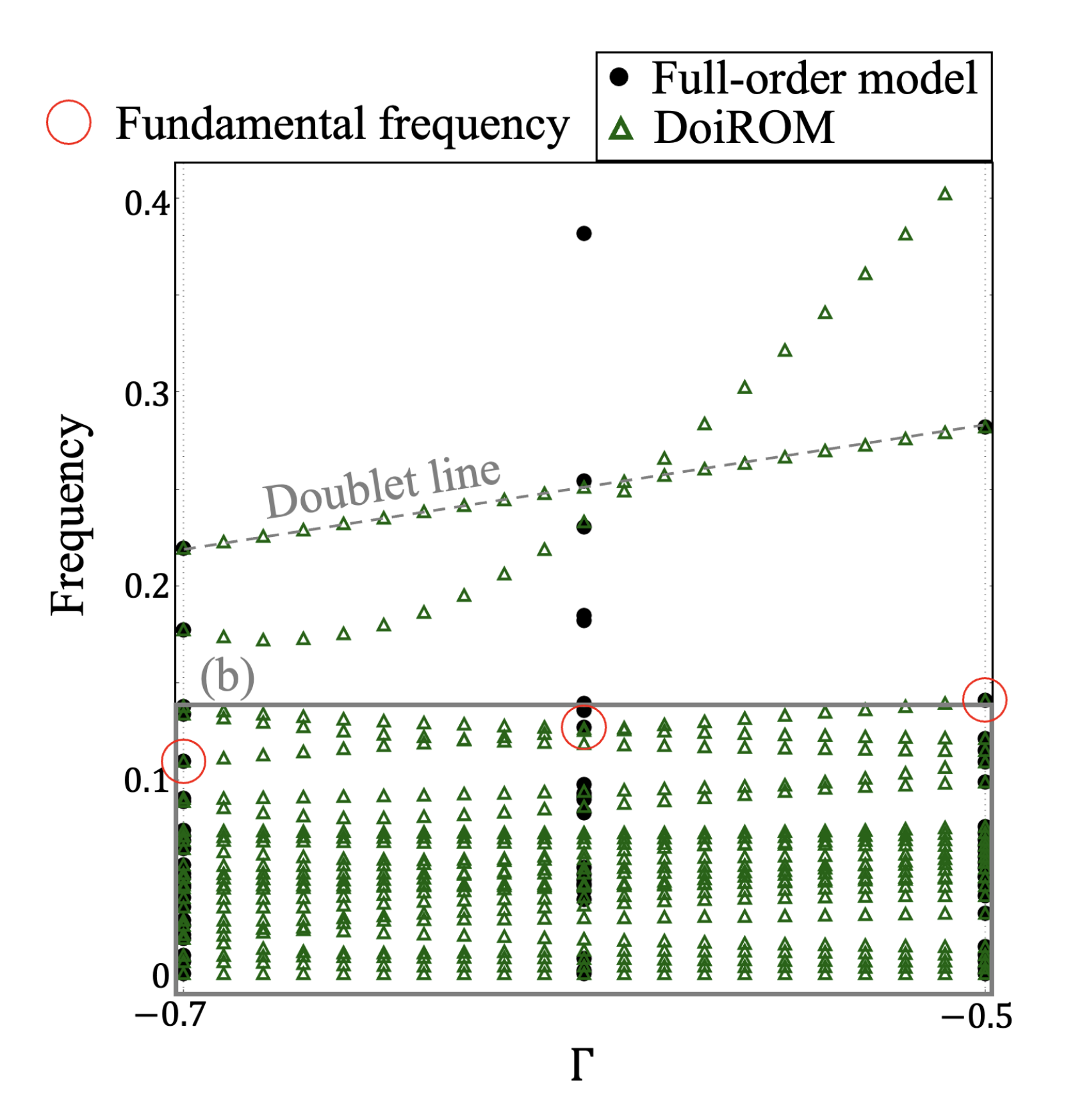}
  \end{minipage}
  &
  \begin{minipage}[b]{0.48\linewidth}
          \centering\includegraphics[height=6.5cm,keepaspectratio]{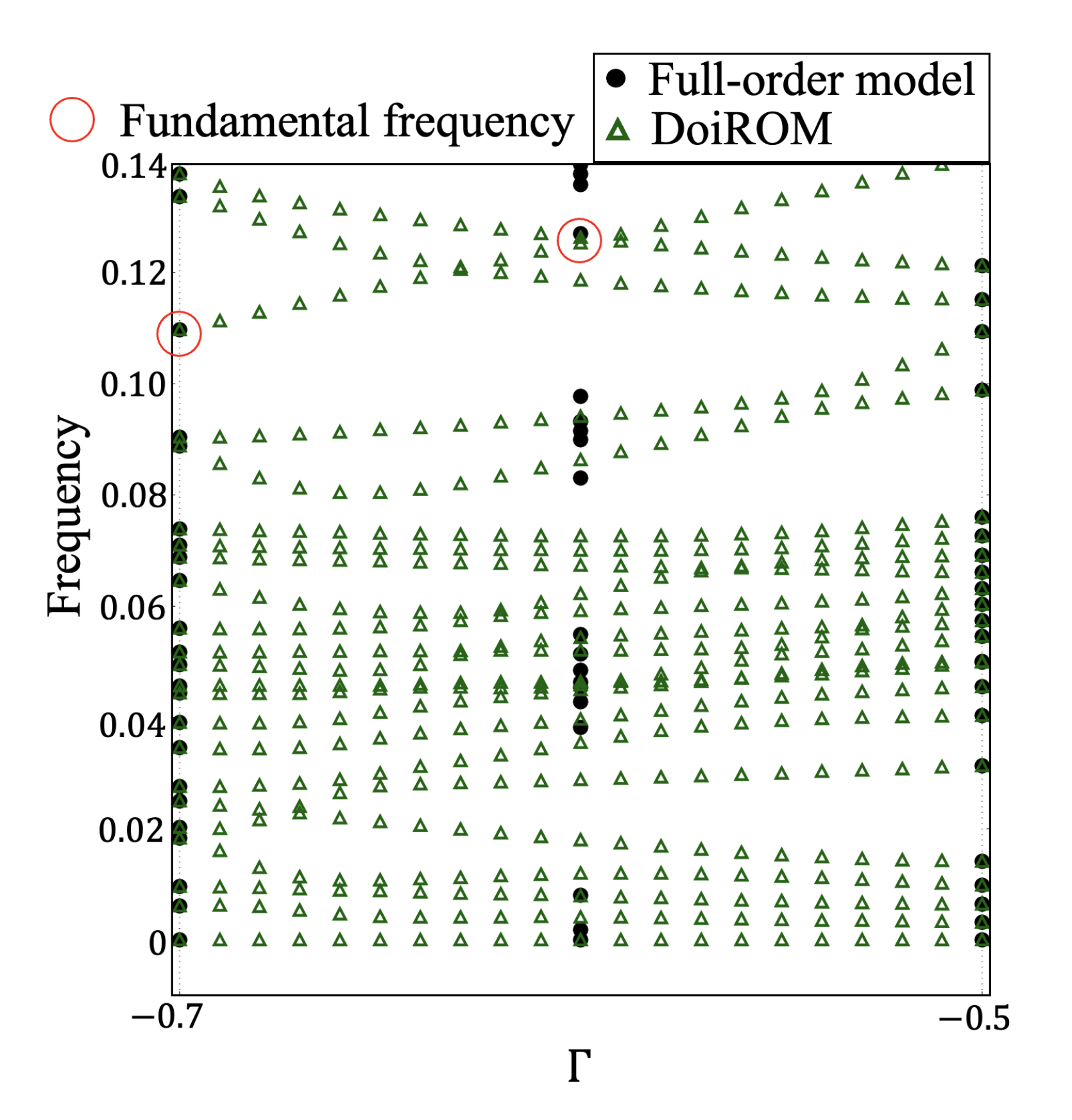}
  \end{minipage}\\
    \multicolumn{1}{l}{(c)}  &  \multicolumn{1}{l}{}\\
  \multicolumn{2}{c}{
    \begin{minipage}[b]{1\linewidth}
        \centering\includegraphics[width=10cm,keepaspectratio]{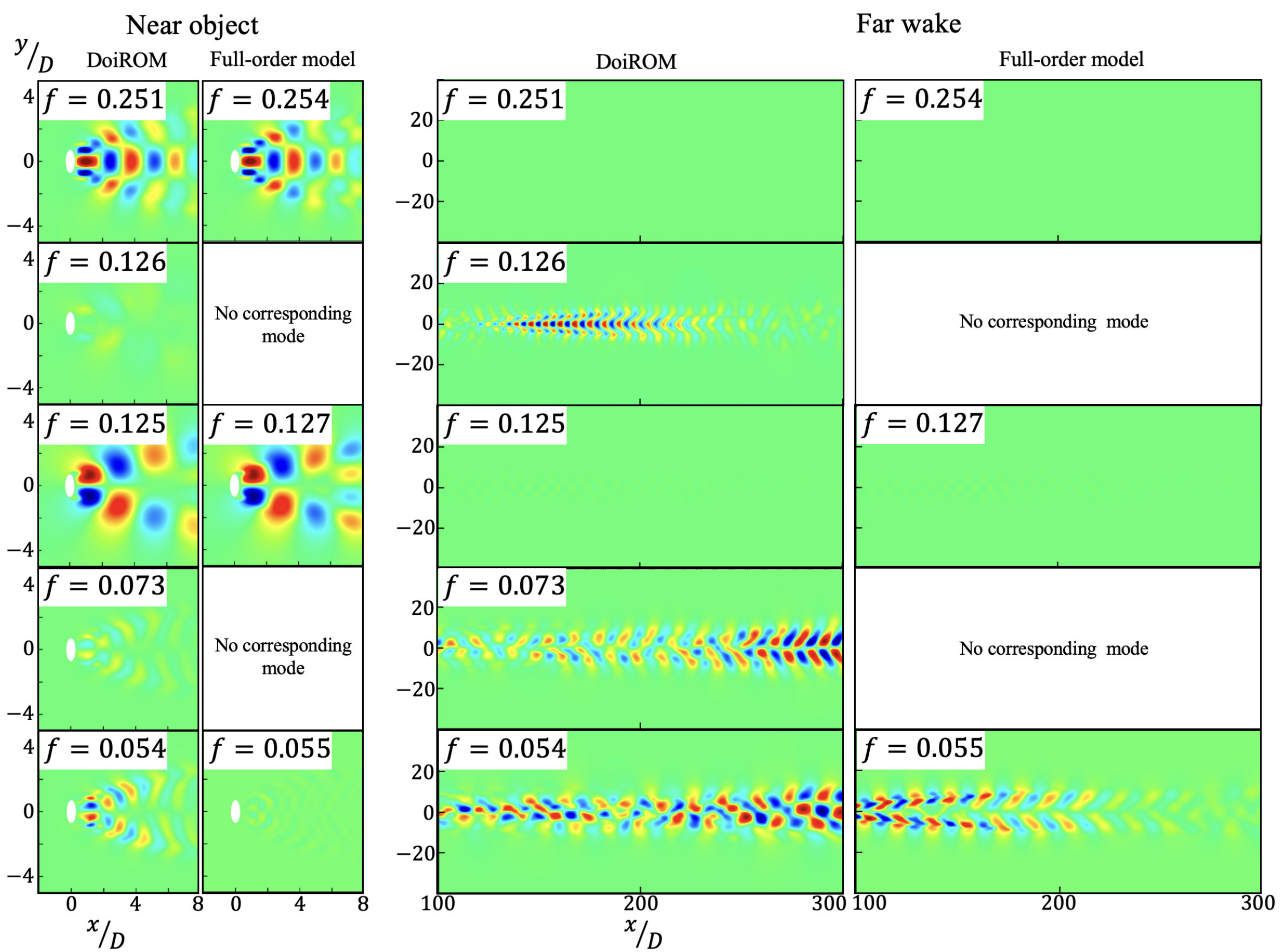}
    \end{minipage}
  }\\
  \end{tabular}
  \captionsetup{justification=raggedright,singlelinecheck=false}
\caption{DoiROM prediction results for secondary vortex streets at various $\Gamma$, with subspace dimension set to $r = 50$.
(a) Frequencies obtained from the eigenvalues of DoiROM and the full-order model (DMD).
(b) Close-up of the low-frequency region in (a), showing smooth variation of low-frequency modes across $\Gamma$.
(c) Spatial distributions of eigenmodes at $\Gamma = -0.6$ from the full-order model and DoiROM. Modes at $f \approx 0.25$ and $f \approx 0.125$, associated with Kármán vortex shedding, are well reproduced by DoiROM. In contrast, modes corresponding to far-wake vortex streets exhibit irregular spatial structures and limited agreement with the full-order results.}
 \label{fig:figure_ellipse_noperi_freq}
\end{figure}

Operator interpolation using DoiROM was performed with $r = 50$ modes to ensure the presence of the second-harmonic frequency at all reference conditions. Figure~\ref{fig:figure_ellipse_noperi_freq} presents the frequencies obtained from the eigenvalues computed using DoiROM, along with those derived via DMD from the full-order model, where the DMD rank is set to $r = 50$. All other conditions match those used for the interpolation shown in Fig.~\ref{fig:figure_ellipse_r20_full}.
Notably, even the second-harmonic frequencies, indicated by the gray dotted lines in Fig.~\ref{fig:figure_ellipse_noperi_freq} (a), are successfully captured by DoiROM. Figure~\ref{fig:figure_ellipse_noperi_freq} (b) provides a close-up of the low-frequency region in Fig.~\ref{fig:figure_ellipse_noperi_freq} (a). Although numerous low-frequency components are present, their frequencies vary smoothly and continuously between the two reference conditions.

Figure~\ref{fig:figure_ellipse_noperi_freq} (c) shows the spatial distributions of eigenmodes obtained from both the DMD results of the full-order model and DoiROM at $\Gamma = -0.6$. Correspondence between DoiROM and the DMD results is identified based on similarities in frequency and spatial structure. The DoiROM eigenmodes at approximately $f \approx 0.25$ and $f \approx 0.125$ capture fluctuations associated with Kármán vortex shedding in the near wake, indicating that DoiROM accurately reproduces the spatial characteristics of the primary shedding eigenmode.
By contrast, eigenmodes representing the secondary vortex street in the far wake exhibit irregular spatial patterns in DoiROM. In particular, those at $f = 0.126$ and $f = 0.073$ show no clear correspondence with the full-order DMD eigenmodes in either frequency or structure. Although the DoiROM eigenmode at $f = 0.055$ has a frequency close to one of the full-order modes, the spatial patterns differ considerably. These results suggest that capturing secondary vortex street dynamics is more challenging than reproducing the primary shedding near the cylinder.

A comparison of the spanwise vorticity fields at $\Gamma = -0.7$ and $-0.5$, shown in Fig.~\ref{fig:figure_ellipse_vol}, reveals that although the near-wake Kármán vortices remain relatively consistent, the onset position and structure of the secondary vortex streets vary significantly. This pronounced sensitivity of the far-wake flow features to changes in $\Gamma$ likely contributes to the reduced accuracy of DoiROM in predicting the associated eigenmodes.

\begin{figure}[htbp]
\begin{tabular}{cc}
\multicolumn{1}{l}{(a)}  &  \multicolumn{1}{l}{(b)}\\
  \begin{minipage}[b]{0.48\linewidth}
          \centering\includegraphics[height=6.5cm,keepaspectratio]{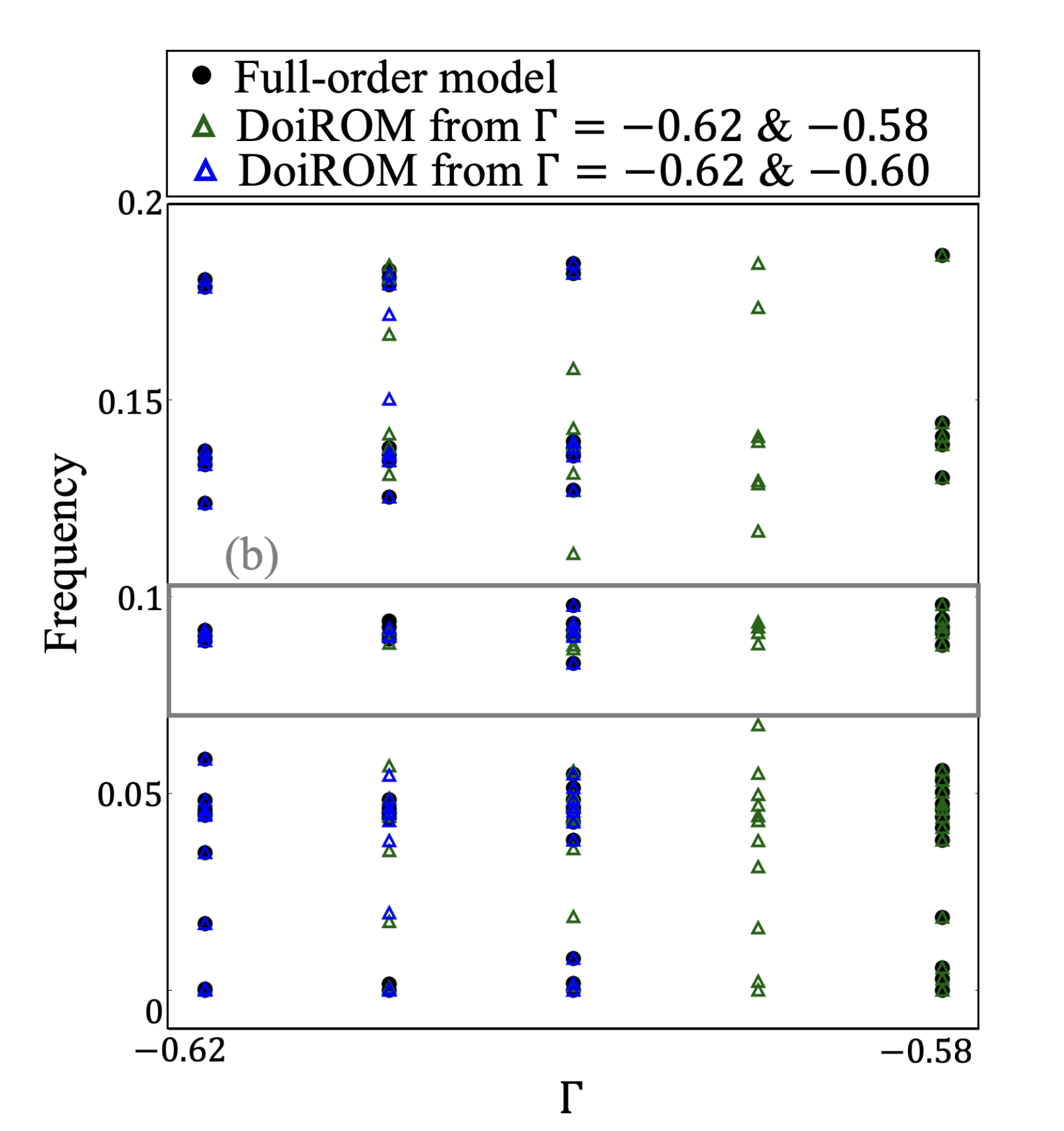}
  \end{minipage}
  &
  \begin{minipage}[b]{0.48\linewidth}
          \centering\includegraphics[height=6.5cm,keepaspectratio]{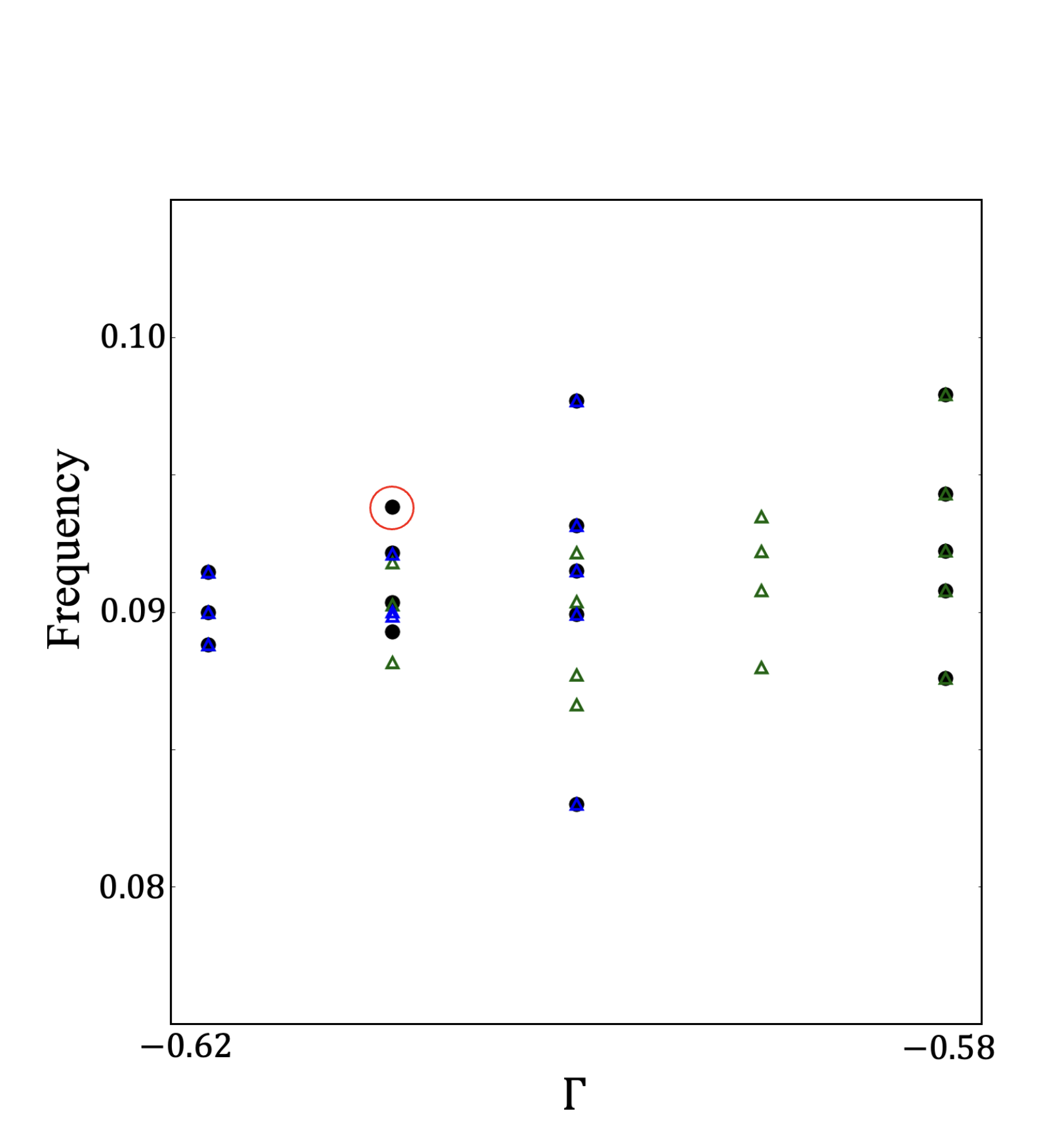}
  \end{minipage}\\
  \end{tabular}
  \captionsetup{justification=raggedright,singlelinecheck=false}
\caption{DoiROM frequency predictions using closer reference conditions than those in Fig.~\ref{fig:figure_ellipse_noperi_freq}.
(a) Predicted frequencies using $(\Gamma = -0.58,\ -0.62)$ and $(\Gamma = -0.60,\ -0.62)$ show improved agreement with the full-order results compared to Fig.~\ref{fig:figure_ellipse_noperi_freq}.
(b) Close-up view at $\Gamma = -0.61$. Most frequencies are accurately captured, except for the red-circled mode, which is missing due to low-rank truncation at $\Gamma = -0.62$.}
 \label{fig:figure_ellipse_freq_close}
\end{figure}

\begin{figure}[htbp]
  \centering\includegraphics[width=13cm,keepaspectratio]{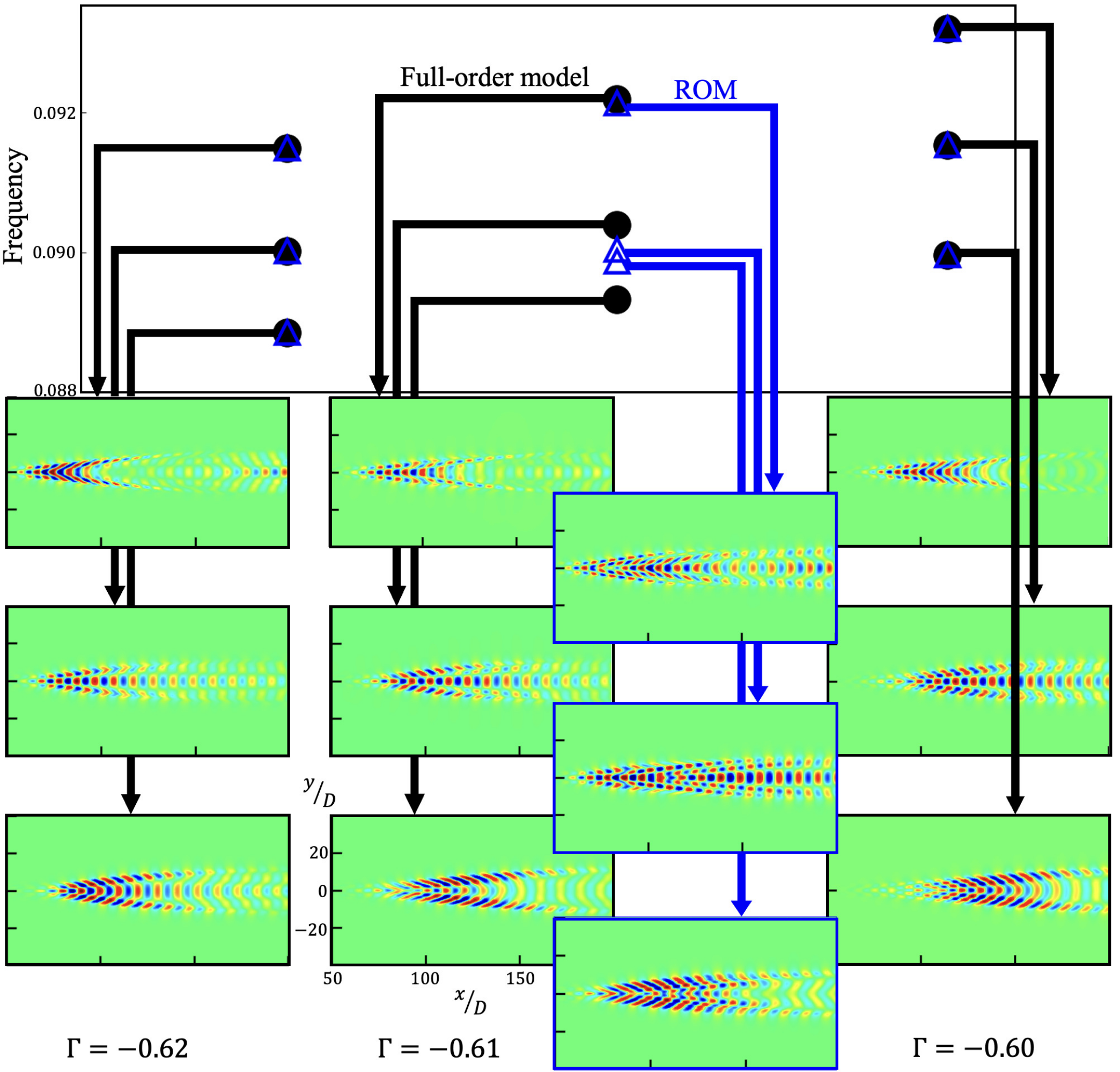}
\caption{Spatial distributions of eigenmodes corresponding to the frequencies predicted by DoiROM using reference conditions $\Gamma = -0.60$ and $-0.62$.}
 \label{fig:figure_ellipse_mode_close}
\end{figure}

To improve the prediction of secondary vortex street eigenmodes using DoiROM, the reference conditions for interpolation were adjusted. Figure~\ref{fig:figure_ellipse_freq_close} shows the frequencies predicted by DoiROM using the reference conditions $\eta_0: (\Gamma = -0.60\ \text{or}\ -0.58)$ and $\eta_1: (\Gamma = -0.62)$. Compared to the previous results shown in Fig.~\ref{fig:figure_ellipse_noperi_freq}, which used $\eta_0: (\Gamma = -0.70)$ and $\eta_1: (\Gamma = -0.50)$ as reference conditions, the new predictions exhibit significantly better agreement with the DMD results of the full-order model.
This improvement is particularly evident in the close-up view at $\Gamma = -0.61$ in Fig.~\ref{fig:figure_ellipse_freq_close} (b), where most of the predicted frequencies align well with those from the DMD results of the full-order model. However, not all frequencies observed in the DMD results of the full-order model at $\Gamma = -0.61$ are successfully captured. In particular, the frequency highlighted by a red circle in Fig.~\ref{fig:figure_ellipse_freq_close} (b) is not reproduced. This discrepancy arises because the corresponding frequency component is absent at the reference condition $\Gamma = -0.62$. The absence may result from the truncation inherent in the low-rank approximation used in the model reduction process.

Figure~\ref{fig:figure_ellipse_mode_close} shows the spatial distributions of eigenmodes corresponding to the frequencies predicted by DoiROM using the reference conditions $\eta_0: (\Gamma = -0.62)$ and $\eta_1: (\Gamma = -0.60)$, as presented in Fig.~\ref{fig:figure_ellipse_freq_close} (b). Since these eigenmodes represent secondary vortex streets, the near-body region is omitted.
Examining the DMD eigenmodes from the full-order model reveals that, as $\Gamma$ increases, the eigenmode distribution progressively shifts downstream. The DoiROM eigenmodes preserve more physical features than those in Fig.~\ref{fig:figure_ellipse_noperi_freq} (c), showing better agreement with the DMD eigenmodes of the full-order model. However, the DoiROM eigenmodes appear to resemble a linear superposition of the eigenmodes at $\Gamma = -0.60$ and $-0.62$, rather than exhibiting a continuous downstream shift.
This observation suggests that DoiROM, which relies on linear interpolation, struggles to capture spatial translations of eigenmode structures. Simple superposition cannot account for the downstream displacement of eigenmodes, especially when they are highly sensitive to small variations in $\Gamma$. Consequently, the difficulty in accurately predicting the spatial structure of secondary vortex eigenmodes likely arises from this strong sensitivity, which induces significant shifts in the $x$ direction.

\begin{figure}[htbp]
\begin{tabular}{cc}
\multicolumn{1}{l}{(a)}  &  \multicolumn{1}{l}{(b)}\\
  \begin{minipage}[b]{0.48\linewidth}
          \centering\includegraphics[height=6cm,keepaspectratio]{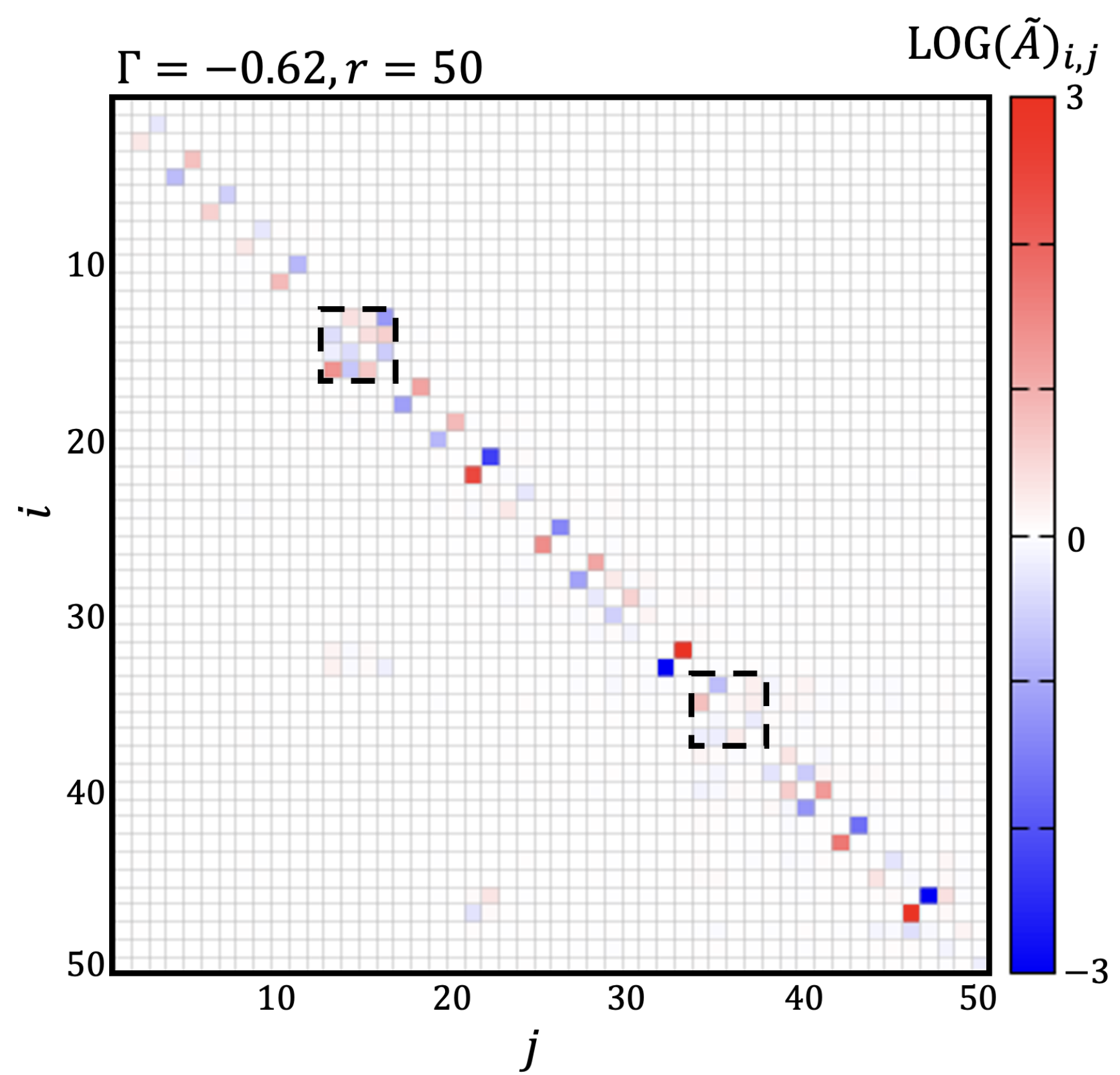}
  \end{minipage}
  &
  \begin{minipage}[b]{0.48\linewidth}
          \centering\includegraphics[height=6cm,keepaspectratio]{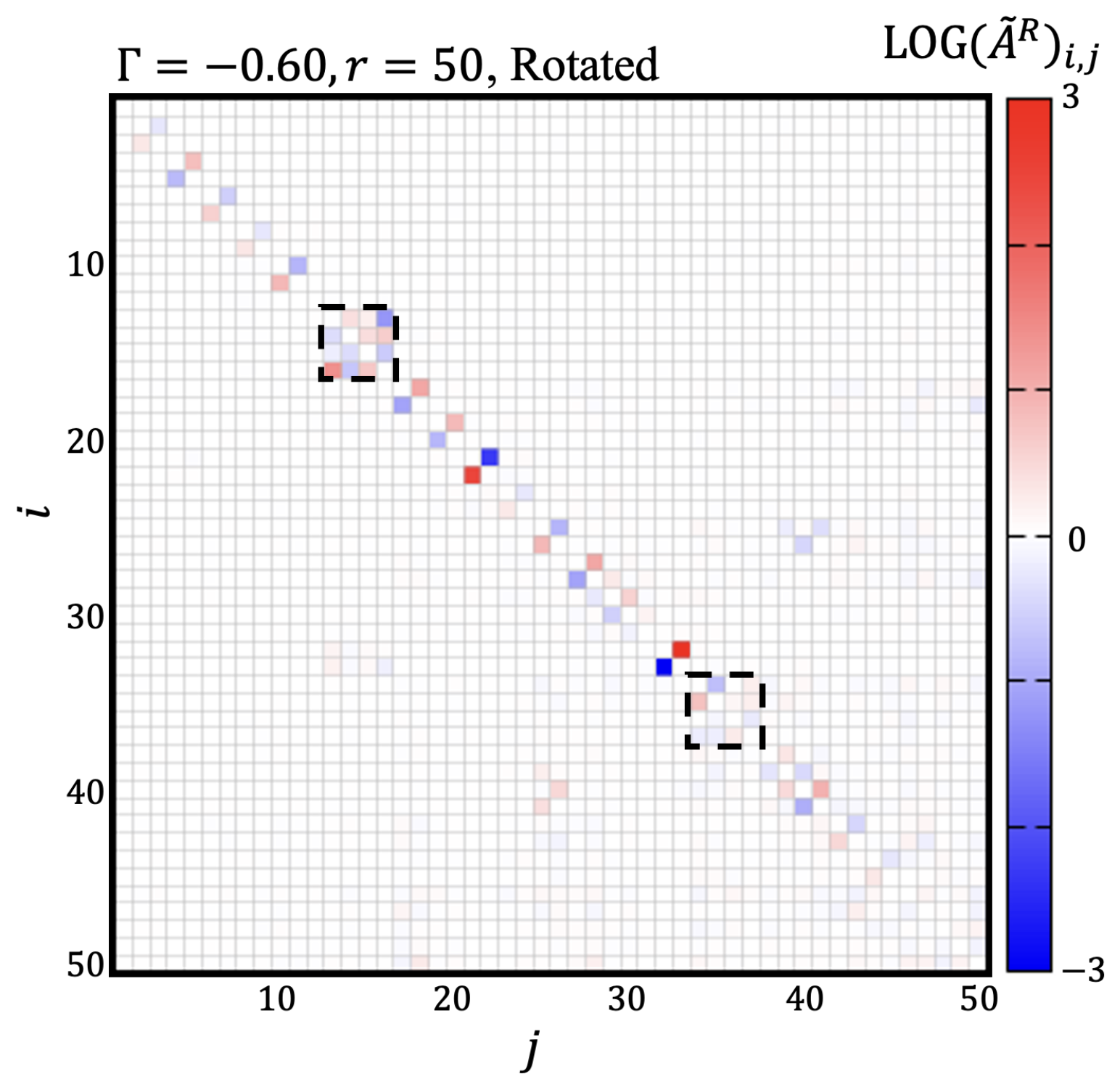}
  \end{minipage}
  \end{tabular}
  \captionsetup{justification=raggedright,singlelinecheck=false}
\caption{Visualization of the matrix elements of the linear operator for flow fields exhibiting secondary vortex streets.}
 \label{fig:figure_ellipse_last}
\end{figure}

Focusing on the frequency variation of secondary vortex street eigenmodes with respect to $\Gamma$, it is observed that, despite the linear interpolation of the operator, the resulting frequencies do not vary linearly with $\Gamma$. To investigate the cause of this nonlinearity, Fig.~\ref{fig:figure_ellipse_last} visualizes the elements of the interpolated linear operator matrix.
As seen in the periodic flow case shown in Fig.~\ref{fig:figure_peri_last} (a), where the frequencies were interpolated linearly, most nonzero elements are arranged in a banded pattern near the diagonal. As discussed earlier, when the nonzero elements are confined to a $2 \times 2$ block around the diagonal, the eigenvalues (i.e., frequencies) depend directly on the corresponding matrix values. Since each matrix element is interpolated linearly, the resulting frequencies also vary linearly in such cases.

However, in the regions enclosed by black dashed lines in Fig.~\ref{fig:figure_ellipse_last}, nonzero elements appear outside the $2 \times 2$ structure. In these cases, the eigenvalues depend on higher-order interactions involving products of multiple matrix elements. As a result, even if the individual elements vary linearly, the corresponding eigenvalues can exhibit nonlinear behavior.
This situation typically arises when a single conjugate pair of oscillatory eigenmodes is represented by more than two POD basis vectors, that is, by three or more POD basis vectors. For secondary vortex streets, multiple eigenmodes with similar frequencies are often observed, indicating limitations of POD in separating frequency components with similar values. Therefore, when these similar-frequency components coexist, the accuracy of operator interpolation does not necessarily ensure accurate prediction of the associated frequencies.

\section{Conclusion}
This study presented operator-based estimation methods for parameterized hydrodynamic systems and evaluated their ability to predict physically relevant features. Two ROMs were constructed: GapiROM, which employs Galerkin projection combined with subspace interpolation on the Grassmann manifold, and DoiROM, which interpolates linear operators derived from DMD. These models are designed to estimate the eigenvalues and eigenmodes of governing operators at previously unseen physical parameters, such as Reynolds number or geometric aspect ratio, using operator information obtained at a limited set of reference conditions.

The first test case considered flow around a circular cylinder with varying Reynolds numbers, where unsteady Kármán vortices emerge from a steady base flow. GapiROM, incorporating Grassmann manifold interpolation, successfully captured Reynolds-number-dependent variations in spatial structures, growth rates, and frequencies. The interpolated basis vectors evolved smoothly with Reynolds number and exhibited spatial features consistent with those of the full-order model. Consequently, the predicted eigenvalues varied continuously and traced nearly linear trajectories across the interpolation range. This behavior indicates that GapiROM preserves the physical continuity of the operator structure under parameter variation.

A decomposition of the eigenvalues into viscous and non-viscous components revealed that the viscous contributions were accurately reproduced by the interpolation, while residual errors were primarily associated with the non-viscous components, which are highly sensitive to basis vector variations. Furthermore, we demonstrated that without Grassmann manifold interpolation of the basis, a ROM cannot accurately estimate the parameter dependence of eigenvalues, as it fails to capture the changes in both viscous and non-viscous components arising from mode variation. These findings underscore the critical role of Grassmann manifold interpolation in constructing ROMs that retain physical fidelity across varying flow conditions.

DoiROM was also applied to this linear-growth scenario from the steady base flow. Unlike GapiROM, which relies on subspace interpolation and Galerkin projection, DoiROM performs operator interpolation directly within the reduced-order space. The predicted growth rates and frequencies, computed from the interpolated eigenvalues, varied linearly between the reference Reynolds numbers, reflecting the inherent linearity of the interpolation scheme.

The second case focused on fully developed periodic flows. GapiROM accurately predicted the fundamental frequency but underestimated the growth rates of its harmonics, often yielding negative values. This limitation arose from the use of Galerkin projection, which constructs linearized operators around the steady base flow.
DoiROM, by contrast, exhibited superior performance in predicting both modal frequencies and eigenvalue distributions, particularly in capturing the harmonic structure characteristic of periodic flows. While GapiROM was limited to the fundamental frequency, DoiROM successfully reproduced multiple harmonics and showed good agreement with the full-order model. The eigenvalues predicted by DoiROM varied linearly with the interpolation parameter, reflecting the linearity of the operator interpolation and contributing to the method’s interpretability and robustness, especially for flows with smoothly varying dynamics. However, when the subspace dimension was excessively large and contaminated by numerical errors, the harmonic structure represented by the operator was no longer preserved through interpolation.

The third case extended the applicability of the proposed framework to geometric variations, such as changes in the aspect ratio of an elliptical cylinder. For geometries in which the flow field consisted of a fundamental frequency and its harmonics, DoiROM successfully connected the frequencies of the reference conditions in a linear fashion, even when the geometries differed. Operator interpolation using DoiROM was also applied to cases in which a secondary vortex street developed far downstream of the body, in addition to the presence of fundamental and harmonic frequencies. In these cases, the fundamental frequency and its harmonics continued to vary linearly with the interpolation parameter. In contrast, the low-frequency eigenmodes associated with the far-wake secondary vortex street could be predicted only when the reference conditions were sufficiently close.

Because multiple eigenmodes with similar frequencies can exist within the secondary vortex street, the frequencies of these eigenmodes do not vary linearly with respect to the parameter, even when the operator is interpolated linearly. This nonlinearity arises from the fact that multiple POD basis vectors are required to represent a coherent structure. As a result, frequency prediction for such eigenmodes becomes inherently nonlinear.
Future work will address the extension of this framework to strongly nonlinear regimes and will focus on developing adaptive strategies for selecting reference parameters and tuning the reduced subspace dimension.

\section*{Acknowledgments}
The Sasakawa Scientific Research Grant from The Japan Science Society, and JST SPRING, Grant Number JPMJSP2114, Japan support this work.
The computational resource is provided by the Advanced Fluid Information Research Center, Institute of Fluid Science, Tohoku University with a Supercomputer system ``AFI-NITY II'', and JAXA Supercomputer System Generation 3 (JSS3).

\appendix
\section{\label{apena}Derivation of basis transformations using exponential and logarithmic maps on the Grassmann manifold}

We derive the relationship between the basis matrix after mapping $U_{\text{out}} \in \mathbb{R}^{N \times r}$ and the basis matrix before mapping $U_{\text{in}} \in \mathbb{R}^{N \times r}$ when the logarithmic and exponential maps on the Grassmann manifold are applied at a reference point $U_o \in \mathbb{R}^{N \times r}$ to $U_{\text{in}}$. Specifically, we consider the operation
\begin{equation}
\mathrm{Exp}^{\mathrm{Gr}}_{[U_o]} \circ \mathrm{Log}_{[U_o]}^\mathrm{Gr} ([U_{\text{in}}]) = [U_{\text{out}}].
\end{equation}

We first compute the logarithmic map of $U_{\text{in}}$. The following singular value decomposition (SVD) is used:
\begin{equation}
(I - U_o U_o^\top ) U_{\text{in}} (U_o^\top  U_{\text{in}})^{-1} \stackrel{\rm{SVD}}{=} V_{\text{in}} \Sigma_{\text{in}} W_{\text{in}}^\top ,
\label{eq:GrLog_SVD0_appen}
\end{equation}
where the subscript ``in'' indicates that the corresponding matrices depend on $U_{\text{in}}$.
The tangent vector is given by
\begin{equation}
\Delta_{\text{in}} = V_{\text{in}}\mathrm{arctan}(\Sigma_{\text{in}})W^\top _{\text{in}}.
\label{eq:GrLog_mat_appen}
\end{equation}
Here, we recall the orthonormality properties from the SVD:
\begin{equation}
\begin{split}
V_{\text{in}}^\top  V_{\text{in}} = I, \quad
W_{\text{in}}^\top  W_{\text{in}} = I, \quad
W_{\text{in}} W_{\text{in}}^\top  = I,
\end{split}
\end{equation}
which will be used in the subsequent derivation.
Next, we apply the exponential map to the tangent vector $\Delta_{\text{in}}$. Using the SVD of $\Delta_{\text{in}}$, we obtain
\begin{equation}
\Delta_{\text{in}} = V_{\text{in}}\mathrm{arctan}(\Sigma_{\text{in}})W^\top _{\text{in}}
\stackrel{\rm{SVD}}{=} \check{V}_{\text{in}} \check{\Sigma}_{\text{in}} \check{W}_{\text{in}}^\top .
\end{equation}

Here, it can be readily confirmed that the diagonal elements of $\mathrm{arctan}(\Sigma_{\text{in}})$ are the singular values of $\Delta_{\text{in}}$. Moreover, since $\mathrm{arctan}(\theta)$ is a monotonically increasing function of $\theta$, the eigenvalues corresponding to the diagonal elements of $\mathrm{arctan}(\Sigma_{\text{in}})$ are arranged in ascending order. That is, 
\begin{equation}
  \check{\Sigma}_{\text{in}} = \mathrm{arctan}(\Sigma_{\text{in}}).
\end{equation}
The resulting set of bases $U_{\text{out}}$ is
\begin{equation}
\begin{split}
  U_{\text{out}} 
  &= U_o \check{W}_{\text{in}} \cos(\check{\Sigma}_{\text{in}}) \check{W}_{\text{in}}^\top  + \check{V}_{\text{in}} \sin(\check{\Sigma}_{\text{in}}) \check{W}_{\text{in}}^\top  \\
  &= \left\{ U_o \check{W}_{\text{in}}  + \check{V}_{\text{in}} \sin(\check{\Sigma}_{\text{in}}) \cos^{-1}(\check{\Sigma}_{\text{in}}) \right\} \cos(\check{\Sigma}_{\text{in}}) \check{W}_{\text{in}}^\top  \\
  &= \left\{ U_o \check{W}_{\text{in}}  + \check{V}_{\text{in}} \tan(\check{\Sigma}_{\text{in}}) \right\}  \cos(\check{\Sigma}_{\text{in}}) \check{W}_{\text{in}}^\top  \\
  &= \Bigr\{ U_o \check{W}_{\text{in}} + \check{V}_{\text{in}} \tan\left\{ \arctan(\Sigma_{\text{in}}) \right\} \Bigr\} \cos\left\{ \arctan(\Sigma_{\text{in}}) \right\} \check{W}_{\text{in}}^\top  \\
  &= \left( U_o \check{W}_{\text{in}} + \check{V}_{\text{in}} \Sigma_{\text{in}} \right) \cos\left\{ \arctan(\Sigma_{\text{in}}) \right\} \check{W}_{\text{in}}^\top .
\end{split}
\end{equation}
Considering Eq.~(\ref{eq:GrLog_SVD0_appen}), the singular value decomposition still holds when the left and right singular vectors are replaced with $\check{V}_{\text{in}}$ and $\check{W}_{\text{in}}$, respectively, as shown below:
\begin{equation}
\check{V}_{\text{in}} \Sigma_{\text{in}} \check{W}_{\text{in}}^\top 
= (I - U_o U_o^\top ) U_{\text{in}} (U_o^\top  U_{\text{in}})^{-1}
= U_{\text{in}} (U_o^\top  U_{\text{in}})^{-1} - U_o.
\end{equation}
Therefore,
\begin{equation}
\begin{split}
  &( U_o \check{W}_{\text{in}} + \check{V}_{\text{in}} \Sigma_{\text{in}}) \cos\left\{ \arctan(\Sigma_{\text{in}}) \right\} \check{W}_{\text{in}}^\top  \\
  &= (U_o \check{W}_{\text{in}} + \check{V}_{\text{in}} \Sigma_{\text{in}} \check{W}_{\text{in}}^\top  \check{W}_{\text{in}}) \cos\left\{ \arctan(\Sigma_{\text{in}}) \right\} \check{W}_{\text{in}}^\top  \\
  &= \left\{ U_o \check{W}_{\text{in}} + U_{\text{in}} (U_o^\top  U_{\text{in}})^{-1} \check{W}_{\text{in}} - U_o \check{W}_{\text{in}} \right\}
     \cos\left\{ \arctan(\Sigma_{\text{in}}) \right\} \check{W}_{\text{in}}^\top  \\
  &= U_{\text{in}} (U_o^\top  U_{\text{in}})^{-1} \check{W}_{\text{in}} \cos\left\{ \arctan(\Sigma_{\text{in}}) \right\} \check{W}_{\text{in}}^\top .
\end{split}
\end{equation}

Here, for a non-negative diagonal matrix $D$, we define $Y=\mathrm{cos}\{\mathrm{arctan}(D)\}$. By algebraic manipulation, $Y$ satisfies
\begin{equation}
\begin{split}
Y &= \cos\left\{\arctan(D)\right\} \\
&\Leftrightarrow \frac{1}{Y^2} = \frac{1}{\cos^2\left\{\arctan(D)\right\}} \\
&\Leftrightarrow \frac{1}{Y^2} = I + \tan^2\left\{\arctan(D)\right\} \\
&\Leftrightarrow \frac{1}{Y^2} = I + D^2 \\
&\Leftrightarrow Y^2 = \frac{1}{I + D^2}.
\end{split}
\end{equation}
That is,
\begin{equation}
\cos^2\left\{\arctan(\Sigma_{\text{in}})\right\} = \frac{1}{I + \Sigma_{\text{in}}^2}.
\end{equation}

From Eq.~(\ref{eq:GrLog_SVD0_appen}), the squared singular value matrix $\Sigma^2_{\text{in}}$ can be written as
\begin{equation}
\begin{split}
\Sigma_{\text{in}}^2
&= W_{\text{in}}^\top  \left( W_{\text{in}} \Sigma_{\text{in}} V_{\text{in}}^\top  \right)\left( V_{\text{in}} \Sigma_{\text{in}} W_{\text{in}}^\top  \right) W_{\text{in}} \\
&= W_{\text{in}}^\top  \left( V_{\text{in}} \Sigma_{\text{in}} W_{\text{in}}^\top  \right)^\top \left( V_{\text{in}} \Sigma_{\text{in}} W_{\text{in}}^\top  \right) W_{\text{in}} \\
&= W_{\text{in}}^\top  \left\{ (I - U_o U_o^\top ) U_{\text{in}} (U_o^\top  U_{\text{in}})^{-1} \right\}^\top \left\{ (I - U_o U_o^\top ) U_{\text{in}} (U_o^\top  U_{\text{in}})^{-1} \right\} W_{\text{in}} \\
&= W_{\text{in}}^\top  \left\{ (U_o^\top  U_{\text{in}})^{-1} \right\}^\top U_{\text{in}}^\top  (I - U_o U_o^\top )^\top  (I - U_o U_o^\top )U_{\text{in}} (U_o^\top  U_{\text{in}})^{-1} W_{\text{in}}.
\end{split}
\end{equation}
Here, since we can simplify the following expression:
\begin{equation}
\begin{split}
(I - U_o U_o^\top )^\top  (I - U_o U_o^\top )\\
&= (I - U_o U_o^\top )(I - U_o U_o^\top ) \\
&= I - 2 U_o U_o^\top  + U_o U_o^\top  U_o U_o^\top  \\
&= I - 2 U_o U_o^\top  + U_o U_o^\top  \\
&= I - U_o U_o^\top ,
\end{split}
\end{equation}
we obtain 
\begin{equation}
 \begin{split}
 &W^\top _{\text{in}}\{(U_o^\top U_{\text{in}})^{-1}\}^\top U^\top _{\text{in}}(I-U_oU_o^\top )^\top (I-U_oU_o^\top )U_{\text{in}}(U_o^\top U_{\text{in}})^{-1}W_{\text{in}}\\
 =&W^\top _{\text{in}}\{(U_o^\top U_{\text{in}})^{-1}\}^\top U^\top _{\text{in}}(I-U_oU_o^\top )U_{\text{in}}(U_o^\top U_{\text{in}})^{-1}W_{\text{in}}\\
 =&W^\top _{\text{in}}\{(U_o^\top U_{\text{in}})^{-1}\}^\top (U^\top _{\text{in}}U_{\text{in}}-U^\top _{\text{in}}U_oU_o^\top U_{\text{in}})(U_o^\top U_{\text{in}})^{-1}W_{\text{in}}\\
 =&W^\top _{\text{in}}\{(U_o^\top U_{\text{in}})^{-1}\}^\top (I-U^\top _{\text{in}}U_oU_o^\top U_{\text{in}})(U_o^\top U_{\text{in}})^{-1}W_{\text{in}}\\
 =&W^\top _{\text{in}}\{(U_o^\top U_{\text{in}})^{-1}\}^\top \left\{(U_o^\top U_{\text{in}})^{-1}-U^\top _{\text{in}}U_o\right\}W_{\text{in}}\\
  =&W^\top _{\text{in}}\Bigr\{\{(U_o^\top U_{\text{in}})^{-1}\}^\top (U_o^\top U_{\text{in}})^{-1}-\{(U_o^\top U_{\text{in}})^{-1}\}^\top U^\top _{\text{in}}U_o\Bigr\}W_{\text{in}}.
       \label{eq:deformate_appen}
     \end{split}
\end{equation}
Moreover, it can be readily confirmed that
 \begin{equation}
 \begin{split}
\{(U_o^\top U_{\text{in}})^{-1}\}^\top U^\top _{\text{in}}U_o&=\Bigr\{(U^\top _{\text{in}}U_o)^\top \{(U_o^\top U_{\text{in}})^{-1}\}\Bigr\}^\top \\
&=\Bigr\{(U^\top _oU_{\text{in}})\{(U_o^\top U_{\text{in}})^{-1}\}\Bigr\}^\top \\
&=I^\top =I.
     \end{split}
\end{equation}
Therefore, Eq.~(\ref{eq:deformate_appen}) becomes
 \begin{equation}
 \begin{split}
W^\top _{\text{in}}&\Bigr\{\{(U_o^\top U_{\text{in}})^{-1}\}^\top (U_o^\top U_{\text{in}})^{-1}-I\Bigr\}W_{\text{in}}\\
\,\,\,\,\,\,\,\,\,\,\,\,\,\,\,&=\Bigr\{W^\top _{\text{in}}\{(U_o^\top U_{\text{in}})^{-1}\}^\top (U_o^\top U_{\text{in}})^{-1}W_{\text{in}}-I\Bigr\},
     \end{split}
\end{equation}
and hence,
 \begin{equation}
 \begin{split}
\mathrm{cos}\{\mathrm{arctan}(\Sigma_{\text{in}})\}=\Bigr\{W^\top _{\text{in}}\{(U_o^\top U_{\text{in}})^{-1}\}^\top (U_o^\top U_{\text{in}})^{-1}W_{\text{in}}\Bigr\}^{-\frac{1}{2}}.\\
     \end{split}
\end{equation}
The matrix after mapping can be written in a simplified form as
\begin{equation}
\begin{split}
  U_{\text{out}} &=U_{\text{in}}(U_o^\top U_{\text{in}})^{-1}\check{W}_{\text{in}}\mathrm{cos}\{\mathrm{arctan}(\Sigma_{\text{in}})\}\check{W}_{\text{in}}^\top \\
  &=U_{\text{in}}(U_o^\top U_{\text{in}})^{-1}\check{W}_{\text{in}}\Bigr\{W^\top _{\text{in}}\{(U_o^\top U_{\text{in}})^{-1}\}^\top (U_o^\top U_{\text{in}})^{-1}W_{\text{in}}\Bigr\}^{-\frac{1}{2}}\check{W}_{\text{in}}^\top .
  \end{split}
\end{equation}
Here, we consider the SVD of $U_o^\top U_{\text{in}}$, given by
\begin{equation}
  (U_o^\top U_{\text{in}}) \stackrel{\rm{SVD}}{=} V_{p} \Sigma_{p} W_{p}^\top .
\end{equation}
Based on these matrices, the inverse can be written as
\begin{equation}
  (U_o^\top U_{\text{in}})^{-1} = W_{p} \Sigma^{-1}_{p} V^\top _{p}.
\end{equation}
Therefore, we obtain
\begin{equation}
\begin{split}
  U_{\text{out}} &=U_{\text{in}}W_{p} \Sigma^{-1}_{p} V^\top _{p}\check{W}_{\text{in}}\{W^\top _{\text{in}}(W_{p} \Sigma^{-1}_{p} V^\top _{p})^\top W_{p} \Sigma^{-1}_{p} V^\top _{p}W_{\text{in}}\}^{-\frac{1}{2}}\check{W}_{\text{in}}^\top \\
  &=U_{\text{in}}W_{p} \Sigma^{-1}_{p} V^\top _{p}\check{W}_{\text{in}}(W^\top _{\text{in}}V_{p}\Sigma^{-2}_{p} V^\top _{p}W_{\text{in}})^{-\frac{1}{2}}\check{W}_{\text{in}}^\top .
  \end{split}
\end{equation}
Since the matrix $W^\top _{\text{in}}V_{p}$ is regular matrix, the exponent $-\frac{1}{2}$ acts element-wise on the diagonal matrix as follows:
\begin{equation}
\begin{split}
  U_{\text{out}} &=U_{\text{in}}W_{p} \Sigma^{-1}_{p} V^\top _{p}\check{W}_{\text{in}}(W^\top _{\text{in}}V_{p}\Sigma_{p} V^\top _{p}W_{\text{in}})\check{W}_{\text{in}}^\top .
  \end{split}
\end{equation}
When we chose the singular vectors satisfying below,
\begin{equation}
\begin{split}
  \check{W}_{\text{in}}=W_{\text{in}},\\
    \check{V}_{\text{in}}=V_{\text{in}},
  \end{split}
\end{equation}
the matrix after mapping is 
\begin{equation}
\begin{split}
  U_{\text{out}} &=U_{\text{in}}W_{p} \Sigma^{-1}_{p} V^\top _{p}\check{W}_{\text{in}}(W^\top _{\text{in}}V_{p}\Sigma_{p} V^\top _{p}W_{\text{in}})\check{W}_{\text{in}}^\top \\
  &=U_{\text{in}}W_{p} \Sigma^{-1}_{p} V^\top _{p}V_{p}\Sigma_{p} V^\top _{p}\\
  &=U_{\text{in}}W_{p} \Sigma^{-1}_{p}\Sigma_{p} V^\top _{p}\\
  &=U_{\text{in}}W_{p}V^\top _{p}.  
  \end{split}
\end{equation}
We note that the matrix $W_{p}V^\top _{p}$ is regular. This matrix $R=W_{p}V^\top _{p}$ is equals to the result of procrastes problem of $U_o$ and $U_{\text{in}}$ presented below
\begin{equation}
\begin{split}
  R = \underset{\Omega  \in \mathbb{R}^{r \times r}} {\operatorname{argmin}} \left\| U_o - U_{\text{in}}\Omega \right\|.
  \end{split}
\end{equation}

\section{\label{apenb}Data preparation for ROM}
\subsection{Numerical method for solving governing equation}\label{numerical}
This paper uses the numerical schemes for incompressible Navier--Stokes equations for all cases. The incompressible governing equations are presented below.
\begin{equation}
\nabla\cdot\boldsymbol{u}=0,
   \label{eqcont}	
\end{equation}
\begin{equation}
\frac{\partial{\boldsymbol{u}}}{\partial{t}}=-(\boldsymbol{u}\cdot\nabla)\boldsymbol{u}-\frac{1}{\rho}\nabla{p}+\frac{1}{Re}{\nabla^2}\boldsymbol{u},
   \label{eqnavi}
\end{equation}
where $\boldsymbol{u}$ represents the velocity vector (bold symbols represent vectors), ${p}$ is the pressure, and $\rho$ is the fluid density.

The governing equations are solved by the fractional step method proposed by \cite{RKincomp}. The time step size is determined based on our previous validation \cite{12ICCFD}. 
The second-order central difference\cite{poisson} and the QUICK method \cite{QUICK} were used for evaluating the spatial differences. The details of these numerical procedures are described in the previous work\cite{mypaper2}. 

\subsection{Global stability analysis for first case}
The purpose of global stability analysis is finding the eigenmodes growing or decaying from basic state. Here, we consider the general governing equation denoted by  
\begin{equation}
\frac{\partial \boldsymbol{u}}{\partial t}= \mathcal{F}(\boldsymbol{u}),
   \label{general_eq}
\end{equation}
where $\mathcal{F}(\boldsymbol{u})$ is the generalized operator.
Global stability analysis considers the time evolution of the perturbation $\boldsymbol{\hat{u}}(t,\boldsymbol{x})$ around the given basic state $\boldsymbol{u}_b$ (base flow) governed by 
\begin{equation}
\frac{\partial \boldsymbol{\hat{u}}}{\partial t}=\left. \frac{\partial \mathcal{F}}{\partial \boldsymbol{u}}\right|_{\boldsymbol{u}=\boldsymbol{u}_{b}}\boldsymbol{\hat{u}}.
   \label{perturb_general}
\end{equation}
In this paper, the base state is chosen as a fixed point of the governing equations; for the Navier–Stokes equations, this corresponds to a time-independent (steady) solution. 

In discrete form, the global stability analysis considers time evolution of perturbation $\boldsymbol{\hat{u}}$ by the linearized evolution matrix $F$ derived from the Navier--Stokes equations around a base flow as follows
\begin{equation}
\frac{\partial \boldsymbol{\hat{u}}}{\partial t}=F\boldsymbol{\hat{u}}.
   \label{evoGSA}
\end{equation}

Assuming that the perturbation is represented by eigenvalues $\lambda$ and corresponding spatial distribution $\boldsymbol{\varphi}$ as follows 
\begin{equation}
\boldsymbol{\hat{u}}=\boldsymbol{\varphi}e^{\lambda t}.
   \label{eigen}
\end{equation}
In this manner, Eq. (\ref{evoGSA}) is
\begin{equation}
\lambda \boldsymbol{\varphi}=F\boldsymbol{\varphi}.
   \label{GSAeigenvalueproblem}
\end{equation}
Therefore, $\boldsymbol{\varphi}$ is the eigenvector (or eigenmode) of the operator $F$ characterized by a specific frequency and growth rate.

To compute the eigenvectors of the linearized operator, we use the time-stepping method\cite{Ohmichi_D,mypaper_7,Ranjan_2020}, which approximates the operator by numerically computing the time evolution of the perturbation. To approximate the time evolution matrix, the time-stepping method obtains the time progression of various perturbation fields from numerical simulations. A detailed derivation of our method is provided in our previous work\cite{mypaper_6, mypaper_7}.

We briefly describe the process of dataset acquisition in our time-stepping global stability analysis. First, an initial perturbation $\boldsymbol{\hat{u}}_0$ is prepared by 
\begin{equation}
\boldsymbol{\hat{u}}_{0} = {\epsilon}_0|\boldsymbol{u}_{b}|\frac{\boldsymbol{r}_0}{|\boldsymbol{r}_0|},
   \label{initialdist}
\end{equation}
where $|\cdot|$ represents the $L^2$ norm of the vector, $\boldsymbol{r}_0$ is the random disturbance, and $\epsilon_0$ is a parameter that can be set to any value, representing the ratio of the base flow and perturbation norms. The time evolution of the initial perturbation is obtained by numerical simuration presented below
\begin{equation}
\boldsymbol{u}'_{n}=\boldsymbol{\mathcal{F}}_{\text{CFD}}(\boldsymbol{u}_n),
   \label{evoCFD}
\end{equation}
where $\boldsymbol{\mathcal{F}}_{\text{CFD}}$ represents the time evolution matrix of numerical simulation by $\Delta T$, and the superscript $'$ means the variables after time evolution. $\boldsymbol{u}_n$ is represented using $\boldsymbol{u}_{b}$ as follows
\begin{equation}
\boldsymbol{u}_n = \boldsymbol{u}_{b}+\boldsymbol{\hat{u}}_n.
   \label{evoGSA2}
\end{equation}
The perturbation after time evolution $\boldsymbol{\hat{u}}'_n$ is obtained from
\begin{equation}
\boldsymbol{\hat{u}}'_n = \boldsymbol{u}'_n- \boldsymbol{u}_{b}.
   \label{evoGSA3}
\end{equation}
For the $n$th ($n > 0$) time evolution of numerical simulation, the perturbation $\boldsymbol{\hat{u}}_n$ to the $\boldsymbol{u}_{b}$ is computed from the flow field of $\boldsymbol{u}'_{n-1}$ presented below
\begin{equation}
\boldsymbol{\hat{u}}_{n} = {\epsilon}_0|\boldsymbol{u}_{b}|\frac{\boldsymbol{u}'_{n-1}-\boldsymbol{u}_{b}}{|\boldsymbol{u}'_{n-1}-\boldsymbol{u}_{b}|}.
   \label{redist}
\end{equation}
The iterative updating of the perturbation yields two matrices $X$ and $X'$, related by operators linearized around the basic state presented below
\begin{eqnarray}
X&=&[\boldsymbol{\hat{u}}_{Ns},\,\boldsymbol{\hat{u}}_{Ns+1},\,\cdots,\,\boldsymbol{\hat{u}}_{M+Ns-1}] \in \mathbb{R}^{N \times M},
   \label{matrixX}
\end{eqnarray}
\begin{eqnarray}
X'&=&[\boldsymbol{\hat{u}}'_{Ns},\,\boldsymbol{\hat{u}}'_{Ns+1},\,\cdots,\,\boldsymbol{\hat{u}}'_{M+Ns-1}]  \in \mathbb{R}^{N \times M},
   \label{matrixX'}
\end{eqnarray}
where $Ns$ is the starting position of the DMD data and is set to remove the effect of snapshots with initial randam disturbance. In this study, $N_s$ set to $150D/(U_{\infty}\Delta T)$. The number of snapshot $M$ set such that $NsM-1 = 50/(U_{\infty}\Delta T)$ based on our previous work\cite{mypaper2}. From these two dataset matrices $X$ and $X'$, eigenvectors and eigenvalues of the operators can be computed by using DMD. 

Numerical simulations to approximate the operator $\boldsymbol{\mathcal{F}}_{\text{CFD}}$ were performed using the numerical methods described in \ref{numerical}. The computational grid is the same as that used in our previous study\cite{mypaper_6}, and grid convergence results were also provided. In the wall-normal direction, the number of grid points was $240$, and in the wall-parallel direction, it was $590$. The height of the first layer adjacent to the cylinder surface was set to $1.0 \times 10^{-3}D$ based on the direct numerical simulation of Jiang et al.\cite{Jiang_2019,cylinder4}. 

\subsection{Periodic flow around a circular cylinder for the second case}
The numerical method and computational grid used for the flow around a two-dimensional circular cylinder are the same as those used in the first case. For all Reynolds numbers, a uniform flow is employed as the initial condition, and time-series data are collected after the flow has evolved sufficiently to reach a periodic state. The periodic state is confirmed by observing that the growth rates obtained from DMD applied to the time-series data are nearly zero. 

\subsection{Flow past an elliptic cylinder for the third case}\label{ellipse}
The grid around the elliptic cylinder is generated by applying a conformal mapping to the grid constructed around a circular cylinder. Since secondary vortex streets are known to form far downstream of the body\cite{Jiang_2019, Johnson_2004, shi2020wakes,variousshapes2}, the far-field boundary is placed at a distance of $650D$, where $D$ denotes the diameter of the circular cylinder before mapping. Following the grid design used in a previous study on secondary vortices behind a circular cylinder\cite{Jiang_2019}, the wake grid spacing is set to be less than $0.5D$ at $y=0$ line up to a distance of $500D$ from the cylinder. 

The conformal mapping employed in this study is defined by the complex function
\begin{equation}
T(z) = z + \frac{a^2}{z},
\label{eqjoukow}
\end{equation}
where $z=x+iy$ represents a complex number in the plane of the original grid, and $a$ is a complex constant with $|a| \leq 0.5$.
The shape of the resulting ellipse depends on the parameter $a$, which is determined by the signed aspect ratio $\Gamma$ as follows:
\begin{equation}
a^2 = \left(\frac{D}{2}\right)^2\frac{-|\Gamma|}{|\Gamma| - 2}.
   \label{sign_Gamma_a}
\end{equation}
where $\Gamma$ is the signed aspect ratio.
Once the aspect ratio $\Gamma$ and the original cylinder diameter $D$ are specified, the size and shape of the resulting ellipse are uniquely determined, regardless of the specific grid configuration around the cylinder before mapping. Therefore, we adopt the cylinder diameter $D$ before mapping as the characteristic length in this study\cite{mypaper2}. The other computational parameters are the same as periodic circular cylinder cases.

\bibliographystyle{elsarticle-num}

\end{document}